\documentclass{emulateapj}
\usepackage{graphicx,float}

\def\aap{A\&A}

\def\apjl{ApJ}

\def\apjs{ApJS}

\def\chandra{{\it Chandra}}
\def\cm{{\rm\thinspace cm}}

\def\erg{{\rm\thinspace erg}}

\def\km{{\rm\thinspace km}}

\def\Lsun{\hbox{$\rm\thinspace L_{\odot}$}}

\def\Mpc{{\rm\thinspace Mpc}}
\def\Msun{\hbox{$\rm\thinspace M_{\odot}$}}

\def\s{{\rm\thinspace s}}

\def\ergps{\hbox{$\erg\s^{-1}\,$}}

\def\kmps{\hbox{$\km\s^{-1}\,$}}
\def\kmpspMpc{\hbox{$\km\s^{-1}\Mpc^{-1}\,$}}

\def\pcmsq{\hbox{$\cm^{-2}\,$}}

\shorttitle{\sc{X-ray properties of MASSIVE Galaxies}}
\shortauthors{{\sc A.D.~Goulding et al.}}

\begin{document}


\title{The MASSIVE Survey IV.: \\The X-ray halos of the most massive
  early-type galaxies in the nearby Universe}


\author{Andy~D. Goulding\altaffilmark{1},
  Jenny~E. Greene\altaffilmark{1}, Chung-Pei Ma\altaffilmark{2},
  Melanie Veale\altaffilmark{2}, Akos Bogdan\altaffilmark{3}, Kristina
  Nyland\altaffilmark{4}, John~P. Blakeslee\altaffilmark{5},
  Nicholas~J.  McConnell\altaffilmark{5} \& Jens
  Thomas\altaffilmark{6}}

\email{Email:- goulding@astro.princeton.edu}

\affil{$^1$Department of Astrophysical Sciences, Princeton University, Princeton, NJ}
\affil{$^2$Department of Astronomy, University of California, Berkeley, CA}
\affil{$^3$Harvard-Smithsonian Center for Astrophysics, Cambridge, MA}
\affil{$^4$National Radio Astronomy Observatory, 520 Edgemont Road, Charlottesville, VA}
\affil{$^5$Herzberg Astronomy and Astrophysics, Victoria, BC, Canada}
\affil{$^6$Max-Planck Institute for Extraterrestrial Physics, Giessenbachstr. 1, Garching, Germany}


\begin{abstract}
  Studies of the physical properties of local elliptical galaxies are
  shedding new light on galaxy formation. Here we present the hot gas
  properties of 33 early-type systems within the MASSIVE galaxy survey
  that have archival {\it Chandra} X-ray observations, and use these
  data to derive X-ray luminosities ($L_{\rm X,gas}$) and plasma
  temperatures ($T_{\rm gas}$) for the diffuse gas components. We
  combine this with the ATLAS$^{\rm 3D}$ survey to investigate the
  X-ray--optical properties of a statistically significant sample of
  early-type galaxies across a wide range of environments. When X-ray
  measurements are performed consistently in apertures set by the
  galaxy stellar content, we deduce that all early-types (independent
  of galaxy mass, environment and rotational support) follow a
  universal scaling law such that
  $L_{\rm X,gas}\propto T_{\rm gas}^{\sim 4.5}$. We further
  demonstrate that the scatter in $L_{\rm X,gas}$ around both $K$-band
  luminosity ($L_K$) and the galaxy stellar velocity dispersion
  ($\sigma_e$) is primarily driven by $T_{\rm gas}$, with no clear
  trends with halo mass, radio power, or angular momentum of the
  stars. It is not trivial to tie the gas origin directly to either
  stellar mass or galaxy potential. Indeed, our data require a steeper
  relation between $L_{\rm X,gas}$, $L_K$, and $\sigma_e$ than
  predicted by standard mass-loss models. Finally, we find
  $T_{\rm gas}$ is set by the galaxy potential inside the optical
  effective radius. We conclude that within the inner-most 10--30~kpc
  region, early-types maintain pressure-supported hot gas, with a
  minimum $T_{\rm gas}$ set by the virial temperature, but the
  majority show evidence for additional heating.

\end{abstract}


\section{Introduction} \label{sec:intro}

Despite decades of active research, the assembly history and evolution
of the most massive elliptical galaxies in the Universe remains a
widely unanswered question. The current consensus is that the most
massive early-type galaxies formed the majority of their stars at
high-redshift ($z > 2$;
\citealt{Blakeslee:2003aa,Thomas:2005aa,vandokkumetal2010}), and have
since grown via mergers (e.g.,
\citealt{bell04,faber07,cowie08,bundyetal2010}). These mergers,
coupled with more minor accretion events of satellite systems, imparts
substantial quantities of gaseous material into the parent galaxy halo
(e.g., \citealt{vandokkumetal2008}). Elliptical galaxies are known to
possess relatively low star-formation rates, typically orders of
magnitude below those found in their spiral counterparts. Hence, it is
unclear which mechanisms prevent the gas from cooling, and forming new
stars, in early-type galaxies.

Massive elliptical galaxies are surrounded by a hot interstellar
medium (ISM) that may have initially formed due to shock heating of
the cooling gas during its entrapment within the galaxy's deep
potential (\citealt{Forman:1985aa, Trinchieri:1985aa}) and/or from
stellar mass that was lost from evolved stars (e.g.,
\citealt{Mathews:1990aa,Brighenti:1997aa,Mathews:2003aa}). These gas
halos are often extended well beyond the stellar distribution of the
galaxy, and can be used to probe the dark halo mass by assuming that
the gas is in virial equilibrium out to large radii
(\citealt{Canizares:1987aa}).

Two of the most fundamental measurables of the X-ray emitting thermal
ISM are its luminosity ($L_{\rm X,gas}$) and gas temperature
($T_{\rm gas}$). These observables reflect the physics related to the
baryons within the system, and are directly linked to the total
entropy and gas mass. With the advent of the Einstein and ROSAT X-ray
telescopes, it became possible to study the hot ISM of statistically
significant samples of ellipticals. Thus, it is now well established
that $L_{\rm X,gas}$ and $T_{\rm gas}$ are strongly correlated in
massive virialized clusters, going as
$L_{\rm X,gas} \propto T_{\rm gas}^{2-3}$ (e.g.,
\citealt{Mitchell:1979aa,Mushotzky:1984aa,Edge:1991aa,Markevitch:1998aa,Ettori:2004aa,Mittal:2011aa,Maughan:2012aa}). On
group/poor-cluster scales, the relation steepens
($L_{\rm X,gas}\propto T_{\rm gas}^{3-4}$;
\citealt{Wu:1999aa,Helsdon:2000aa,Mulchaey:2000ab,Osmond:2004aa}) as
baryon physics (e.g., feedback from active galactic nuclei;
galaxy-scale winds; cooling flows) begins to dominate over gravity,
causing deviations from the self-similar scaling relation
($L_{\rm X,gas} \propto T_{\rm gas}^2$) expected in fully
gravitationally collapsed systems. The $L_{\rm X,gas}$--$T_{\rm gas}$
relation may further steepen in the poorest-groups and individual
galaxy halos as the hot gas surrounding the galaxies is no longer
virialized and/or is dominated by other effects. Previous studies have
sought to link these differing structural scales to provide a coherent
picture of the X-ray properties of the most massive systems in the
Universe (e.g.,
\citealt{OSullivan:2003aa,Khosroshahi:2007aa,Sun:2007aa,Diehl:2008ac,Sun:2009aa,Mulchaey:2010aa,Vajgel:2014aa}). However,
observational evidence for the hot gas origin is still in question. It
has yet to be well established to what extent the gas properties are
influenced by the effects of cooling versus external agents (such as
accretion, stripping, sloshing, shocks), and at a given stellar mass,
the dependence of the gas luminosity and/or temperature on environment
(e.g.,
\citealt{White:1991ab,Brown:2000aa,OSullivan:2001aa,Helsdon:2001aa}).

Given their substantially lower $L_{\rm X,gas}$, the hot gas
properties of individual elliptical galaxies are less well
known. However, successive generations of X-ray telescopes, with
improved sensitivity, larger effective areas, and superior spatial
resolution are now beginning to allow a more in-depth study of the hot
gas component in unprecedented detail, which in turn sheds new light
on the formation history of galaxies. Using the excellent angular
resolution of the {\it Chandra} X-ray telescope, studies are directly
detecting and isolating the X-ray emission from stellar processes such
as the low mass X-ray binary population that have plagued the previous
generation of X-ray telescopes, particularly in the low stellar-mass
regime (e.g., \citealt{Sivakoff:2004aa,Boroson:2011aa,Lehmer:2014aa}),
as well as exploring the X-ray halos in early-type galaxies as a
function of cosmic time (e.g.,
\citealt{Lehmer:2007aa,Civano:2014aa,Paggi:2015aa}), and probing the
thermal gas content (e.g.,
\citealt{Diehl:2007aa,Jeltema:2008aa,Memola:2009aa,Bogdan:2012aa,Bogdan:2012ab,Kim:2013aa,Su:2015aa,Kim:2015aa}).

On a galaxy-wide scale, insight into the sources and retention of the
hot gas may be gained by correlating the X-ray properties and optical
properties of the galaxies. Given the expected origins (e.g., stellar
mass loss; stellar feedback) of the hot gas, to zeroth order, one
might imagine that $L_{\rm X,gas}$ would correlate with
the stellar mass ($M_*$) of the galaxy. However, the observed scatter
in the $L_{\rm X,gas} - M_*$ relation is larger than this simple expectation,
requiring subsequent studies to assess if additional physical
parameters can explain the scatter in the relation. Obvious physical
properties that may drive the scatter include the large-scale
environment, the presence of an AGN, and the rotation properties of
the host galaxy.

In general, early-type galaxies are thought to fall into two families
(e.g.,
\citealt{Bender:1989aa,Kormendy:1996aa,Kormendy:2012aa}). Typically
lower-mass galaxies tend to have net rotation (fast rotators;
\citealt{emsellemetal2007}), disky isophotes, and a power-law light
distribution that continues all the way to their centers (e.g.,
\citealt{Faber:1997aa,Lauer:2007aa}). By contrast, the most massive
elliptical galaxies tend to be completely dispersion dominated (slow
rotators), have boxy isophotes, and have a light deficit in their
center known as a core (e.g.,
\citealt{Ferrarese:1994aa,Lauer:1995aa,Ferrarese:2006aa,Kormendy:2009ab}). Generally,
it is thought that the fast rotators have dissipation in their most
recent merger history, while slow rotators have suffered a number of
dry mergers, and binary black hole scouring has left a core in their
center (e.g., \citealt{Krajnovic:2013aa} and references
there-in). Integral field unit (IFU) observations provide
high-fidelity measurements of the rotation/angular-momentum of
early-type galaxies (\citealt{Emsellem:2011aa}).

Earlier work has focused on the ATLAS$^{\rm 3D}$ sample (e.g.,
\citealt{Emsellem:2011aa,Cappellari:2013ab,Krajnovic:2011aa,Young:2011aa}),
a volume-limited sample ($D < 42$~Mpc) of all nearby early-type
galaxies with $M_K < -21.5$~mags. Using the ATLAS$^{\rm 3D}$ galaxies,
there has been suggestive evidence that galaxies with net rotation
(i.e., fast rotators) have lower $L_X$ than slow rotators at a given
stellar mass. However, due in part to the limited volume, the
ATLAS$^{\rm 3D}$ sample contains few slow rotating systems with
available high-resolution X-ray data, and virtually no galaxies with
stellar masses above $M_* > 10^{11} \Msun$.  Complementary to these
previous multi-wavelength studies, here we focus our study on the hot
gas emission present in the volume-limited MASSIVE survey
(\citealt{Ma:2014aa}; M14 hereafter) of the 116 most massive
early-type galaxies within 108 Mpc. Similar to ATLAS$^{\rm 3D}$,
MASSIVE uses integral-field spectroscopy (IFS) combined with uniform
$K$-band imaging to study the stellar populations
\citep{greeneetal2015} and stellar kinematics (Veale et al., in prep.)
of the MASSIVE galaxies.

In this paper, we investigate the connection between the X-ray and
optical properties of massive early-type galaxies in the nearby
Universe by combining the volume-limited MASSIVE sample with the
nearby sample of ellipticals in the ATLAS$^{\rm 3D}$ survey. In
section 2 we present the MASSIVE sample, the optical spectral
measurements from our integral field unit spectroscopy, and the
comparison sample derived from the ATLAS$^{\rm 3D}$ survey. In section
3, we provide the data reduction and spectral fitting analyses of the
{\it Chandra} X-ray data for the 33 MASSIVE galaxies with appropriate
X-ray data. In section 4, we discuss the X-ray and optical properties
of the combined MASSIVE and ATLAS$^{\rm 3D}$ samples, and show
evidence that when the X-ray luminosity and gas temperatures are
measured within physically-motivated apertures, set by the galaxy
stellar content, these measurables correlate significantly, and are
driven primarily by the small-scale galaxy potential. Finally, in
section 5 we summarize our findings. Throughout the manuscript we
adopt a flat $\Lambda$CDM cosmology with $H_0 = 71 \kmpspMpc$ and
$\Omega_M = 0.3$.

\section{The MASSIVE Galaxy Sample and Auxiliary Data} \label{sec:sample}

\subsection{The MASSIVE Survey} \label{sec:massive}

MASSIVE is a volume-limited sample of the 116 most massive early-type
galaxies within 108~Mpc. The selection and sample properties are
described in detail in \citet{Ma:2014aa}; we summarize salient
details here for completeness. To emulate a stellar mass selection, we
select galaxies from the 2MASS Redshift Survey
\citep[2MRS;][]{huchraetal2012} that are brighter than a total
$K-$band magnitude limit of $M_K < -25.3$ mag
($\approx 10^{11.5}$~$M_{\odot}$).  We additionally apply a
morphological cut based on the Hyperleda database
\citep{patureletal2003} to remove large spiral and interacting
galaxies. The large-scale environment, group numbers and halo masses
are taken from the 2MASS-based high-density contrast (HDC) group
catalog of \citet{crooketal2007}. The HDC catalog is constructed using
a Friends-of-Friends algorithm; it contains lists of galaxy membership
in groups that have a density contrast of $\geq 80$, and is complete
to $\sim L^*$ galaxy mass at $\approx 70$~Mpc. The resulting 116
early-type galaxies span a range of stellar velocity dispersion
($\sim 180-400$~km/s) and are found in a wide array of environments,
from ``field'' galaxies with no $\sim L^*$ companions to rich clusters
(Coma, Perseus, and Virgo).

The core science goals of the MASSIVE survey are driven by the need
for wide-format integral field unit spectroscopy taken with the
Mitchell Spectrograph (\citealt{hilletal2008b}) on the 2.7m Harlan
J. Smith telescope at the McDonald Observatory. The large field of
view of the Mitchell Spectrograph (107$\times$107\arcsec) is
well-suited to the study of massive nearby galaxies.  To date, we have
IFU spectroscopy for 70 of the 116 galaxies in the MASSIVE
survey. With these data we are measuring stellar population gradients
\citep{greeneetal2013,greeneetal2015}, molecular gas masses and
star-formation efficiencies \citep{Davis:2016aa}, stellar kinematic
fields (Veale et al.\ in prep), and when combined with modeling, the
enclosed dark matter halo mass (e.g. \citealt{thomasetal2011}).

\begin{figure}[!t]
\centering
\includegraphics[width=\linewidth]{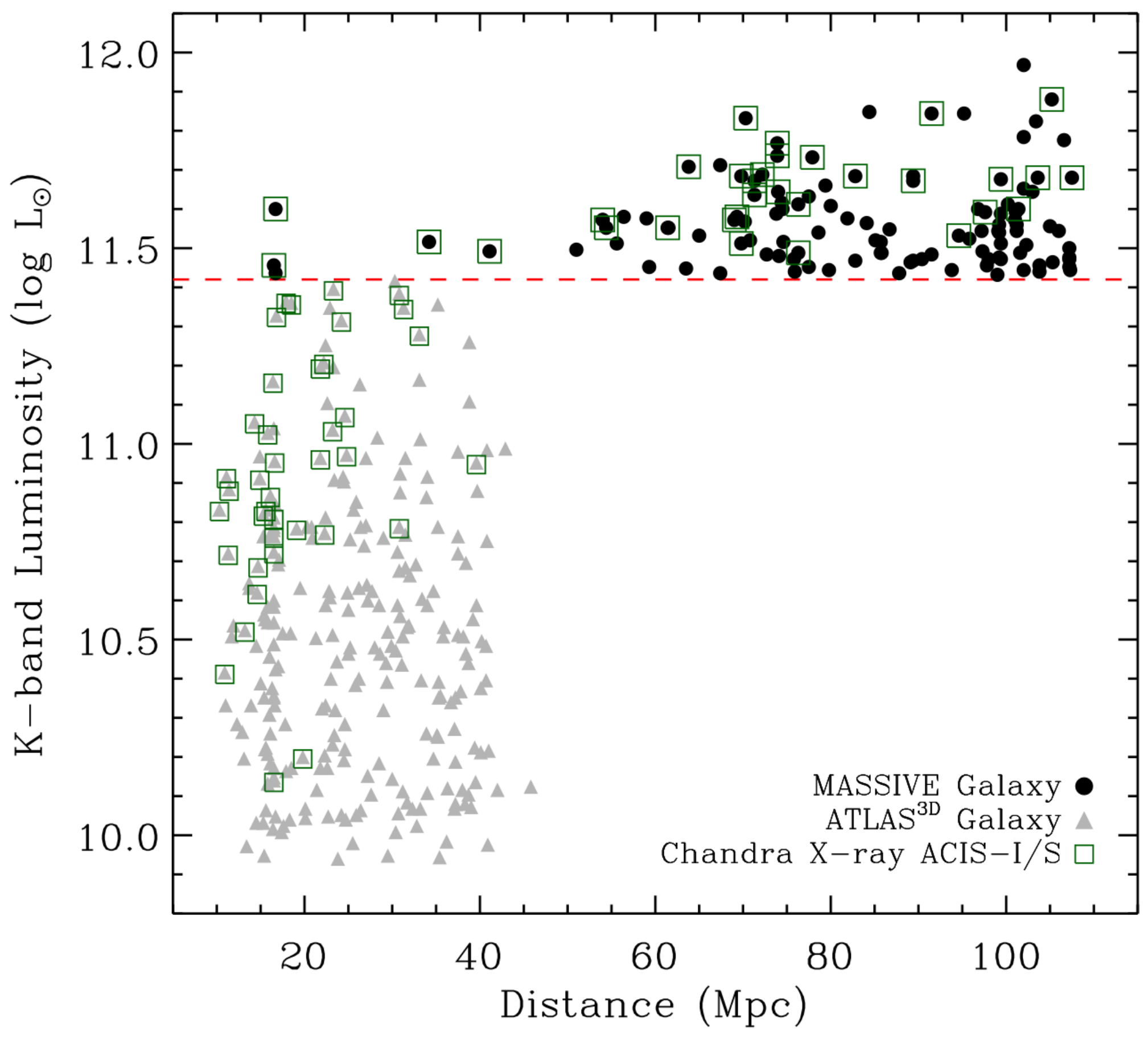}
\caption{Logarithm of 2MASS total $K$-band luminosity in units of
  solar luminosities versus distance for the MASSIVE galaxy sample
  (filled circles). The red dashed line represents the total $K$-band
  cut of $M_K < −25.3$ magnitudes, which was used to define the
  initial MASSIVE sample selection presented in Ma et al. (2014). As a
  comparison, the ATLAS$^{\rm 3D}$ galaxy sample is shown with gray
  triangles. The sub-sample of galaxies with publicly available
  archival {\it Chandra} X-ray ACIS-I or ACIS-S observations are
  additionally highlighted with open green squares.}
\label{fig:sample}
\end{figure}

\subsection{Effective Radii Measurements}
\label{sec:re}

We adopt single \citet{sersic63} fits to the light profiles to derive
the half-light radii. For roughly two thirds of our sample, we have
Sloan Digital Sky Survey \citep[SDSS;][]{yorketal2000} imaging. For
these galaxies, we use the NASA-Sloan Atlas \citep{blanton09} catalog
effective-radius ($R_e$) measurements. For the remaining galaxies that
do not have SDSS imaging, we adapt a size measurement from 2MASS
\citep{jarrettetal2003}. We fit a linear conversion relation to
correct the 2MASS measurements to match the SDSS measurements, such
that
$\rm{log}_{10} (R_{\rm e,2MASS}) = 0.8 \rm{log}_{10} (R_{\rm e,SDSS})
- 0.076$
\citep[][their Eq. 4]{Ma:2014aa}, where $R_{\rm e,2MASS}$ is the
median effective radius measured from the $JHK$ band 2MASS
imaging.\footnote{We note that it is well established that the 2MASS
  size measurement typically underestimates the galaxy size relative
  to the SDSS measurement \citep[][]{Lauer:2007aa,Kormendy:2009ab}}

\begin{table*}
\footnotesize
\begin{center}
\setlength{\tabcolsep}{0.95mm}
\caption{The MASSIVE X-ray sample\label{tbl:sample}}
\begin{tabular}{lcccrrrrrccccccc}
\tableline\tableline
\multicolumn{1}{c}{Common} &
\multicolumn{1}{c}{$\alpha_{\rm J2000}$} &
\multicolumn{1}{c}{$\delta_{\rm J2000}$} &
\multicolumn{1}{c}{$z$} &
\multicolumn{1}{c}{$D_L$} &
\multicolumn{1}{c}{$L_K$} &
\multicolumn{1}{c}{$N_{H,gal}$} &
\multicolumn{1}{c}{$\sigma_e$} &
\multicolumn{1}{c}{$R_e$} &
\multicolumn{1}{c}{$\epsilon$} &
\multicolumn{1}{c}{$\lambda_{R_e}$} &
\multicolumn{1}{c}{X-ray} &
\multicolumn{1}{c}{\# of } &
\multicolumn{1}{c}{Cent.} &
\multicolumn{1}{c}{$M_{\rm Halo}$} & 
\multicolumn{1}{c}{$P_{\rm 1.4GHz}$} \\

\multicolumn{1}{c}{Name} &
\multicolumn{2}{c}{(degrees)} &
\multicolumn{1}{c}{} &
\multicolumn{1}{c}{(Mpc)} &
\multicolumn{1}{c}{($L_{\odot}$)} &
\multicolumn{1}{c}{$\times 10^{20}$} &
\multicolumn{1}{c}{(km/s)} &
\multicolumn{1}{c}{($''$)} &
\multicolumn{1}{c}{} &
\multicolumn{1}{c}{} &
\multicolumn{1}{c}{AGN?} &
\multicolumn{1}{c}{Neighbor} &
\multicolumn{1}{c}{Gal?} & 
\multicolumn{1}{c}{(log $M_{\odot}$)} &
\multicolumn{1}{c}{(log W/Hz)} \\

\multicolumn{1}{c}{(1)} &
\multicolumn{2}{c}{(2)} &
\multicolumn{1}{c}{(3)} &
\multicolumn{1}{c}{(4)} &
\multicolumn{1}{c}{(5)} &
\multicolumn{1}{c}{(6)} &
\multicolumn{1}{c}{(7)} &
\multicolumn{1}{c}{(8)} &
\multicolumn{1}{c}{(9)} &
\multicolumn{1}{c}{(10)} &
\multicolumn{1}{c}{(11)} &
\multicolumn{1}{c}{(12)} &
\multicolumn{1}{c}{(13)} &
\multicolumn{1}{c}{(14)} &
\multicolumn{1}{c}{(15)} \\

\tableline

NGC0057	& 3.8787   &	17.3284  &	0.0181	& 	76.3	& 11.61	& 	3.86	& 	253	& 	27.0	& 	0.17	& 	0.02	& - &	1	& 	- & $<$11.0	& 20.30 \\
NGC0315	& 14.4538  &	30.3524  & 	 0.0165	& 	70.3	& 11.83	& 	5.83	& 	341	& 	25.1	& 	0.28	& 	0.06	& $\checkmark$ &	6	& $\checkmark$	 & 13.4	& 23.66 \\
NGC0383	& 16.8540  &	32.4126  &	 0.0170	& 	71.3	& 11.63	& 	5.40	& 	257	& 	20.5	& 	0.14	& 	0.25	& $\checkmark$ &	29	& 	- & 14.3	& 23.82 \\
NGC0410	& 17.7453  &	33.1520	 &	0.0177	& 	71.3	& 11.67	& 	5.34	& 	246	& 	31.6	& 	0.25	& 	0.04	& - &	29	& 	$\checkmark$ & 14.3	& 21.54 \\
NGC0499	& 20.7978  &	33.4601	 &	0.0147	& 	69.8	& 11.51	& 	5.33	& 	266	& 	15.6	& 	-	& 	-	& - &	35	& 	- & 14.5	& 20.22 \\
NGC0507	& 20.9164  &	33.2561	 &	0.0165	& 	69.8	& 11.68	& 	5.25	& 	257	& 	38.4	& 	0.09	& 	0.05	& $\checkmark$ &	35	& 	$\checkmark$ & 14.5	& 20.22 \\
NGC0533	& 21.3808  &	1.7590	 &	 0.0185	& 	77.9	& 11.73	& 	3.00	& 	259	& 	40.7	& 	0.26	& 	0.04	& - &	3	& 	$\checkmark$ & 13.8	& 22.31 \\
NGC0547	& 21.5024  &	-1.3451	 &	 0.0178	& 	74.0	& 11.64	& 	4.03	& 	232	& 	19.7	& 	0.14	& 	0.07	& - &	32	& 	- & 14.3	& 22.59$\dagger$ \\
NGC0708	& 28.1937  &	36.1518	 &	 0.0162	& 	69.0	& 11.57	& 	5.30	& 	209	& 	65.1	& 	0.40	& 	0.05	& - &	39	& 	$\checkmark$ & 14.5	& 22.57 \\
NGC0741	& 29.0874  &	5.6289	 &	 0.0185	& 	73.9	& 11.74	& 	4.50	& 	291	& 	26.9	& 	0.17	& 	-	& - &	5	& 	$\checkmark$ & 13.7	& 21.91$\dagger$ \\
NGC0777	& 30.0622  &	31.4294	 &	 0.0167	& 	72.2	& 11.69	& 	4.80	& 	292	& 	18.6	& 	0.17	& 	0.05	& - &	7	& 	$\checkmark$ & 13.4	& 21.64 \\
NGC1129	& 43.6141  &	41.5796	 &	 0.0173	& 	73.9	& 11.77	& 	9.91	& 	261	& 	30.2	& 	0.04	& 	0.07	& - &	33	& 	$\checkmark$ & 14.7	& 20.27 \\
NGC1132	& 43.2159  &	-1.2747	 &	 0.0231	& 	97.6	& 11.59	& 	5.46	& 	215	& 	30.9	& 	0.37	& 	0.05	& - &	3	& 	$\checkmark$ & 13.6	& 21.78 \\
NGC1600	& 67.9161  &	-5.0861	 &	 0.0156	& 	63.8	& 11.71	& 	4.80	& 	303	& 	55.3	& 	0.26	& 	0.02	& - &	16	& 	$\checkmark$ & 14.2	& 22.47 \\
NGC1700	& 74.2347  &	-4.8658	 &	 0.0130	& 	54.4	& 11.55	& 	4.00	& 	231	& 	31.9	& 	0.28	& 	0.14	& - &	4	& 	$\checkmark$ & 12.6	& 20.00 \\
NGC2340	& 107.7950 &	50.1747	 &	 0.0198	& 	89.4	& 11.67	& 	7.41	& 	231	& 	51.7	& 	0.44	& 	0.03	& - &	18	& 	- & 14.1	& 20.92$\dagger$ \\
NGC2783	& 138.4145 &	29.9929	 &	0.0225	& 	101.4	& 11.60	& 	2.10	& 	268	& 	38.2	& 	0.39	& 	0.05	& - &	3	& 	$\checkmark$ & 12.3	& 22.47 \\
NGC2832	& 139.9453 &	33.7498	 &	0.0232	& 	105.2	& 11.88	& 	1.59	& 	295	& 	21.2	& 	0.31	& 	0.05	& - &	4	& 	$\checkmark$ & 13.8	& 20.58 \\
NGC3209	& 155.1601 &	25.5050	 &	0.0209	& 	94.6	& 11.53	& 	1.95	& 	303	& 	29.4	& 	-	& 	-	& - &	3	& 	$\checkmark$ & 11.9	& 21.07$\dagger$ \\
NGC3842	& 176.0090 &	19.9498	 &	 0.0211	& 	99.4	& 11.67	& 	2.00	& 	230	& 	24.2	& 	0.22	& 	0.03	& - &	42	& 	$\checkmark$ & 14.7	& 22.15 \\
NGC4073	& 181.1128 &	1.8960	 &	0.0196	&  	91.5	& 11.84	& 	1.80	& 	289	& 	23.0	& 	0.32	& 	0.02	& - &	10	& 	$\checkmark$ & 14.0	& 20.46 \\
NGC4472	& 187.4450 &	8.0004	 &	0.0033	& 	16.7	& 11.60	& 	1.66	& 	289	& 	181.8	& 	0.17	& 	0.08	& - &	205	& 	$\checkmark$ & 15.0	& 21.86 \\
NGC4555	& 188.9216 &	26.523	 &	0.0223	& 	103.6	& 11.68	& 	1.36	& 	268	& 	29.8	& 	0.20	& 	0.13	& - &	1	& 	- & $<$11.0	& 21.40$\dagger$ \\
NGC4649	& 190.9167 &	11.5526	 &	 0.0037	& 	16.5	& 11.46	& 	2.20	& 	340	& 	44.1	& 	0.16	& 	0.13	& - &	205	& 	- & 15.0	& 20.98 \\
NGC5129	& 201.0417 &	13.9765	 &	 0.0230	& 	107.5	& 11.68	& 	1.70	& 	222	& 	21.8	& 	0.37	& 	0.41	& - &	1	& 	- & $<$11.0	& 21.99 \\
NGC5322	& 207.3133 &	60.1904	 &	 0.0059	& 	34.2	& 11.52	& 	1.64	& 	239	& 	20.1	& 	0.33	& 	0.07	& $\checkmark$ &	8	& 	$\checkmark$ & 13.5	& 22.04 \\
NGC5353	& 208.3613 &	40.2831	 &	 0.0078	& 	41.1	& 11.49	& 	0.98	& 	290	& 	27.8	& 	-	& 	-	& - &	12	& 	$\checkmark$ & 13.3	& 21.91 \\
NGC6482	& 267.9534 &	23.0719	 &	 0.0131	& 	61.4	& 11.55	& 	7.74	& 	301	& 	22.4	& 	0.36	& 	0.10	& - &	3	& 	$\checkmark$ & 13.0	& 20.11 \\
NGC7052	& 319.6377 &	26.4469	 &	 0.0156	& 	69.3	& 11.58	& 	13.60	& 	284	& 	35.8	& 	0.50	& 	-	& $\checkmark$ &	1	& 	- & $<$11.0	& 22.96 \\
NGC7265	& 335.6145 &	36.2098	 &	 0.0170	& 	82.8	& 11.68	& 	12.90	& 	214  	& 	39.8	& 	0.22	& 	0.05	& - &	21	& 	$\checkmark$ & 14.6	& 20.37 \\
NGC7618	& 349.9468 &	42.8526	 &	 0.0173	& 	76.3	& 11.49	& 	11.90	& 	298	& 	21.6	& 	-	& 	-	& $\checkmark$ &	10	& 	$\checkmark$ & 13.6	& 22.42 \\
NGC7619	& 350.0605 &	8.2063	 &	 0.0125	& 	54.0	& 11.57	& 	5.00	& 	273	& 	34.6	& 	0.23	& 	0.13	& - &	12	& 	$\checkmark$ & 13.9	& 21.84 \\
NGC7626	& 350.1772 &	8.2170	 &	 0.0114	& 	54.0	& 11.57	& 	5.03	& 	251	& 	26.7	& 	0.14	& 	0.03	& $\checkmark$ &	12	& 	- & 13.9	& 22.96 \\				

\tableline
\end{tabular}
\end{center}
{\bf Notes:-} $^{1}$Common galaxy name;
$^{2}$J2000 positional co-ordinates from the NASA Extra-galactic Database (NED);
$^{3}$Spectroscopic redshift;
$^{4}$Luminosity distance in Megaparsecs assuming $H_0 = 71 \kmpspMpc$ and $\Omega_{\Lambda} = 0.70$ and corrected for non-cosmological flows;
$^{5}$Extinction-corrected total absolute $K$-band luminosity;
$^{6}$Galactic Neutral hydrogen column density in units of $10^{20} \pcmsq$;
$^{7}$Stellar velocity dispersion in units of km/s (Veale et al. in prep.);
$^{8}$Effective radius from optical and near-IR imaging in units of arc-seconds;
$^{9}$Ellipticity, such that $\epsilon = 1 - b/a$;
$^{10}$2-D galaxy rotation parameter, such that $\lambda_R \sim \langle R |V| \rangle / \langle R \sqrt{V^2 +
  \sigma_{*}^2}$ measured in our IFU data (Veale et al. in prep.). Those sources marked with `-' currently lack sufficient IFU data for an accurate determination of their rotation parameter;
$^{11}$Evidence for an X-ray point source in the nuclear region in the 4--7~keV band;
$^{12}$Number of neighbors present in the galaxy group/cluster;
$^{13}$Is the source considered to be at the center of the group/cluster?;
$^{14}$Logarithmic total mass of the large environment halo in solar masses (Crook et al. 2007);
$^{15}$Logarithmic radio power at 1.4~GHz in units of W/Hz from the NVSS or FIRST ($\dagger$) catalog.\\
\normalsize
\end{table*}

\subsection{Spectral Measurements}
\label{sec:kinematics}

The galaxy kinematics are derived from the optical IFU spectra. Full
details of our kinematic measurements will be described in Veale et
al.\ in preparation, but for completeness we summarize the data and
analysis methodology here. Veale et al. bin the fibers in the galaxies
to ensure a S/N$>20$ per spatial bin. In conjunction with the MILES
(\citealt{Sanchez-Blazquez:2006aa,Falcon-Barroso:2011aa}) stellar
template library, they use the penalized pixel-fitting (pPXF) method
\citep{cappellariemsellem2004} to measure the line-of-sight velocity
distribution, which is parameterized using Gauss-Hermite
expansions. Six Gauss-Hermite moments are fit for each spectrum to
measure the velocity ($V$), velocity dispersion ($\sigma_{*}$), and
the subsequent higher-order kinematical moments $h_3, h_4, h_5, h_6$.

In this work, we will use the luminosity-weighted $\sigma_{*}$
measurements within $R_e$ from Veale et al. We are also interested in
the angular momentum content of the galaxies. As in the
ATLAS$^{\rm 3D}$ project, we use a two-dimensional version of
$V/\sigma$,
$\lambda_R \sim \langle R |V| \rangle / \langle R \sqrt{V^2 +
  \sigma_{*}^2}$
\citep{binney05,emsellemetal2007}, to quantify the angular momentum
content of the galaxies as a function of
radius. \citet{emsellemetal2013} define the threshold between slow and
fast rotators, as $\lambda_R({\rm slow}) < 0.3 \sqrt{\epsilon}$, and
we adopt their definition here for ease of comparison with the
ATLAS$^{\rm 3D}$ (see Table 1). At least eighteen out of the 33 MASSIVE
galaxies studied here are unequivocally slow rotators.

\subsection{The Archival \chandra\ X-ray Sample}
\label{sec:chandrasample}

The Advanced CCD Imaging Spectrometer (ACIS) on-board the {\it
  Chandra} X-ray Observatory provides an exquisite view of the X-ray
universe to an unprecedented angular resolution
($\sim 0.49''/{\rm pixel}$). For example, the 75\% encircled energy
fraction (EEF) of XMM-{\it Newton} is $\sim 20''$, which corresponds
to $\sim 7$~kpc at 70~Mpc. By contrast, \chandra\ has a 90\% EEF
within $1''$ at $E\sim0.5$--2~keV. Thus, {\it Chandra} has been used
to great effect to accurately measure the X-ray properties of local
early-type systems, and directly remove contaminants (e.g., from
stellar sources) from measurements of the diffuse gas halo. To ensure
a more homogeneous dataset, and to provide the most accurate
measurements of the hot ISM, here we choose to harness the {\it
  Chandra} observatory to provide the X-ray imaging and spectroscopy
for the MASSIVE galaxies.

To date, 39 of the 116 galaxies in the MASSIVE sample have publicly
available {\it Chandra} ACIS-I and/or ACIS-S observations. These
observations were performed as part of a multitude of different
scientific programs. These sources cover a wide range of total
exposure times ($\sim 2$--300~ks), with some observations occurring
more than a decade apart. While heterogeneous in construction, this
X-ray sample still provides a representative subset ($\approx$ 1/4) of
the whole MASSIVE survey. Of the 39 MASSIVE galaxies with available
data, we determine that the {\it Chandra} observations for three
galaxies (NGC~3862, NGC~5252, M87) are heavily contaminated by a
central point-source that dominates the X-ray emission, and are thus
subject to significant pile-up effects. Hence, we choose to remove
these three sources from the sample, given the significant measurement
uncertainties introduced due to modeling of the pile-up effects and
separation of the AGN and diffuse emission.\footnote{We note that two
  further galaxies (NGC~315; NGC~383) have prominent X-ray emission
  produced by central AGN. These observations are not subject to
  significant pile-up and do not contribute significantly to the total
  X-ray emission. Hence, we accurately include the AGN contribution to
  the diffuse ISM component in our spectral fitting analysis. Thus,
  our MASSIVE X-ray galaxy sample does not include an explicit bias
  against systems that are actively growing their central supermassive
  black holes.} We also do not include satellites within the Coma
cluster due to the difficulties in separating the galaxy ISM from the
intra-cluster medium at $\sim 100$~Mpc (e.g., NGC~4874; NGC~4889) or
the galaxy ISM being subject to ram-pressure stripping (e.g.,
NGC~4839). Our final X-ray sample consists of 33 galaxies (see Figure 1).

\subsection{ATLAS$^{\rm 3D}$: A representative lower mass comparison
  sample of nearby early-type galaxies}
\label{sec:atlas3d}

To place the MASSIVE galaxy sample into the wider context of galaxy
evolution, we require a comparison sample of lower mass early-type
systems that have similar quality optical and X-ray data. In order to
mitigate systematic effects that can dominate samples that have been
gathered from multiple datasets and/or investigations, this comparison
sample should be derived from a single homogeneous survey. The
ATLAS$^{\rm 3D}$ project is a volume-limited sample of all nearby
early-type galaxies within 42~Mpc selected from the 2MRS with
$M_K < -21.5$~mags (see \citealt{Cappellari:2011aa}). The
ATLAS$^{\rm 3D}$ sources have optical IFU spectroscopy using the
SAURON instrument (Bacon et al. 2001) on the William Herschel
Telescope. The near-infrared selection of systematically lower mass
systems combined with their optical ancillary data makes the
ATLAS$^{\rm 3D}$ project the ideal comparison sample for MASSIVE. Most
recently, \cite{Kim:2015aa} presented constrained X-ray temperature
and luminosity measurements for 49 (of 61) early-type galaxies in
ATLAS$^{\rm 3D}$ that have observations with the {\it Chandra} X-ray
telescope. These X-ray measurements were performed using large
extraction apertures irrespective of the angular size of the galaxy
observed in the optical or near-IR. Complementary to the study of
\cite{Kim:2015aa}, \cite{Su:2015aa} present X-ray measurements for a
sub-sample of the \cite{Kim:2015aa} objects with X-ray measurements
performed within apertures defined by the effective radius of the
galaxy (see Section~3 of this work). As a comparison sample to our
MASSIVE survey, we include measurements from both of these studies, as
well as additional X-ray measurements for other ATLAS$^{\rm 3D}$
galaxies, and updated measurements for a further subset of the objects
presented in \cite{Su:2015aa} (see Appendix). After removal of objects
that have X-ray measurements adversely affected by AGN emission and
source overlap between MASSIVE and ATLAS$^{\rm 3D}$, our comparison
ATLAS$^{\rm 3D}$ galaxy sample contains 41 sources.

\section{Chandra X-ray analysis}

In this section we describe the data reduction, imaging analysis, and
spectral analysis of the 33 MASSIVE galaxies with publicly available
archival {\it Chandra} X-ray observations. Of the 33 galaxies, 32 have
observations performed in the ACIS-S mode, while NGC~1129 was
performed in ACIS-I. The individual observation identification numbers
(ObsIDs), and the basic observation information are provided in Table
2.

\begin{figure*}[!ht]

  \centering
  \includegraphics[width=0.85\linewidth]{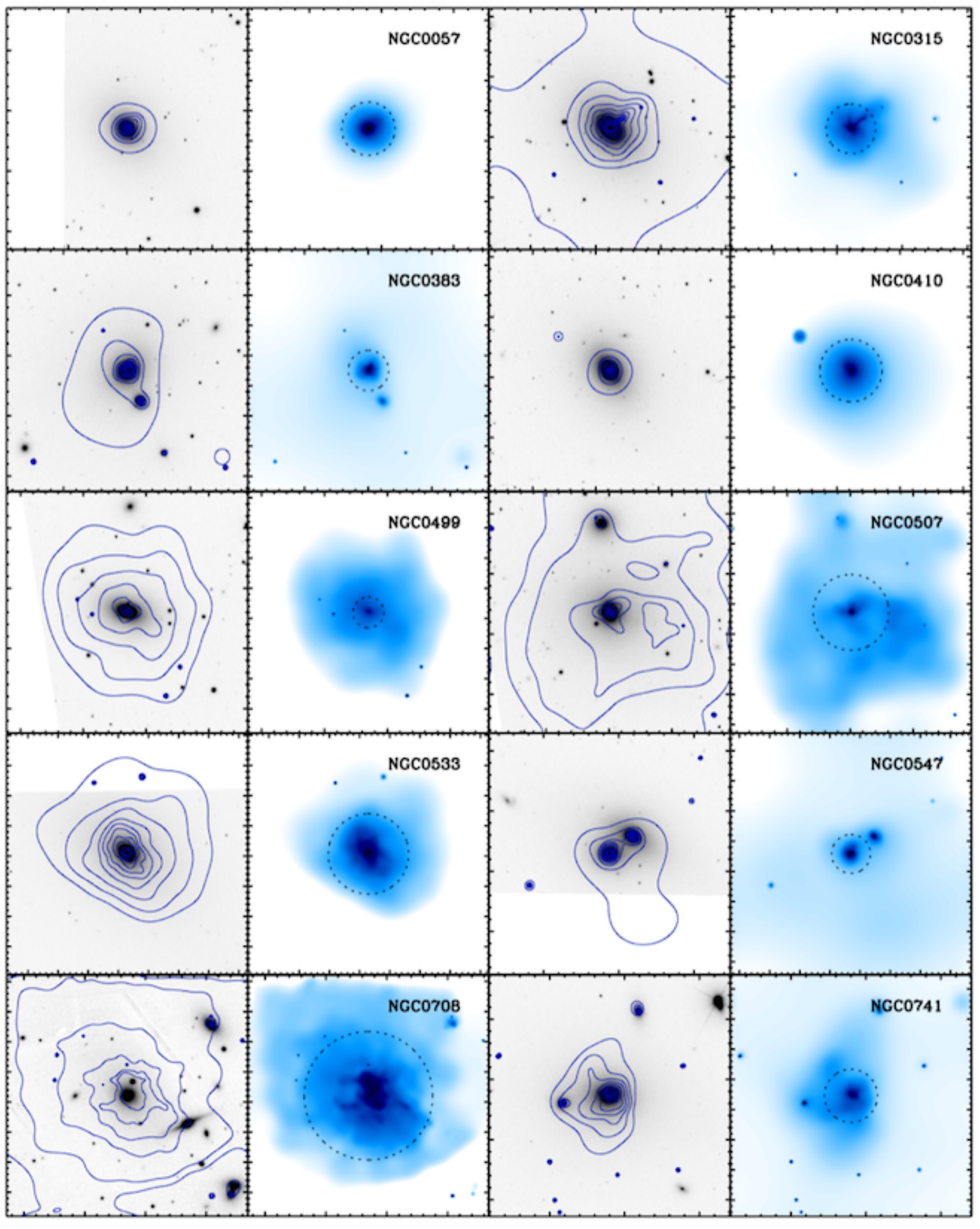}
  \caption{4$\times$4 arc-minute cutouts of the MASSIVE galaxies with
    publicly available archival {\it Chandra} X-ray observations
    performed with the ACIS instrument. Left column shows the optical
    $i$-band SDSS DR12 Atlas image, logarithmic X-ray contours are
    overlaid in solid blue. $g$-band PANSTARRS images are shown for
    NGC 708, 1600, 2340, 6482, 7052, 7265 and 7618 as they do not fall
    within the SDSS survey footprint. The logarithm of the
    adaptively-smoothed, vignetting and exposure-corrected {\it
      Chandra} ACIS-S (NGC 1129: ACIS-I) images are shown to the right
    of the optical images. Dashed circles represent one effective
    radius derived from optical and near-infrared imaging. }
  \label{fig:cxoimgs01}
\end{figure*}

\begin{figure*}[!ht]
  \centering
  \includegraphics[width=0.85\linewidth]{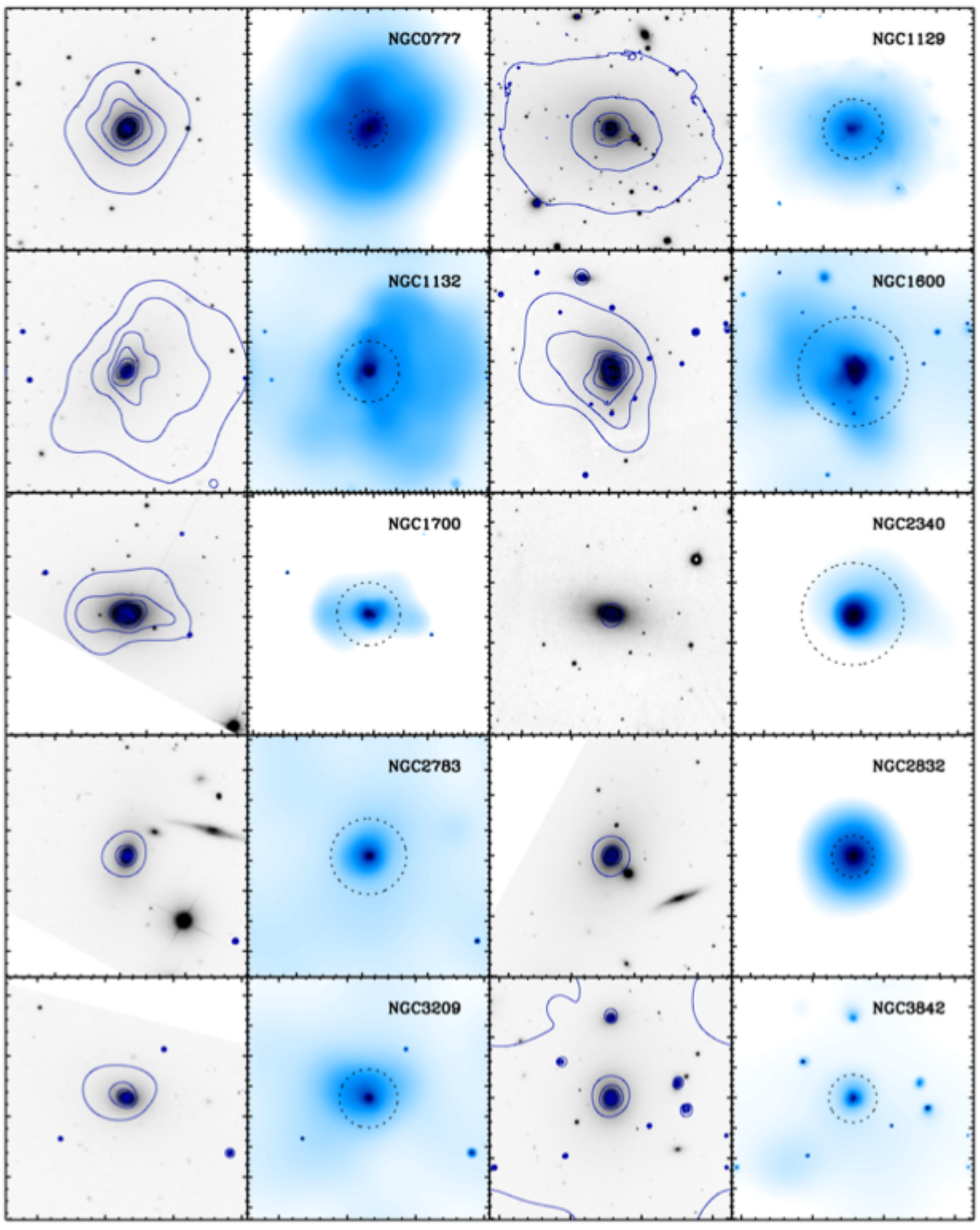}
  \caption{Figure 2 continued...}
\end{figure*}

\begin{figure*}[!ht]
  \centering
  \includegraphics[width=0.85\linewidth]{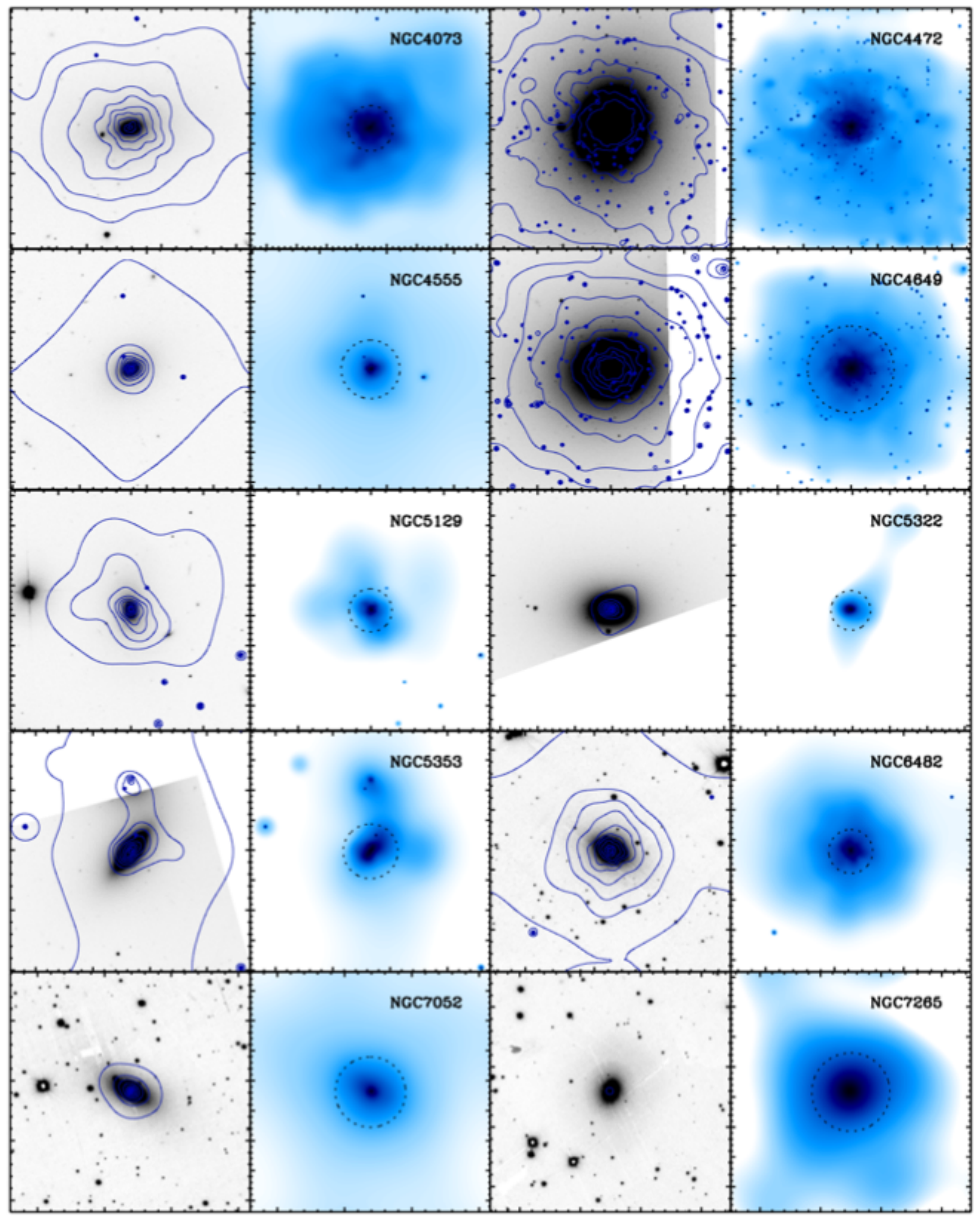}
  \caption{Figure 2 continued...}
\end{figure*}

\begin{figure*}[!ht]
  \centering
  \includegraphics[width=0.85\linewidth]{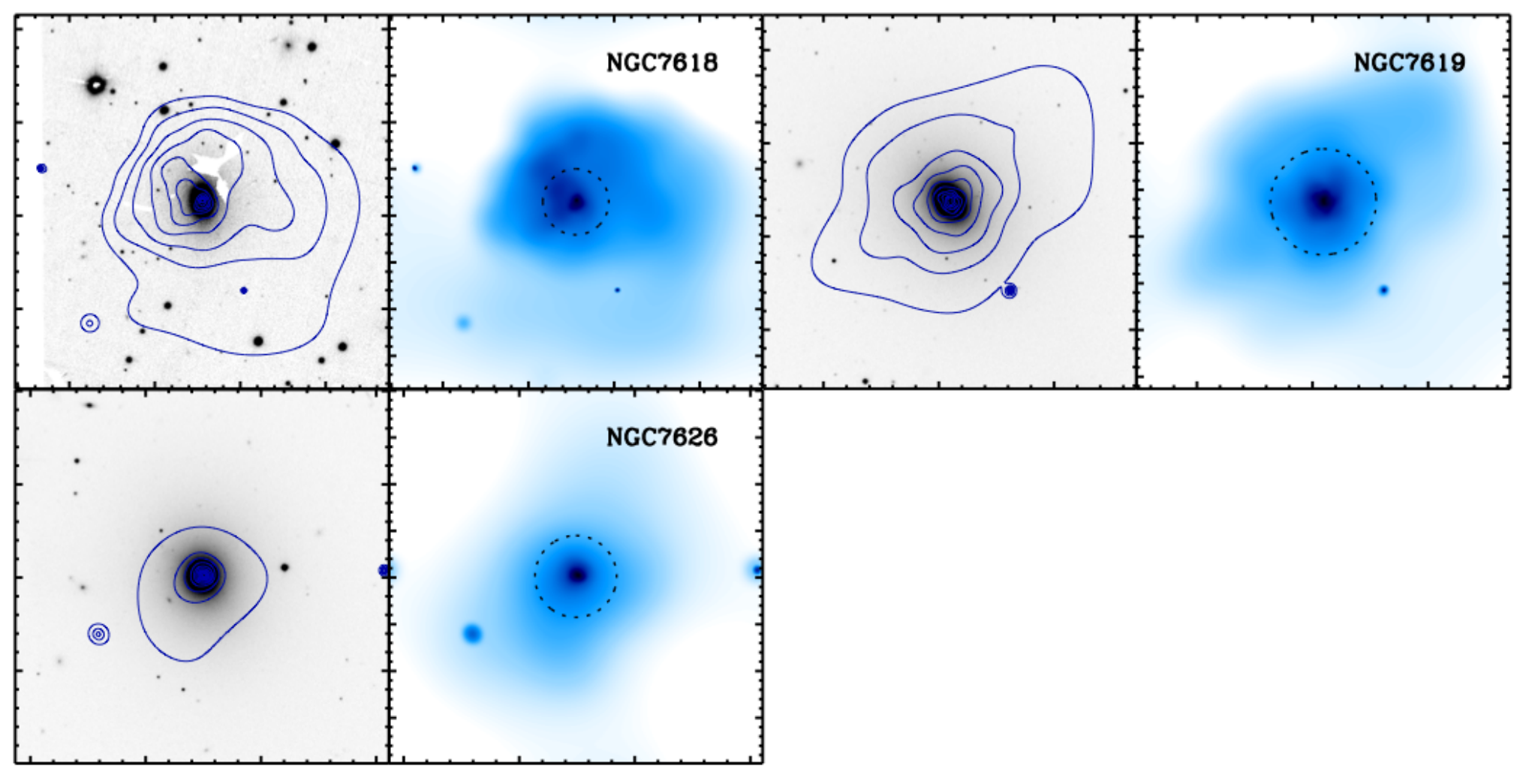}
  \caption{Figure 2 continued...}
  \vspace{0.5cm}
\end{figure*}

\subsection{Data reduction and point source removal}
\label{sec:xray_datreduc}

Basic data processing was carried out using the CXC pipeline software
packages available in {\sc ciao} v4.7. Each ACIS observation was
handled separately, and the latest calibration files ({\sc caldb}
v4.6.7) were applied using the {\tt chandra\_repro} script to produce
new and up-to-date level-1 events files. We used the {\tt
  acis\_process\_events} tool to remove the standard pixel
randomization, and streak events, bad pixels and cosmic rays were
removed by setting ${\rm STATUS} = 0$. Events files were screened
using a typical grade set (grade $=$ 0, 2, 3, 4, 6) and background
flares that were detected at $3\sigma$ above the mean count rate were
removed using {\tt lc\_clean} to create new Level-2 events
files. Subsequent spectral images were constructed for each ObsID from
the Level-2 events. Aspect histograms were constructed using the {\tt
  aspecthist} tool available in the {\sc chav} software package. The
aspect histograms were then convolved with the appropriate ACIS
chip-map to generate the necessary exposure maps. The total exposure
in good-time intervals are provided in Table 2. The median exposure
time was $\sim$40~ks. Exposure corrected images in the 0.3--5keV band
were constructed from the band images and exposure maps (see Figure
2). For those sources with multiple {\it Chandra} ObsIDs, the events
images and exposure maps were combined into single contiguous images
using the method outlined in \cite{goulding12b}. However, this merging
was performed for presentation purposes only. All subsequent spectral
extraction and spectral fitting (see below) was performed on the
individual ObsID data.

In-depth studies of the Local Group galaxies have shown that the X-ray
emission detected in these systems is a complex mixture of emission
arising from a diffuse thermal plasma (the hot ISM), low-mass X-ray
binaries (LMXBs), active binaries (ABs), cataclysmic variables (CVs),
and mass accretion onto the supermassive black hole at the galaxy
center
(\citealt{Fabbiano:2006ab,Revnivtsev:2007aa,Bogdan:2008aa}). Indeed,
observations of M31 and M32 have resolved the full population of LMXBs
to provide a more complete view of the expected contributions to the
total X-ray emission from the unresolved stellar systems (e.g.,
\citealt{Gilfanov:2004aa,Boroson:2011aa}). To remove a significant
fraction of the X-ray emission due to those LMXBs that are resolved as
point-sources (typically 5--10 per galaxy) within our data, we use the
{\sc ciao} tool {\tt wavdetect} to detect and subsequently mask these
stellar systems. Point sources were detected with {\tt wavdetect} in
the 0.3--7~keV band with a significance threshold of $10^{-7}$ using
native pixel wavelet scales of 1, $\sqrt{2}$, 2, $2\sqrt{2}$ and
4. After source detection, the point source position was iterated up
to five times to more accurately determine the centroid. The region
covered by 90\% of the predicted encapsulated energy by a point-source
at 1.5~keV at the appropriate chip position of the source was then
masked from further spectral analyses. In the case of the presence of
an active galactic nucleus, this masking procedure included the
central region of the galaxy.

\begin{turnpage}
\begin{table*}
\footnotesize
\begin{center}
\setlength{\tabcolsep}{1.6mm}
\caption{X-ray Data \& Properties\label{tbl:xray}}
\begin{tabular}{lccccrcccrcc}
\tableline\tableline
\multicolumn{1}{c}{Common} &
\multicolumn{1}{c}{Date} &
\multicolumn{1}{c}{Obs-ID} &
\multicolumn{1}{c}{$t_{\rm exp}$} &
\multicolumn{1}{c}{$T_{\rm X,gas,R_e}$} &
\multicolumn{1}{c}{$f_{\rm X,gas,R_e}$} &
\multicolumn{1}{c}{$L_{\rm X,gas,R_e}$} &
\multicolumn{1}{c}{$\chi^{2}_{\rm R_e}$ (d.o.f.)} &
\multicolumn{1}{c}{$T_{\rm X,gas,tot}$} &
\multicolumn{1}{c}{$f_{\rm X,gas,tot}$} &
\multicolumn{1}{c}{$L_{\rm X,gas,tot}$} &
\multicolumn{1}{c}{$\chi^{2}_{\rm tot}$ (d.o.f.)} \\

\multicolumn{1}{c}{Name} &
\multicolumn{1}{c}{} &
\multicolumn{1}{c}{} &
\multicolumn{1}{c}{(ks)} &
\multicolumn{1}{c}{(keV)} &
\multicolumn{1}{c}{(erg s$^{-1}$ cm$^{-2}$)} &
\multicolumn{1}{c}{($10^{40} \ergps$)} &
\multicolumn{1}{c}{} &
\multicolumn{1}{c}{(keV)} &
\multicolumn{1}{c}{(erg s$^{-1}$ cm$^{-2}$)} &
\multicolumn{1}{c}{($10^{40} \ergps$)} &
\multicolumn{1}{c}{} \\

\multicolumn{1}{c}{(1)} &
\multicolumn{1}{c}{(2)} &
\multicolumn{1}{c}{(3)} &
\multicolumn{1}{c}{(4)} &
\multicolumn{1}{c}{(5)} &
\multicolumn{1}{c}{(6)} &
\multicolumn{1}{c}{(7)} &
\multicolumn{1}{c}{(8)} &
\multicolumn{1}{c}{(9)} &
\multicolumn{1}{c}{(10)} &
\multicolumn{1}{c}{(11)} &
\multicolumn{1}{c}{(12)} \\

\tableline

NGC0057	& 2008-10-29 & 10547      & 9.9 &	        0.92 $\pm$ 0.05 & -12.548 $\pm$ 0.019 &  19.8 & 185.6 (171) & 0.92 $\pm$ 0.05 &	-12.529 $\pm$ 	0.025	 &  20.6 & 339.7 (378) \\
NGC0315	& 2003-02-22 & 4156	  & 55.0 &	        0.66 $\pm$ 0.04 & -12.574 $\pm$ 0.089 &	 16.2 & 470.3 (327) & 0.57 $\pm$ 0.05 &	-12.192 $\pm$ 	0.042	 &  38.0 & 563.1 (433) \\
NGC0383	& 2000-11-06 & 2147	  & 44.4 &	        0.82 $\pm$ 0.05 & -13.091 $\pm$ 0.021 &	  5.0 & 259.1 (226) & 1.09 $\pm$ 0.09 &	-12.861 $\pm$ 	0.048	 &   8.4 & 495 (458) \\
NGC0410	& 2004-11-30 & 5897	  & 2.6 &	        0.80 $\pm$ 0.06 & -12.194 $\pm$	0.027 &  39.6 & 141.1 (108) & 0.86 $\pm$ 0.06 &	-12.129 $\pm$ 	0.031	 &  45.2 & 204.2 (255) \\
NGC0499	& 2009-02-12 & 10536	  & 18.4 &		0.83 $\pm$ 0.05 & -12.512 $\pm$	0.015 &  18.2 & 208.4 (183) & 0.75 $\pm$ 0.04 &	-11.561 $\pm$ 	0.006	 & 160.2 & 548.6 (454) \\
NGC0507	& 2000-10-11 & 317	  & 26.9 &	        1.13 $\pm$ 0.05 & -12.173 $\pm$	0.081 &  39.7 & 670.7 (400) & 1.29 $\pm$ 0.04 &	-11.452 $\pm$ 	0.070	 & 205.9 & 669.1 (455) \\
NGC0533	& 2002-07-28 & 2880	  & 37.6 &		0.92 $\pm$ 0.04 & -11.998 $\pm$	0.056 &  75.5 & 771.0 (330) & 1.11 $\pm$ 0.09 &	-11.802 $\pm$ 	0.063	 & 114.6 & 1116.1 (454) \\
NGC0547	& 2007-09-07 & 7823	  & 64.8 &		0.74 $\pm$ 0.05 & -12.998 $\pm$	0.015 &   7.5 & 431.3 (281) & 0.70 $\pm$ 0.05 &	-12.834 $\pm$ 	0.018	 &   9.6 & 1078.3 (458) \\
NGC0708	& 2006-11-20 & 2215,7921  & 139.5 &		1.56 $\pm$ 0.04 & -11.312 $\pm$	0.080 &	254.5 & 6687.6 (982) & 1.91 $\pm$ 0.07 &	-10.915 $\pm$ 	0.011	 & 693.0 & 4833.6 (908) \\
NGC0741	& 2001-01-28 & 2223	  & 30.3 &		0.80 $\pm$ 0.04 & -12.523 $\pm$	0.015 &	 19.7 & 302.9 (243) & 1.04 $\pm$ 0.13 &	-12.301 $\pm$ 	0.017	 &  32.7 & 704.3 (454) \\
NGC0777	& 2004-12-23 & 5001$\dagger$	  & 10.0 &	0.96 $\pm$ 0.05 & -12.106 $\pm$	0.017 &	 49.9 & 216.7 (158) & 0.79 $\pm$ 0.04 &	-11.689 $\pm$ 	0.011	 & 127.7 & 427.7 (398) \\
NGC1129	& 2009-11-05 & 11717$\dagger$,12017$\dagger$  & 79.9 &		2.91 $\pm$ 0.36 & -11.499 $\pm$	0.035 &	206.7 & 579.6 (442) & 4.23 $\pm$ 0.16 &	-10.158 $\pm$ 	0.026	 & 4542.5 & 665.4 (454) \\
NGC1132	& 2003-11-16 & 801,3576	  & 53.7 &		1.01 $\pm$ 0.04 & -12.530 $\pm$	0.010 &	 32.9 & 535.6 (296) & 1.09 $\pm$ 0.10 &	-11.923 $\pm$ 	0.068	 & 136.1 & 988.2 (454) \\
NGC1600	& 2002-09-20 & 4283,4371  & 53.5 &		1.05 $\pm$ 0.04 & -12.560 $\pm$	0.017 &	 12.6 & 1074.7 (707) & 1.10 $\pm$ 0.05 & -12.457 $\pm$ 	0.020	 &  17.0 & 1537.6 (908) \\
NGC1700	& 2000-11-03 & 2069	  & 42.8 &		0.45 $\pm$ 0.05 & -12.938 $\pm$ 0.015 &	  4.1 & 389.2 (328) & 0.34 $\pm$ 0.04 &	-12.586 $\pm$ 	0.018	 &   9.2 & 207.8 (104) \\
NGC2340	& 2005-02-04 & 5898	  & 1.9 &		0.85 $\pm$ 0.10 & -12.679 $\pm$	0.055 &	 20.2 & 73.4 (75) & 1.15 $\pm$ 0.15 &	-12.687 $\pm$ 	0.124	 &  20.0 & 213.0 (240) \\
NGC2783	& 2005-01-13 & 5789	  & 17.9 &		0.81 $\pm$ 0.09 & -13.323 $\pm$	0.038 &	  5.7 & 242.8 (180) & 0.82 $\pm$ 0.08 &	-13.297 $\pm$ 	0.055	 &   6.2 & 391.9 (375) \\
NGC2832	& 2005-01-02 & 5904$\dagger$	  & 3.0 &	1.16 $\pm$ 0.09 & -12.292 $\pm$	0.060 &	 67.6 & 70.5 (95) & 1.13 $\pm$ 0.12 &	-12.245 $\pm$ 	0.081	 &  75.3 & 359.8 (399) \\
NGC3209	& 2005-02-17 & 5792	  & 21.7 &		0.68 $\pm$ 0.08 & -13.331 $\pm$	0.033 &	  4.7 & 189.1 (159) & 0.70 $\pm$ 0.06 &	-13.245 $\pm$ 	0.039	 &   6.1 & 472.0 (440) \\
NGC3842	& 2003-01-24 & 4189	  & 47.5 &		1.01 $\pm$ 0.05 & -13.212 $\pm$	0.032 &	  7.2 & 437.5 (272) & 1.32 $\pm$ 0.11 &	-13.009 $\pm$ 	0.056	 &  11.6 & 1065.1 (454) \\
NGC4073	& 2002-11-24 & 3234	  & 30.0 &		1.54 $\pm$ 0.05 & -11.806 $\pm$	0.011 &	152.4 & 773.6 (318) & 1.85 $\pm$ 0.04 &	-11.120 $\pm$ 	0.070	 & 760.1 & 1262.1 (454) \\
NGC4472	& 2011-02-21 & 12888,12889  & 294.9 &	 	0.94 $\pm$ 0.04 &	 $>$ -11.610  &	 $>$16.7 & 14851.0 (908) & $>$1.25 &	 	$>$ -11.594 	 &  $>$83.2 & - \\
NGC4555	& 2003-02-04 & 2884	  & 30.0 &		0.91 $\pm$ 0.04 & -12.817 $\pm$ 0.014 &	 20.2 & 323.8 (210) & 1.06 $\pm$ 0.05 &	-12.509 $\pm$ 	0.020	 &  39.8 & 600.2 (449) \\
NGC4649	& 2011-02-24 & 12976	  & 101.0 &	 	0.89 $\pm$ 0.04 & -11.642 $\pm$ 0.002 &	  7.5 & 1161.6 (445) & 0.90 $\pm$ 0.04 &	$>$ -11.507  	 &  $>$12.9 & 998.4 (454) \\
NGC5129	& 2006-05-14 & 6944,7325  & 46.8 &		0.87 $\pm$ 0.05 & -12.695 $\pm$ 0.015 &	 28.0 & 585.0 (373) & 0.92 $\pm$ 0.04 &	-12.163 $\pm$ 	0.011	 &  95.0 & 937.6 (906) \\
NGC5322	& 2006-08-20 & 6787	  & 13.8 &		0.44 $\pm$ 0.08 & -13.444 $\pm$ 0.073 &	  0.4 & 146.3 (147) & 0.44 $\pm$ 0.06 &	-13.433 $\pm$ 	0.068	 &   0.5 & 157.4 (148) \\
NGC5353	& 2014-03-31 & 14903	  & 40.3 &		0.73 $\pm$ 0.05 & -12.744 $\pm$ 0.015 &	  3.5 & 357.9 (253) & 0.66 $\pm$ 0.04 &	-12.362 $\pm$ 	0.013	 &   8.8 & 553.0 (454) \\
NGC6482	& 2002-05-20 & 3218	  & 19.3 &		0.89 $\pm$ 0.04 & -11.984 $\pm$ 0.083 &	 48.1 & 429.0 (202) & 0.71 $\pm$ 0.04 &	-11.676 $\pm$ 	0.072	 &  95.1 & 565.8 (451) \\
NGC7052	& 2002-09-21 & 2931	  & 9.6 &		0.56 $\pm$ 0.06 & -12.535 $\pm$ 0.023 &	 17.6 & 245.6 (157) & 0.61 $\pm$ 0.06 &	-12.439 $\pm$ 	0.024	 &  20.9 & 539.6 (351) \\
NGC7265	& 2006-06-07 & 7058$\dagger$	  & 5.0 &	0.55 $\pm$ 0.10 & -12.759 $\pm$ 0.066 &	 14.3 & 70.0 (76) & 0.86 $\pm$ 0.10 &	-12.092 $\pm$ 	0.052	 &  66.4 & 292.7 (326) \\
NGC7618	& 2007-09-08 & 7895	  & 34.1 &	 	0.99 $\pm$ 0.04 & -12.301 $\pm$ 0.011 &	 35.2 & 479.1 (242) & 0.83 $\pm$ 0.04 &		$>$ -11.747  	 & $>$124.8 & 1825.8 (454) \\
NGC7619	& 2003-09-24 & 3955	  & 37.5 &		0.84 $\pm$ 0.04 & -12.308 $\pm$ 0.007 &	 17.2 & 648.1 (309) & 0.87 $\pm$ 0.04 &	-11.986 $\pm$ 	0.007	 &  36.0 & 1060.9 (454) \\
NGC7626	& 2001-08-20 & 2074$\dagger$	  & 26.7 &	0.69 $\pm$ 0.06 & -12.862 $\pm$ 0.026 &	  4.9 & 220.5 (220) & 0.71 $\pm$ 0.05 &	-12.579 $\pm$ 	0.026	 &   9.2 & 486.9 (454) \\

\tableline
\end{tabular}
\end{center}
{\bf Notes:-} $^{1}$Common galaxy name;
$^{2}${\it Chandra} X-ray observation date;
$^{3}${\it Chandra} observation identification numbers for ACIS-S observations ($\dagger$: ACIS-I);
$^{4}$Cleaned exposure time in kilo-seconds;
$^{5}$Temperature of best-fit {\tt APEC} model for photons extracted within $R_e$ (uncertainties include an additional systematic for the choice of the {\tt APEC} model);
$^{6}$0.3--5keV X-ray flux of best-fit model for photons extracted within $R_e$;
$^{7}$Derived 0.3--5keV gas X-ray luminosity within $R_e$;
$^{8}$Chi-squared value from model fitting for $R_e$ aperture, and the number of degrees of freedom are given in parentheses; 
$^{9}$Temperature of best-fit {\tt APEC} model for photons extracted out to background-level of observation (uncertainties include an additional systematic for the choice of the {\tt APEC} model);
$^{10}$0.3--5keV X-ray flux of best-fit model for photons extracted out to background-level of observation;
$^{11}$Derived 0.3--5keV gas X-ray luminosity for photons extracted out to background-level of observation;
$^{12}$Chi-squared value from model fitting for the large aperture, and the number of degrees of freedom are given in parentheses.  \\
\normalsize
\end{table*}
\end{turnpage}

\subsection{Spectral Extraction}
\label{sec:xray_spec}

The hot X-ray emitting gas extends to tens to hundreds of times the
optical extent of the galaxy. As such, we choose to adopt two
different regions for extracting the diffuse gas emission. The first
(smaller) region is a circular aperture matched to the optical size of
the galaxy, specifically centered at the peak of the X-ray emission
with radius equal to one effective radius of the galaxy ($r=R_e$; see
section~\ref{sec:re}; e.g., \citealt{Diehl:2007aa}). This particular
methodology allows a more direct measurement of the X-ray gas
associated with the galaxy properties. The second (larger) extraction
region follows the more typical methodology for analyzing diffuse
X-ray emission. We extract all of the X-ray photons out to where the
diffuse component becomes equal to the background emission within the
X-ray observation (e.g.,
\citealt{Boroson:2011aa,Bogdan:2012aa,Bogdan:2012ab,Kim:2015aa}).\footnote{For
  NGC~4472 and NGC~4649, the diffuse emission, as measured within a
  circular aperture, extended beyond the coverage of a single
  chip. Hence, for the full aperture diffuse X-ray emission we adopt
  the previous lower-limit measurements of Diehl \& Statler (2007).}

In several cases, the diffuse emission associated with the galaxy,
group and/or cluster filled a significant fraction of the ACIS chip,
and hence prevented a direct extraction of the background photons from
the on-source chip. Hence, we chose to adopt three separate methods to
provide the best appropriate estimate of the X-ray background within
the observation: First, and where possible, we used extraction regions
twice the area of the source aperture on the same chip (typically S3)
as the source, assuming these were free from diffuse source
emission. Second, multiple regions from the S1 (i.e., the second
back-illuminated chip on the ACIS instrument) chip were extracted,
assuming this chip was also free from diffuse emission. Third, we
harnessed the appropriate stowed {\it Chandra} observations normalized
to the hard energy (10--12~keV) count rates of the relevant ObsIDs to
extract background spectra from the same regions used for the full
aperture source spectra. These `stowed' observations are a set of blank-sky
observations performed with {\it Chandra} that provide the appropriate
ACIS background for a set of ObsIDs.\footnote{See
  \url{http://cxc.harvard.edu/ciao/threads/acisbackground/} for
  further details.}

Spectra were extracted from source and background regions using the
{\tt specextract} tool available in {\sc ciao}, and ancillary response
and redistribution matrices were constructed during the spectral
extraction. Extracted spectra were grouped to contain one count per
bin. For sources with weak jet structures (i.e., those not already
excluded due to pile-up effects) visible in the exposure-corrected
images, we additionally masked the jet region using a rectangular box
during spectral extraction oriented along the direction of the jet
axis.

As noted previously, the high-energy emission observed in elliptical
galaxies arises due to several phenomena. We followed
\cite{Boroson:2011aa} to characterize the dominant emission sources
contributing to the X-ray spectroscopy. After the direct subtraction
of the bright LMXB population, we model the diffuse hot gas as a
thermal plasma ({\tt APEC}) component, with the metallicity fixed at
solar; for the remaining LMXB population that lies below the
sensitivity threshold of the X-ray observations we use a thermal
Bremsstrahlung model fixed to a temperature of 7~keV; a powerlaw
component with $\Gamma=1.6$ is used to model the AB/CV population; and
for those galaxies with obvious nuclear activity (i.e., significant
emission at $E > 4$~keV), we include an absorbed powerlaw with
$\Gamma$ allowed to vary within the fitting. Additionally, we include
a photoelectric absorption component fixed to the Galactic column
density ($N_{\rm H}$) along the line of sight to each source. In {\sc
  xspec} language our adopted model is given as {\tt
  phabs}$\times$({\tt apec}$+${\tt bremss}$+${\tt pow}$+$[{\tt
  phabs}$\times${\tt pow}]) (e.g.,
\citealt{Boroson:2011aa,Kim:2015aa}). During spectral fitting we limit
the fitted energy range to $E = 0.3$--7~keV, which is typical for {\it
  Chandra} ACIS observations. Typical reduced $\chi^2$ statistics for
the model fits were in the range 0.8--2.5 (see columns 8 and 12 of
Table 2). Some of the fits are formally not acceptable, this is
particularly evident for sources with large numbers of detected
photons. Based on testing of our assumed model and comparison to
previous literature (e.g., Diehl \& Statler 2008; Kim \& Fabbiano
2015; see also Appendix in this work), we determine that a combination
of our assumed fixed solar metallicity (see below), the requirement of
a second temperature component in some sources, as well as marginal
systematic differences in the {\it Chandra} calibration at low
energies, drive the large chi-squared values in these
sources. However, we find within this subset of high-count sources
that while the fit may be formally improved with a more complex model
that is fit in a restricted energy range, it has no effect on our
measured $T_{\rm gas}$ or $f_{\rm X,gas}$, particularly within the
1$R_e$ apertures, which are adopted throughout Section 4.

In the majority of cases, the {\tt powerlaw} model, motivated to
represent the AB/CV population, was substantially sub-dominant
compared to the other emission components. However, given the
relatively high stellar mass (high $K$-band luminosity) of the MASSIVE
galaxies, it should not be surprising given that the expected X-ray
luminosity for the AB/CV population is over an order of magnitude
lower than the gas components measured here based on the typical
$L_X / L_K$ ratio for the AB/CV population located in Local Group
galaxies (e.g., \citealt{Revnivtsev:2007aa}). For those sources with
multiple observations, the spectra were not combined into one single
spectrum, but instead the spectra were modeled simultaneously. The
final unabsorbed X-ray fluxes at $E\sim 0.3$--5~keV ($f_{\rm X,gas}$),
derived luminosities ($L_{\rm X,gas}$), and temperatures
($T_{\rm gas}$) quoted in Table 2 are for the {\tt APEC} thermal gas
component derived from the X-ray fitting process for the emission
detected within $r=R_e$ and $r<R_{\rm bkg}$.

\begin{figure*}[ht]
\centering
\includegraphics[width=0.95\textwidth]{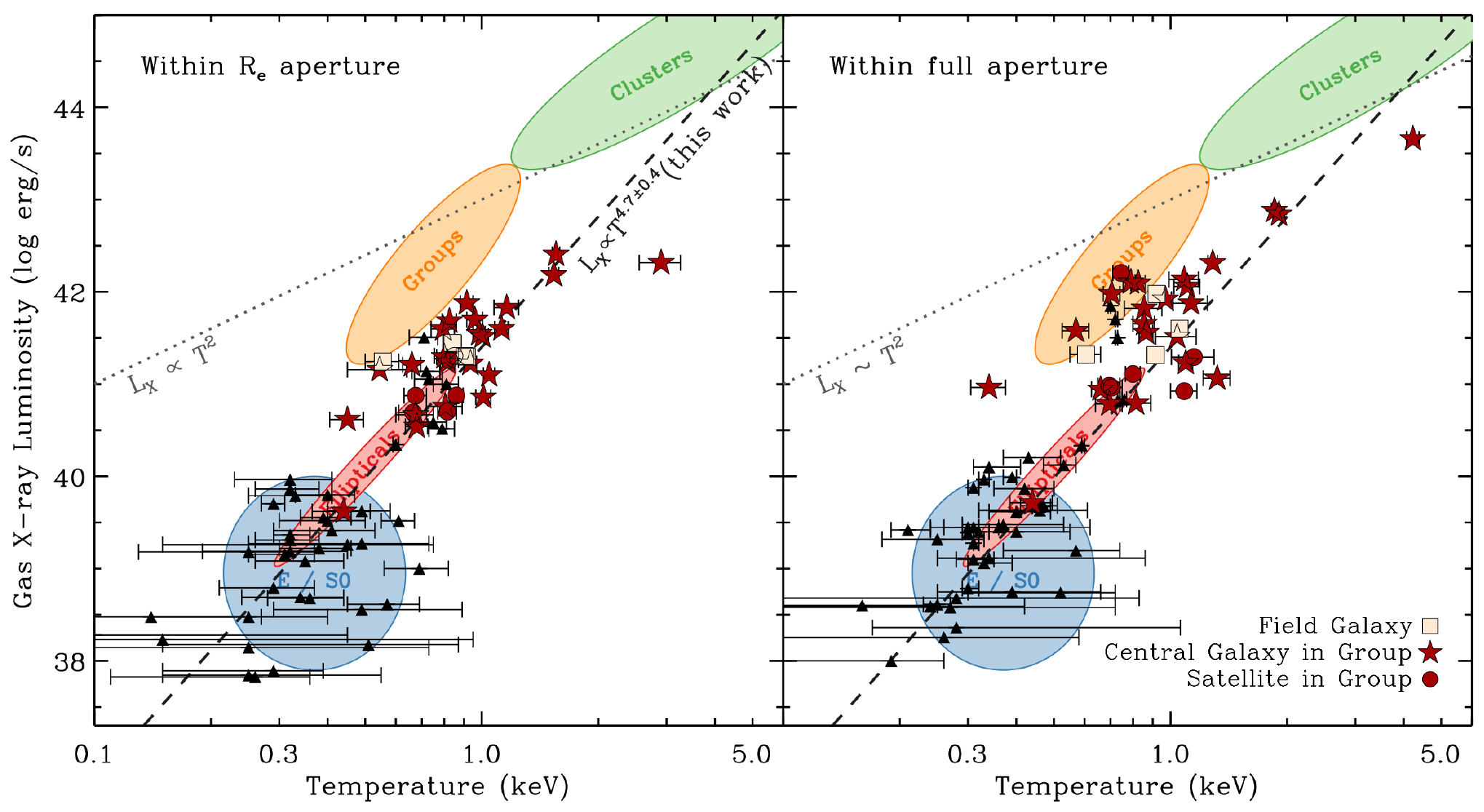}
\caption{{\bf Common:} Logarithm of 0.3--5.0~keV X-ray luminosity of
  the diffuse gas {\tt APEC} component versus the X-ray temperature of
  the gas (in units of keV).  Stars represent those MASSIVE galaxies
  that are determined to lie at the center of their group environment
  in the catalog of Crook et al. (2007), satellite galaxies in groups
  are shown with circles, and field galaxies ($\leq 3$ neighbors) are
  shown with light-shaded square symbols. Black triangles represent
  the ATLAS$^{\rm 3D}$ comparison sample (see
  Section~\ref{sec:atlas3d} and Appendix). Background ellipses are
  reproduced and adapted from Kim \& Fabbiano (2015). These regions
  are used to represent the previously observed positions of the
  (coreless/disky) E/S0 galaxies (blue shading),
  core-dominated/slow-rotating ellipticals (red shading), hot gaseous
  emission associated with groups (orange shading) and clusters (green
  shading). The dashed line shows our best-fit relation
  ($L_{\rm X,gas} \propto T^{4.7 \pm 0.4}$;
  $\epsilon_{\rm intrinsic} \sim 0.42-0.50$~dex) to the MASSIVE and
  ATLAS$^{\rm 3D}$ galaxies with constrained $T_{\rm gas}$
  measurements performed within $R_e$ apertures. The dotted line
  represents the self-similar $L_{\rm x,gas} \sim T_{\rm gas}^2$
  relation. In general, the uncertainties in the X-ray luminosities
  are small (e.g., see Table 2 and Figure~\ref{fig:lx_lk_sigma}), and
  are not shown in the figure for presentation purposes only. {\bf
    Left:} X-ray luminosity and gas temperature measured within one
  effective optical radius of the galaxy {\bf Right:} X-ray luminosity
  and gas temperature measured out to large radii where the X-ray
  emission associated with the galaxy/halo meets the X-ray background
  level in the observation.}
\vspace{0.5cm}
\label{fig:lx_T}
\end{figure*}

While our assumption of a solar metallicity for the thermal component
allows a homogeneous temperature measurement (i.e., one that is not
subject to a degeneracy between abundance and temperature in the
low-count regime), it may not provide the most accurate description of
the data for some sources. To test for a systematic bias in the
measured temperature and normalization (flux) of the {\tt APEC}
component, we re-fit the 23 MASSIVE galaxies that had $>$1000 counts
in their X-ray spectra, but with the metallicity as a free parameter
during the fitting process. The metallicities for these sources were
found to be in the range, 0.1--2.8 solar consistent with previous
analyses (e.g., \citealt{Buote:2000ab}). However, for the majority of
the systems, we found that the metallicity was still poorly
constrained, and often consistent with solar metallicity. Particularly
for the highest $T_{\rm gas}$ sources, we found that allowing the gas
metallicity to vary within the fit improved the overall $\chi^2$
significantly ($\Delta \chi^2 \sim 5-10$\%). However, with the
exceptions of NGC~708 and NGC~4073, within $R_e$, a fixed solar
metallicity had no significant impact on the temperature or flux at
the $<3$\% level. For NGC~708 and NGC~4073, we found an increase in
the observed temperature of $\sim 0.2$~keV, and a $\sim 10$\% decrease
in the flux, though neither of these have any significant impact on
our results.

In order for us to better compare to previous literature, we also
tested for underlying systematics in our determination of the
luminosity and temperature measured through the use of {\tt APEC}
compared to a {\tt MEKAL} model (for example) which has also been
used. {\tt APEC} now includes the most up-to-date atomic libraries
with the most recent photoionization and recombination rates for a
variety of elements while the {\tt MEKAL} model is no longer
maintained. Indeed, we determine that, on average, the plasma
temperatures measured for the MASSIVE sample with a {\tt MEKAL} model
are $\sim 7$--12\% lower than those found using the more recent {\tt
  APEC} model. However, we found no appreciable difference between the
fluxes produced using these two models. Within our MASSIVE X-ray
sample, there are 14 sources in common with the analyses of the
structure of the hot ISM in ellipticals of
\cite{Diehl:2008ac}. Adopting our above method with a {\tt MEKAL}
model produces perfectly consistent temperatures for these 14 sources,
despite Diehl \& Statler adopting an extraction radius that was a
factor $\sim 3$ larger than that considered here. However, a
comparison between the most recent {\tt APEC} model and that used in
Diehl \& Statler, produced temperatures that were consistent only to
within a factor $\sim$1.1 of the quoted temperatures, which is several
sigma larger than the statistical uncertainties found during the
fitting process. We thus adopt an additional 10\% systematic
uncertainty into the measured temperatures with {\tt APEC} to allow us
to more readily compare to the previous measured values for some of
these sources derived using {\tt MEKAL} or older ionization tables
used with previous versions of the {\tt APEC} model.

Finally, we also tested for the presence of a systematic bias in our
flux and temperature measurements within $R_e$ due to the assumption
of a single homogeneous gas temperature. For larger radius
measurements, well-studied galaxies are known to have radial
temperature gradients due to the large scale halo. Given much of the
following analyses focus on the X-ray measurements within $R_e$ it is
important to test whether our $T_{\rm gas}$ measurements are biased by
line-of-sight gas, associated with the halo, rather than the central
galaxy. We refit each of the $R_e$ aperture X-ray spectra by including
an additional {\tt APEC} component with fixed solar metallicity. For
the majority of the MASSIVE galaxies (28/33), we found no significant
improvement in the calculated $\chi^2$ value (F-test reveals $<0.95$
confidence). Five of the MASSIVE galaxies showed a significant
improvement in the fit. We found that within the $R_e$ aperture, the
dominant APEC component had a temperature consistent with that
measured with the single temperature model. We determined that the
overall improvement to the fit was driven by a sub-dominant higher
temperature component, consistent with that expected from
line-of-sight emission. The one exception was NGC~708, where we
determined that a single temperature model was insufficiently fitting
two distinct temperature gas components. The two temperature model for
NGC~708 has $T_{\rm gas,Re,1} \sim 2.0 \pm 0.1$~keV and
$T_{\rm gas,Re,2} \sim 1.1 \pm 0.1$~keV, where the cooler component is
a factor $5-7\times$ fainter than the hot component. For consistency,
with the remainder of the MASSIVE sample, we adopt the single
temperature model measurement. Overall, we conclude that a single
temperature model provides an accurate measurement of the dominant
X-ray gas properties within $R_e$ with no systematic bias.

\section{Results}

\subsection{The X-ray properties of the high mass early-type galaxy population}
\label{sec:lx_t}

Two of the most fundamental measurables of the X-ray emitting thermal
ISM are its luminosity and temperature.  In Figure~\ref{fig:lx_T} we
present the X-ray luminosity at 0.3--5~keV versus temperature measured
for the thermal gas component of the 33 sources in the MASSIVE X-ray
galaxy sample, along with the ATLAS$^{\rm 3D}$ sample described in
detail in Section~\ref{sec:atlas3d} (also see Appendix). To fully
harness the available X-ray data and build the most coherent picture
of the hot gas properties of early-type galaxies, we present two
measurements of $L_{\rm X,gas}$ and $T_{\rm gas}$: (1) measured within
an aperture defined by the optical radius of the galaxy ($r=R_e$) to
accurately characterize the X-ray emission directly associated with
the galaxy and the galaxy potential; and (2) measured in large
apertures out to radii where the source X-ray photons approximate the
observed X-ray background to provide a good measure of the total X-ray
luminosity of the gas associated with the galaxy and the galaxy’s
wider environment. See Section~\ref{sec:xray_spec} for further
details.

With the exception of NGC~5322, all of the MASSIVE galaxies have
$L_{\rm X,R_e} \gtrsim 10^{40} \ergps$ and hence, can be considered
hot-gas--rich systems.  The full MASSIVE X-ray galaxy sample covers a
wider range in gas temperature ($T_{\rm gas}\sim0.3-4$~keV) than
previously observed for individual early-type galaxies. Although the
majority (27/33; $\sim$80\%) exhibit temperatures of
$T_{\rm gas}\sim0.5-1.2$~keV, this is the poorly-sampled region of
$T_{\rm gas}$-space in previous studies. The observed temperatures are
almost independent of the aperture radius used to extract the source
photons.

On average, we observe a shift of $\sim$0.1~dex towards higher
$T_{\rm gas}$ in the large aperture measurements, suggesting that the
inner X-ray halos of massive galaxies are systematically (marginally)
cooler than the larger scale gas. We tested this systematic difference
between the temperatures within $R_e$ and out to $R_{\rm tot}$ by
additionally performing an X-ray spectral analysis that excluded the
photons from the inner $R_e$ aperture for the 13 objects that had
$>$1000 counts within the annulus ($R_{\rm tot} - R_e$)
region\footnote{We did not perform this annular analysis on NGC4472,
  NGC4649 and NGC7618 as the X-ray halos of these galaxies are
  extended well beyond the Chandra field of view}. For seven of the
sources, the temperature measurements within the annulus were entirely
consistent with the total measurements performed in the large
aperture. One source, NGC~6482, showed a decrease in the temperature
within the annulus of 0.1~keV ($\sim$14\%) compared with the large
$R_{\rm tot}$ aperture measurement. This is consistent with our
measurement that $T_{\rm gas}$ is higher at smaller radii, i.e.,
$T_{\rm gas,R_e} > T_{\rm gas,R_{tot}}$. The annulus $T_{\rm gas}$
measurements for five of the most luminous objects were
$\sim$0.1--0.2~keV higher compared with the large aperture
measurement. This suggests that at high luminosities and high
temperatures, our $T_{\rm gas}$ measurements within $R_{\rm tot}$ may
be biased marginally low due to the inclusion of the cooler gas within
$R_e$. This in turn would suggest a marginally larger discrepancy
between the inner gas temperature and the temperature of the wider
halo than the $\sim 0.1$~dex we calculate above.

We perform a linear regression (for the combined ATLAS$^{\rm 3D}$ and
MASSIVE samples) using a Bayesian approach to simultaneously include
the uncertainties in both $L_{\rm X,gas}$ and $T_{\rm gas}$ using the
{\sc idl} routine {\sc linmix\_err}
\citep{Kelly:2007aa}. Specifically, we fit the data for the
galaxy-scale ($R_e$) aperture for all objects with constrained
measurements. We find
${\rm log}L_{\rm X,gas} = (41.4 \pm 0.1) + (4.7 \pm 0.4) {\rm log}
T_{\rm gas}$,
with a well-constrained intrinsic scatter of $\sim$0.42--0.50 dex. The
quoted uncertainties are derived from the central 67\% of the
posterior draws. The data is characterized by a single powerlaw fit
with a reduced $\chi^2$ ($\chi^2$/d.o.f.) of $441.2 / 72 = 6.1$.  This
large reduced chi-squared value is no doubt driven by the amplitude of
the intrinsic scatter (0.42--0.50~dex) present within the data. A
two-powerlaw model marginally improves the fit with
$\Delta \chi^2 \sim 18.6$ for two additional free parameters. An
F-test between the models results in an acceptance of the new 2 power
law model with a confidence of only 0.78 (i.e., $< 2\sigma$).  Thus,
we find no significant evidence to reject a single power law.

The $L_{\rm X,gas}-T_{\rm gas}$ relation calculated from the full
aperture measurements has a consistent slope, but with considerably
higher scatter ($\sim$1.0 dex).  This larger scatter is likely due to
the inclusion of gas from the larger-scale environment for the MASSIVE
galaxies. For the $R_e$ aperture measurements, the scatter about the
$L_{\rm X,gas}-T_{\rm gas}$ relation is relatively constant over a
wide range in $L_{\rm X,gas}$ and $T_{\rm gas}$. Thus, despite the
early-type systems considered here covering a wide-range in stellar
mass, velocity-dispersion, and wide-scale environment, the intrinsic
scatter remains remarkably constant at $\sim 0.5$~dex.  Furthermore,
we also find that there appears to be no dependence of the scatter
about the $L_{\rm X,gas}-T_{\rm gas}$ relation with the
slow/fast-rotators (core/power-law) galaxy morphologies or the radio
power of the central black hole.

Within $1R_e$, the $L_{\rm X,gas}$ for MASSIVE galaxies in poor groups
($N_{\rm neighbors} \leq 5$) versus those in rich groups
($N_{\rm neighbors} > 5$) are also extremely similar, median
$L_{\rm X,poor}\sim 2.0\times 10^{41} \ergps$ and
$L_{\rm X,rich}\sim 1.7\times 10^{41} \ergps$, respectively. However,
when we consider $L_{\rm X,gas}$ measured to large radii, we find a
systematic increase in $L_{\rm X,gas}$ of a factor $\sim 3-5$ for the
rich group galaxies. Indeed, for the rich-group systems in MASSIVE and
ATLAS$^{\rm 3D}$, we show in Figure~\ref{fig:lx_T}b that these sources
occupy a similar locus in $L_{\rm X,gas}-T_{\rm gas}$ to that expected
for group gas halos. Furthermore, we find that the scatter in the
$L_{\rm X,gas}$--$T_{\rm gas}$ relation increases significantly
($\sim 0.95$~dex) for the large aperture measurements. Taken together,
this suggests that the increase in $L_{\rm X,gas}$ for large aperture
measurements is due to significant contributions from the
group/cluster gas as the galaxy X-ray emitting halo extends into the
large-scale environment within which the galaxy is embedded.

Our main conclusions about the $L_{\rm X,gas}-T_{\rm gas}$ relation
are as follows.  First, we see that over a very wide range in galaxy
mass, early-type galaxies obey a single power-law relation between
these two parameters when attention is restricted to the gas within
the effective radius. We argue that an aperture matched to the host
stellar light is the most physically motivated way to compare these
systems. Second, from the data in hand, the intrinsic scatter within
this relation is $\sim 0.5$ dex across the entire relation with no
dependence on galaxy mass ($L_K$). Third, when we focus on the
galaxy-scale aperture measurements, the slopes are found to be
consistent (within 1 sigma) for both the fast-rotating and
slow-rotating galaxies. We further determine that for the high mass
galaxies, there is no distinction between the X-ray luminosities (or
temperatures) of galaxies based on their locale in the large-scale
environment. Specifically, we show that isolated/field galaxies (with
$N_{\rm neighbors} \leq 3$) and those systems considered to be central
or satellite members of their respective galaxy groups, all occupy the
same locus of the $L_{\rm X,gas}-T_{\rm gas}$ relation; there is no
evidence for an offset towards high-$L_{\rm X,gas}$ for central
galaxies in groups.

\subsubsection{Comparison with Previous Work}

To date, the largest study of early-type systems with high-quality
X-ray observations was presented in \cite{Kim:2015aa}. By focusing on
the ATLAS$^{\rm 3D}$ sample, Kim \& Fabbiano made a comprehensive
effort to link the stellar and hot gas properties of a well-defined,
volume-limited sample of early-type galaxies. The main result from Kim
\& Fabbiano is illustrated schematically in
Figure~\ref{fig:lx_T}. They argue that low-mass, fast-rotating,
power-law ellipticals and S0 galaxies occupy a poorly-defined region
of $L_{\rm X,gas}-T_{\rm gas}$ parameter space, with
$L_{\rm X,gas} \sim (0.1-10) \times 10^{39} \ergps$ and $T\sim0.2-0.6$~keV, as
illustrated schematically by the blue circle in Figure~\ref{fig:lx_T}
(see also their Figs.~4 and 8). By contrast, when they focus on the
core galaxies in ATLAS$^{\rm 3D}$ that are isolated field galaxies,
they seemingly form a tight ($\sim$0.2~dex) relation in $L_{\rm X,gas}-T_{\rm gas}$,
following a steep slope of $\sim4.5$ (the red ellipse). However, due
to the limitied 40 Mpc volume of ATLAS$^{\rm 3D}$, their survey is
vastly dominated by relatively low-mass fast-rotating power-law
early-type galaxies.

By adding a much larger sample of high-mass galaxies, we first confirm
(within $1\sigma$) the slope of $\sim$4.5 measured by Kim \& Fabbiano,
although we argue that it is only robustly determined when considering
the gas within $R_e$. If we consider the MASSIVE galaxies in
isolation, we measure a slope of $4.5\pm0.2$ for measurements
performed within $R_e$.\footnote{Due to our adopted parameterization
  of the X-ray spectra, the resultant best-fit temperature measurement
  within $R_e$ for sources with low-$L_{\rm X,gas}$
  ($< 10^{39} \ergps$; low number of source counts) is degenerate
  around $T\sim0.3-0.4$~keV, due to roughly equivalent contributions
  of the ISM, unresolved LMXBs and/or AB/CV components in the X-ray
  spectroscopy. Due to this degeneracy, sources with ISM at
  $T\lesssim 0.3$~keV are biased marginally high, and the measured
  uncertainties on $T_{\rm gas}$ are large ($>$50\%), as shown in Fig
  6a. Hence, we measure a marginally steeper slope of $\sim$4.7 for
  the full (ATLAS$^{\rm 3D}$ $+$ MASSIVE) sample, though this is
  within the 1-$\sigma$ uncertainty of that found by \cite{Kim:2015aa}
  and when only considering the more robust measurements of the
  MASSIVE galaxies.} By contrast, the slope becomes poorly determined
if we consider the measurements in the full aperture, owing to the
significant scatter in $L_{\rm X,gas}$ likely due to mixing and
contamination from gas associated with the group/cluster halo. In
contrast to the previous results of Kim \& Fabbiano, we do not find
evidence that the detailed properties of the early-type galaxies
(e.g., fast/slow rotation; radio power at 1.4 GHz; large scale halo
mass) play a significant role in either the slope or the scatter in
the $L_{\rm X,gas}-T_{\rm gas}$ relation (see Section~\ref{sec:lx_t}),
when the X-ray measurements are performed within physically motivated
$R_e$ apertures. Therefore, we deduce that an
$L_{\rm X,gas} \propto T_{\rm gas}^{\sim4.5}$ relation can be
considered universal across all early-type galaxies when the X-ray
properties are measured on scales set by the stellar light.

Our different lines of investigation strongly suggest that the X-ray
properties of elliptical galaxies can be readily determined by
considering only their X-ray emission within $1R_e$. On these smaller
scales, the gas temperature may be primarily set by the sub-halo of
the galaxy being at (or close to) virial equilibrium with bulk gas
motions due to temperature diffences (so-called cooling flows)
possibly playing a secondary role\footnote{NGC~1129 is a distinct
  outlier from the observed $L_{\rm X,gas}$--$T_{\rm gas}^{4.5}$
  relation. NGC~1129 is the central galaxy of the poor group
  AWM7. There are two clear X-ray peaks within AWM7, one centered on
  the 2MASS position of NGC~1129, and the other offset $\sim 1$~kpc
  eastwards \citep{Furusho:2003aa}. The X-ray surface brightness (SB)
  profile is strongly peaked, with small scale disturbances in the
  core, with evidence for bulk motions and a cooling flow
  \citep{Neumann:1995aa,Furusho:2003aa}. Additionally, there are small
  SB depressions indicative of gas density fluctuations
  \citep{Sanders:2012aa}. Given the short cooling time of the group
  ($\sim 0.4$~Gyr; \citealt{Cavagnolo:2009aa}) as well as evidence for
  a recent merger given the radial twist in the galaxy morphology
  \citep{Peletier:1990aa}, it is likely that the gas within $1R_e$
  will relax and rapidly cool to a gas temperature more consistent
  with the velocity-dispersion of NGC~1129 (see
  section~\ref{sec:t_sigma}).}. The single sequence observed in
$L_{\rm X,gas}-T_{\rm gas}$ for all early-type galaxies suggests that,
despite the vastly different mass and gravitational scales, the gas
$T_{\rm gas}$ and $L_{\rm X,gas}$ of the inner $10-30$~kpc of a galaxy
group or cluster are set solely by the properties of the galaxy
residing at its center, and not by the wider environment. Overall, our
analysis of the X-ray luminosity and temperature distributions of
elliptical galaxies suggests a smooth and uniform scaling of the
intrinsic X-ray properties between low-mass (ATLAS$^{\rm 3D}$) and
high-mass (MASSIVE) early-type systems.

\subsection{Linking the X-ray \& optical properties of massive
  early-type galaxies}
\label{sec:lx_opt}

\begin{figure*}[ht]
\centering
\includegraphics[width=0.95\textwidth]{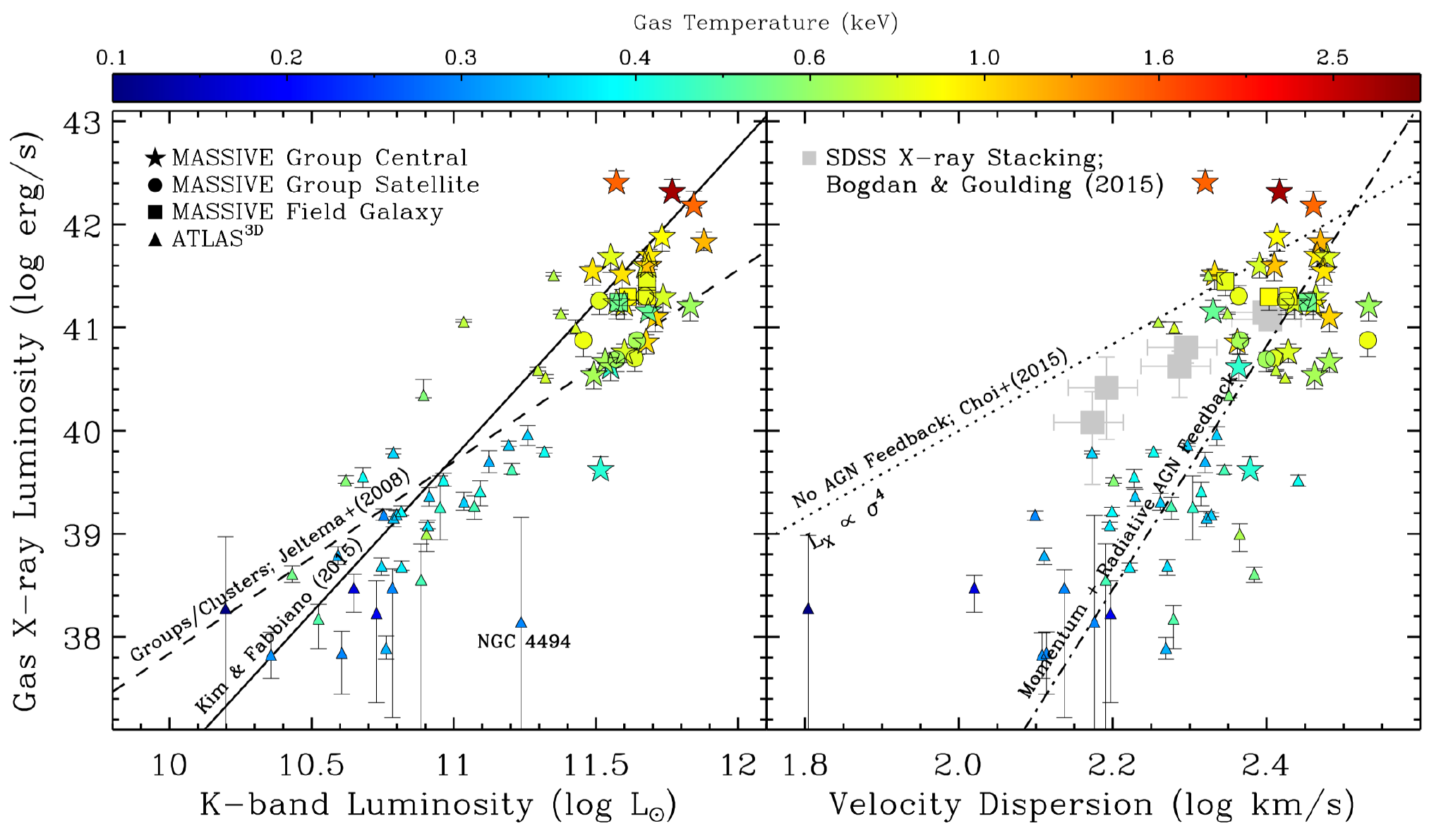}
\caption{{\bf Left:} Logarithm of 0.3--5.0~keV X-ray luminosity (in
  units of erg/s) of the diffuse gas measured within optical radii of
  the galaxy versus $K$-band luminosity (in units of solar
  luminosities). Stars represent the MASSIVE galaxies considered to be
  at the centers of their respective group halos (Crook et
  al. 2007). Circles represent MASSIVE galaxies considered to be
  satellite galaxies within the group halo. Squares represent the
  field galaxies. ATLAS$^{\rm 3D}$ galaxies are shown with filled
  triangles (see Section~\ref{sec:atlas3d} and Appendix). The
  $L_{\rm X,gas}$--$L_K$ relations for individual ellipticals
  \citep{Kim:2015aa} and groups/clusters \citep{Jeltema:2008aa} are
  shown as solid and dashed lines, respectively. Our best-fit relation
  is $L_{\rm X,gas} \propto L_K^{2.5-2.8}$ with intrinsic scatter
  $\epsilon_{int} \sim 0.44-0.74$~dex. {\bf Right:} Logarithm of
  0.3--5.0~keV X-ray gas luminosity (in units of erg/s) versus the
  stellar velocity dispersion measured out to one effective radius (in
  units of km/s). Large filled gray square symbols show the
  $L_{\rm X,gas}$ measurements found from X-ray stacking of SDSS
  elliptical galaxies \citep{Bogdan:2015aa}. Cosmological
  hydrodynamical simulations of the predicted X-ray luminosity arising
  during the formation of galaxies when the effects of mechanical and
  radiation driven feedback (dash-dot line) and no AGN feedback
  (dotted line; i.e., a self-similar
  $L_{\rm X,gas} \propto \sigma_e^4$ relation) are included
  \citep{Choi:2015aa}. Our best-fit relation is
  $L_{\rm X,gas} \propto \sigma_e^{6.5-8.7}$ with intrinsic scatter
  $\epsilon_{int} \sim 0.71-0.97$~dex. {\bf Common:} Color scaling
  represents the gas X-ray temperature derived from the X-ray spectral
  fitting analyses (see Section~\ref{sec:xray_spec} and Appendix).}
\label{fig:lx_lk_sigma}
\end{figure*}

Several origins for the hot gas surrounding early-type galaxies have
been proposed in the previous decades, such as accretion from
extremely large scales (\citealt{Forman:1985aa, Trinchieri:1985aa}) or
from stellar mass loss material due to evolved stars (e.g.,
\citealt{Mathews:1990aa,Brighenti:1997aa,Mathews:2003aa}). The
simplest self-similar expectation for the X-ray luminosity of the gas
is that $L_{\rm X,gas} \sim m_{\rm gas}T_{\rm gas}^2$. Assuming a
galactic origin for the gas, one would then expect a significant
correlation between $L_{\rm X,gas}$ and the galaxy stellar mass and/or
the stellar velocity dispersion. X-ray observations have demonstrated
that early-type galaxies with similar stellar masses can have vastly
different X-ray properties
\citep{Forman:1985aa,OSullivan:2001aa,Mathews:2003aa,Bogdan:2011aa},
resulting in significant scatter in these relations. In the following
sections, we combine the X-ray and optical properties of our sample of
early-type galaxies. Using the large baselines in stellar mass,
velocity dispersion, environment, galaxy rotation and accretion rates
afforded by our sample, we can explore the parameters that may drive
the scatter previously seen in the X-ray--optical relations.

\subsubsection{The $L_{\rm X,gas}$--$L_K$ relation}

Many previous investigations have sought to identify the relationship
between the X-ray luminosity of the hot gas and the galaxy stellar
content, most succinctly probed in the near-IR. In
Figure~\ref{fig:lx_lk_sigma}, we show the relation between the total
$K$-band luminosity ($L_K$), taken directly from the 2MASS extended
source catalog (parameter {\tt k\_m\_ext}), and X-ray luminosity of
the hot gas component ($L_{\rm X,gas}$) extracted within a $1R_e$
aperture\footnote{As we provide clear evidence to in
  Figure~\ref{fig:lx_T}, the X-ray properties of even the most massive
  elliptical galaxies (irrespective of their larger-scale
  environments) when measured at $r<R_e$ are consistent with the same
  scaling as early-type field galaxies. Given our goal is to probe the
  link between the optical and X-ray properties, we will consider only
  the X-ray measurements from the $R_e$ aperture for all subsequent
  analyses.}.

Although the MASSIVE galaxy sample spans a relatively modest range in
$L_K$ ($\sim (3-8) \times 10^{11} \Lsun$) the range covered in
$L_{\rm X,gas}$ is many times larger. Even considering only the hot
gas entrained within 1$R_e$, the range in $L_{\rm X,gas}$ for these
systems is, on average, almost $\sim 2$ orders of magnitude at a given
$L_K$, which increases to $\sim$3 orders of magnitude if we include
the low-luminosity outlier, NGC~5322.

To provide the largest baseline in stellar mass, in
Fig.~\ref{fig:lx_lk_sigma}a we additionally include the X-ray and
$K$-band luminosities for the ATLAS$^{\rm 3D}$ early-type galaxies
(see Section~\ref{sec:atlas3d} and the Appendix). Similar to
Section~\ref{sec:lx_t}, we perform a linear regression using {\sc
  linmix\_err} \citep{Kelly:2007aa} to include the uncertainties in
both $L_{X}$ and $L_{K}$ (for the combined ATLAS$^{\rm 3D}$ and
MASSIVE galaxies). We find that the best fit relation goes as
$L_{\rm X,gas} \propto L_K^{2.5-2.8}$ (for 67\% of the posterior
samples). The intrinsic scatter is determined to be
$\sim 0.44-0.74$~dex for the posterior draws, but characterized by an
asymmetric distribution with a significant tail out to
$\sim$1.1~dex.\footnote{We note that the choice of a 3-Gaussian
  mixture model versus a flat prior does not result in significant
  differences in the slope or the derived intrinsic scatter of the
  fitted relation.}  We find no clear evidence for a break in the
$L_{\rm X,gas}-L_K$ relationship as both galaxy samples, when fit
independently, are characterized by similar slopes
($\propto L_K^{\approx 2.7}$) and both present similar scatter (median
$\approx 0.6$~dex). Hence, the MASSIVE galaxies provide a smooth
continuation of the $L_{\rm X,gas}-L_K$ relationship determined for
the ATLAS$^{\rm 3D}$. Our observed slope for
MASSIVE$+$ATLAS$^{\rm 3D}$ measured within $R_e$ is marginally flatter
than the slope ($3 \pm 0.4$) presented in Kim \& Fabbiano (2015) based
solely on large aperture measurements of the ATLAS$^{\rm 3D}$
galaxies.

We find that the shallower $L_{\rm X,gas}-L_K$ relationship described
by the groups and cluster systems ($L_{\rm X,gas} \propto L_K^{1.9}$)
outlined in Jeltema et al. (2008; see also \citealt{Mulchaey:2010aa})
provides a lower envelope that almost a third of the MASSIVE sample
lie on or close to. Our limited sample size does not allow us to probe
for a divergence into two branches of the $L_{\rm X,gas} - L_K$
relationship at $L_K \gtrsim 2\times 10^{11}\Lsun$. At a given $L_K$,
there appears to be little or no difference between the measured
$L_{\rm X,gas}$ of isolated/field early-type galaxies and those
residing close to the center of a group (central galaxies), suggesting
that large-scale environment does not primarily drive the observed
$L_{\rm X,gas}$ at a given $L_K$.

Overall, due to the smooth transition observed between the
ATLAS$^{\rm 3D}$ and MASSIVE samples, across the full range in $L_K$
considered here, we find little evidence that there is a particular
$L_K$ threshold where early-type galaxies become less effective at
retaining their gas haloes.

\subsubsection{The $L_{\rm X,gas}$--$\sigma_e$ relation}

Some recent studies argue that the stellar velocity dispersion, set by
the galaxy's gravitational potential, may provide a better tracer of
the hot gas. Assuming that the gas entrained in the gravitational
potential is virialized, self-similarity predicts
$L_{\rm X,gas} \propto \sigma^4$. This has previously been observed in
galaxy clusters (e.g.,
\citealt{Mushotzky:1997aa,Mulchaey:1998aa,Wu:1999aa,Helsdon:2000aa,Xue:2000aa,Mahdavi:2001aa,Ortiz-Gil:2004aa}). However,
several relations between total $L_{\rm X,gas}$ and $\sigma$ have been
reported for elliptical galaxies, with exponents ranging from
$\sim 8-14$ (including uncertainties; e.g.,
\citealt{Eskridge:1995aa,Mahdavi:2001aa,Diehl:2007aa}).

In Figure~\ref{fig:lx_lk_sigma}b we show the $L_{\rm X,gas} - \sigma$
distribution for the MASSIVE and ATLAS$^{\rm 3D}$ galaxies. With the
inclusion of the higher-mass early-type galaxy sample from MASSIVE, a
correlation between $L_{\rm X,R_e}$ and $\sigma_e$ clearly continues
towards high values of $L_{\rm X,gas}$, following a steep
$\sigma_e^{6.5-8.7}$ (for 67\% of the posterior samples) relationship,
and an intrinsic scatter of $\sim$0.71--0.97~dex.

In Figure~\ref{fig:lx_lk_sigma}b, we highlight the stacked X-ray
luminosities and average velocity dispersion bins derived from SDSS
galaxies in \cite{Bogdan:2015aa}. Though not fully inconsistent, given
the large uncertainties of the stacked measurements, the
$L_{\rm X,gas}$ derived from stacking are systematically high at
$\sigma_e \sim 160$km/s when compared to the ATLAS$^{\rm 3D}$
sample. While statistically complete, the \cite{Bogdan:2015aa}
analysis performed with the ROSAT X-ray telescope lacked the angular
resolution as well as sufficient signal-to-noise in the stacked X-ray
images to probe ellipticals with low stellar mass or low stellar
velocity dispersions. In the narrow luminosity range,
$L_{\rm X,gas} \sim (0.9-3) \times 10^{40} \ergps$, where the SDSS
stacking results are most complete, our individual detections are
considerably lower than the $L_{\rm X,gas}$ observed in the X-ray
stacks. This excess seen in the stack may be of physical origin due to
contamination at large scales (due to the poor angular resolution of
ROSAT) or may be induced by the presumption of a single gas
temperature for each stack when calculating X-ray luminosities, as we
show in Section~\ref{sec:t_sigma}, galaxies with low-$\sigma_e$ have
systematically lower gas temperatures. Alternately, there may be a
long tail in the luminosity distribution towards high-$L_{\rm X,gas}$
at low- to moderate-$\sigma_e$ that systematically drives up the mean
$L_{\rm X,gas}$ observed in the stacks.

The steeper slope observed for the individual galaxies over that from
the X-ray stacking results is similar to that predicted by recent
hydrodynamical simulations (e.g., \citealt{Choi:2015aa}). These models
predict systematically lower X-ray luminosities than those expected
from a self-similar ($\propto \sigma_e^4$) origin due to the injection
of energy back into the thermal ISM or the removal of hot gas by
mechanical and radiatively-driven AGN feedback. For a given $\sigma_e$
these feedback models produce $L_{\rm X,gas}$ in line with the general
trend we find in Figure~\ref{fig:lx_lk_sigma}b. However, we show that
AGN may play less of a role for high-$\sigma_e$ galaxies, as these
sources do not diverge significantly from a more simple self-similar
model that does not require feedback.

\subsubsection{The scatter in $L_{\rm X,gas}$--$L_K$ versus $L_{\rm X,gas}$--$\sigma_e$}

\begin{figure*}[ht]
\centering
\includegraphics[width=\textwidth]{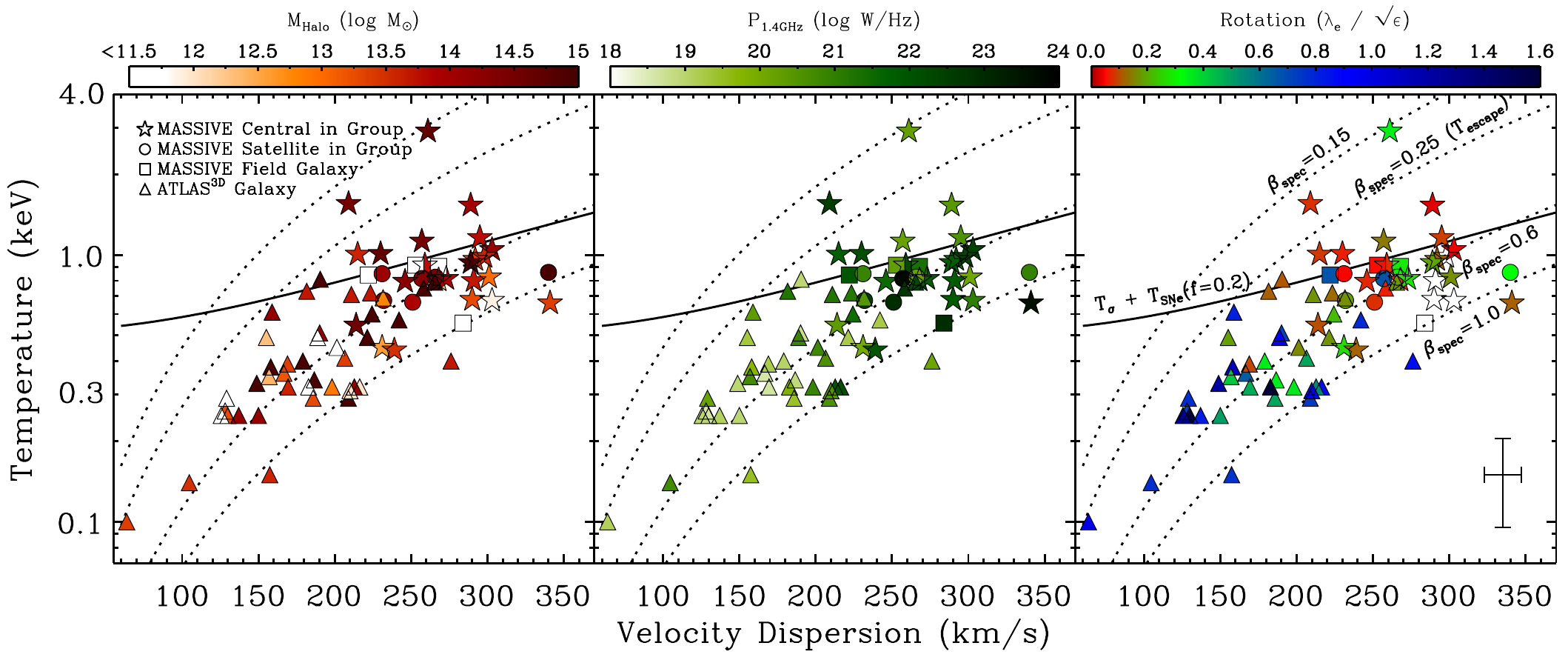}
\caption{X-ray gas temperature (in units of keV) versus the stellar
  velocity dispersion (in units of km/s) measured within
  $R_e$. Symbols have the same meaning as previous figures. The
  best-fit $T-\sigma_e$ relation goes as $\sigma_e^{1.3-1.8}$ with an
  intrinsic scatter of $\sim 0.13-0.16$~dex. The idealized gas virial
  temperature,
  $\beta_{\rm spec} = \mu m_p \sigma_e^2 / kT_{\rm gas} = 1$,
  $\beta_{\rm spec} =0.6$,0.25 and 0.15 are shown with dotted
  lines). The mass-averaged injection temperature produced due to
  perceived supernovae activity (i.e., the sum of the virial
  temperature and the gas temperature from SNe with $f=0.2$; see
  \cite{Pellegrini:2011aa}) is shown with a solid line.  Color
  gradient provides the total mass of the large (group-scale) halo
  (left panel; Crook et al. 2007); logarithm of the 1.4 GHz radio
  power in W/Hz (center panel); and galaxy rotation factor
  ($\lambda_e / \sqrt{\epsilon}$; right panel).}
\label{fig:T_sigma}
\end{figure*}

The combination of the MASSIVE and ATLAS$^{\rm 3D}$ galaxies afford
the opportunity to ascertain the fundamental relationship between the
hot gas and the galaxy properties, as well as place constraints on the
properties that may drive the scatter in these relations. We use a
Bayesian bi-sector estimator to compare the dispersions between
$L_{\rm X,gas}$, $L_K$ and $\sigma_e$ while correctly accounting for
the uncertainties in the (in)dependent variables. From the previous
sections, we find that the intrinsic scatter in these relationships
goes as $\epsilon(L_{\rm X,gas}/\sigma_e) \sim 0.71-0.97$~dex and
$\epsilon(L_{\rm X,gas}/L_K) \sim 0.44-0.74$~dex. To zeroth order,
this could suggest that the hot gas and the galaxy are linked
primarily through the stellar mass and not the gravitational potential
(probed by $\sigma_e$). However, there are individual early-type
galaxies, such as NGC~4342 and NGC~4291, that have values of
$L_{\rm X,gas}$ that are 10--30 times larger than other galaxies with
similar $L_K$ (see \citealt{Bogdan:2012aa,Bogdan:2012ab}), but have
high values of $\sigma_e$ that provides are far better prediction of
$L_{\rm X,gas}$.

There are several observational and physical origins to the scatter in
these relations. Measurement uncertainties in $R_e$ and projection
effects could both contribute to systematics in $L_K$, $\sigma_e$ and
$L_{\rm X,gas}$ on a galaxy-to-galaxy basis. However, these
uncertainties are likely to be small compared to the observed scatter.

In Fig~\ref{fig:lx_lk_sigma} we provide the associated gas
temperatures of both the ATLAS$^{\rm 3D}$ and MASSIVE samples. At a
given $L_K$ or $\sigma_e$, we show in Figure~\ref{fig:lx_lk_sigma}
that the measured $L_{\rm X,gas}$ is strongly dependent on the gas
temperature, and hence, the dominant source of scatter in these
X-ray--optical relations may be due to gas temperature variations. To
quantify the dependence of $L_{\rm X,gas}$ on $T_{\rm gas}$ at fixed
$L_K$, we take the narrow range in
${\rm log} (L_K/L_{\odot}) \sim 11.25-11.75$ defined by our MASSIVE
sample alone and refit the relation between $L_{\rm
  X,gas}$--$T_{\rm
  gas}$
for these objects. We determine that for a fixed value of $L_K$,
$L_{\rm X,gas} \propto T_{\rm gas}^{4.1 \pm 0.4}$, which is consistent
with our earlier finding of a univeral relationship between
$L_{\rm X,gas}$ and $T_{\rm gas}$ in Section~\ref{sec:lx_t}. By
contrast, if we fix $L_{\rm X,gas}$, we find no such dependence on
$T_{\rm gas}$ with $L_K$ ($L_K \propto T_{\rm gas}^{-0.1 \pm 0.4}$)
for those sources with $L_{\rm X,gas} \sim 10^{39}-10^{40} \ergps$.
Furthermore, we determine that those early-type galaxies that lie
below the $L_{\rm X,gas} - L_K$ relation of \cite{Kim:2015aa}, and
hence have relatively low $L_{\rm X,gas}$ (i.e., they are possibly
gas-poor for their stellar mass), typically exhibit the coolest gas
temperatures ($T_{\rm gas} \sim 0.2$--0.4).

Given that the $L_{\rm X,gas}-L_K$ and $L_{\rm X,gas} - \sigma_e$
relations are strongly dependent on $T_{\rm gas}$, we suggest that
when considered in isolation, neither $L_K$ or $\sigma_e$ alone are
capable of accurately predicting $L_{\rm X,gas}$ for an early-type
galaxy. Thus, care must be taken when using these scaling relations
for objects lacking a constrained $T_{\rm gas}$ measurement. The gas
temperature is likely to be driven by the gravitational potential well
of the system (see Section~\ref{sec:t_sigma}) with secondary
contributions from energy injection imparted by stochastic effects
that may differ on a galaxy-to-galaxy basis, such as from a variety of
internal and external processes such as gas/halo/stellar stripping,
galaxy--galaxy mergers, and AGN feedback that imparts energy back into
the thermal ISM and/or the dark matter halo. These differing
heating/cooling/feedback effects likely influence the wide range in
gas temperature for galaxies with similar $L_K$ or $\sigma_e$, and
hence, this results in the high scatter seen in
Figure~\ref{fig:lx_lk_sigma}. In the next section, we use our optical
data to further address how gas temperatures may drive this observed
scatter.

\subsection{$T_{\rm gas} - \sigma_e$: probing the relationship between
  gas temperature and gravitational potential}
\label{sec:t_sigma}

Both $L_{\rm X,gas} - L_K$ and $L_{\rm X,gas} - \sigma_e$ present
large scatter and exhibit a clear dependence on gas temperature. Here
we specifically investigate what drives the scatter in $T_{\rm gas}$
at a given $\sigma_e$, i.e., we ask whether $T_{\rm gas}$ is related
to $\sigma_e$ as expected from virial arguments. On the largest
scales, studies have shown that the relationship between the
intra-cluster gas temperature and the velocity dispersion of the
galaxies is consistent with arising solely due to virial arguments
(e.g., \citealt{Wu:1999aa,Xue:2000aa}), while conflicting results have
been found on whether the virial theorem holds for systems with lower
temperature ($T<1$ keV; e.g.,
\citealt{White:1991ab,Ponman:1996aa,White:1997aa,Mulchaey:1998aa,Xue:2000aa,Pellegrini:2011aa}
and references there-in).

Figure~\ref{fig:T_sigma} presents the $T_{\rm gas}-\sigma_e$
distribution for the MASSIVE and ATLAS$^{\rm 3D}$ galaxy samples. A
Pearson-test reveals a significant correlation (at the $>$99.99\%
level) between $T_{\rm gas}$ and $\sigma_e$, with $r = 0.78$ for a
two-tail test with 72 degrees of freedom. By applying the same
Bayesian estimator as before (see section~\ref{sec:lx_opt}), we find
that $T_{\rm gas} \propto \sigma_e^{1.3 - 1.8}$, with intrinsic
scatter of $\sim 0.13-0.16$~dex. This scaling is marginally flatter
than the simple virial expectation of $\sigma_e^2$, and is similar to
that found by \cite{Davis:1996aa} using ROSAT data. Using the same
Bayesian bi-sector method as in the previous section, we compare the
scatter in $T_{\rm gas}-\sigma_e$ to the $L_{\rm X,gas}-L_K$ and
$L_{\rm X,gas}-\sigma_e$ relations, finding that the scatter in
$T_{\rm gas}-\sigma_e$ is a factor $\sim 2$ and 4 smaller,
respectively. Hence, the scatter observed in the
$T_{\rm gas}-\sigma_e$ relation is significantly smaller than the
scatter found in any relation containing $L_{\rm X,gas}$. This in turn
suggests that the gas temperature is fundamentally connected to the
galaxy potential.

The galaxies considered here cover a significant range in both gas
temperature ($T_{\rm gas} \sim 0.1-3.0$~keV) and velocity dispersion
($\sigma_e \sim 60$--330 km/s). In Figure~\ref{fig:T_sigma} it is
abundantly clear that there are no high velocity dispersion galaxies
($\sigma_e > 250 \kmps$) with low gas temperatures ($T < 0.5$~keV).
This apparent paucity of objects is expected assuming a virial origin
for the heating of the gas. By comparing our sample to the simplest
theoretical prediction for the virial gas temperature, such that
$\beta_{\rm spec} = \mu m_p \sigma^2 / kT_{\rm gas}$, where $m_p$ is
the mass of a proton, $\mu m_p$ is the mean particle mass with
$\mu =0.62$, and $k$ is the Boltzmann constant (e.g.,
\citealt{Loewenstein:1999aa,Loewenstein:2000aa,Sun:2007aa,Jeltema:2008aa,Pellegrini:2011aa}),
we show that within the measurement uncertainties, no galaxies
considered here have $\beta_{\rm spec} > 1$ suggesting that a virial
model provides a lower-boundary to the observed gas temperature in
early-type galaxies (e.g., \citealt{Cox:2006ab}). Hence, we find good
evidence that in all systems, the minimum gas temperature,
$T_{\rm gas,min}$, goes as $\sigma_e^2$.

Consistent with previous results (e.g., \citealt{Loewenstein:1999aa}),
the median $\beta_{\rm spec}$ for the sample is $\sim 0.6$.  A median
$\beta_{\rm spec} < 1$ suggests that, on average, every galaxy is
under-going (or has recently undergone) some form of supplemental
heating from internal/external processes (e.g., SNe feedback;
star-formation; AGN feedback; large-scale environment; galaxy
dynamics), such as those highlighted in some theoretical studies
(e.g., \citealt{Pellegrini:2011aa,Negri:2015aa}) or are undergoing a
rapid cooling phase back to the virial temperature. Such a finding is
also consistent with detailed studies of radial entropy profiles in
some nearby systems, finding excess entropy over that expected from
gravitational heating alone (e.g.,
\citealt{Humphrey:2008ab,Humphrey:2011aa,Humphrey:2012aa,Werner:2012aa}). This
excess heating must occur on time scales shorter than the cooling time
of recently acquired gas (e.g., for NGC~1129
$t_{\rm cooling} < 0.4$~Gyr). A median $\beta_{\rm spec} < 1$ may also
hint at the dark matter halo playing a significant role in heating the
gas above the virial temperature (e.g.,
\citealt{Davis:1996aa,Loewenstein:1999aa}). However, in this study we
measured $T_{\rm gas}$ on $R_e$ scales, and at these radii the stellar
mass dominates the gravitational potential (e.g.,
\citealt{Humphrey:2006aa,Humphrey:2009aa}). As such, the effect of
heating due to the halo should still be small. In
Figure~\ref{fig:T_sigma}, we further utilize available ancillary data
to probe the scatter in $\beta_{\rm spec}$ for our galaxy sample to
place constraints on the heating mechanisms.

\subsubsection{The role of large-scale environment in X-ray gas heating}

A galaxy property that has been readily investigated in the literature
is that of the large-scale environment. There is an expectation that
the large scatter in the X-ray temperature or luminosity at a given
stellar mass is, in part, due to the large scale environment of the
galaxy, i.e., group/cluster-scale effects. Galaxies embedded at the
center of a group should retain their hot gas, and are predicted to
have enhanced X-ray emission due to the group gas. By contrast,
satellite systems are expected to lose gas due to ram-pressure or
viscous stripping. Recent studies have focused on small samples of
isolated early-type galaxies in order to mitigate the effects of the
host entrained as part of the intra-cluster medium. However, results
have been inconclusive, with these investigations determining that
some isolated systems retain their hot ISM, while others do not (e.g.,
Bogdan et al. 2012; Kim \& Fabbiano 2013).

In the left panel of Figure~\ref{fig:T_sigma} we include an estimate
of the total halo mass measured from the motions of galaxies in the
group taken from the 2MRS \citep{crooketal2007}. The nine most
isolated galaxies (i.e., those in the smallest halos) considered here
span relatively wide ranges in temperature ($T \sim 0.2-0.9$~keV) and
velocity dispersion ($\sigma_e \sim 120-290 \kmps$). If the
environment plays a substantial role in the central gas temperature,
then isolated galaxies should have lower temperature and lower
scatter. We find no evidence for such an environment-driven
scenario in this sample; the parameter space occupied by isolated
galaxies at fixed $\sigma_e$ is the same as that for galaxies that
reside in the most massive environments.

\begin{figure}[ht]
\centering
\includegraphics[width=\linewidth]{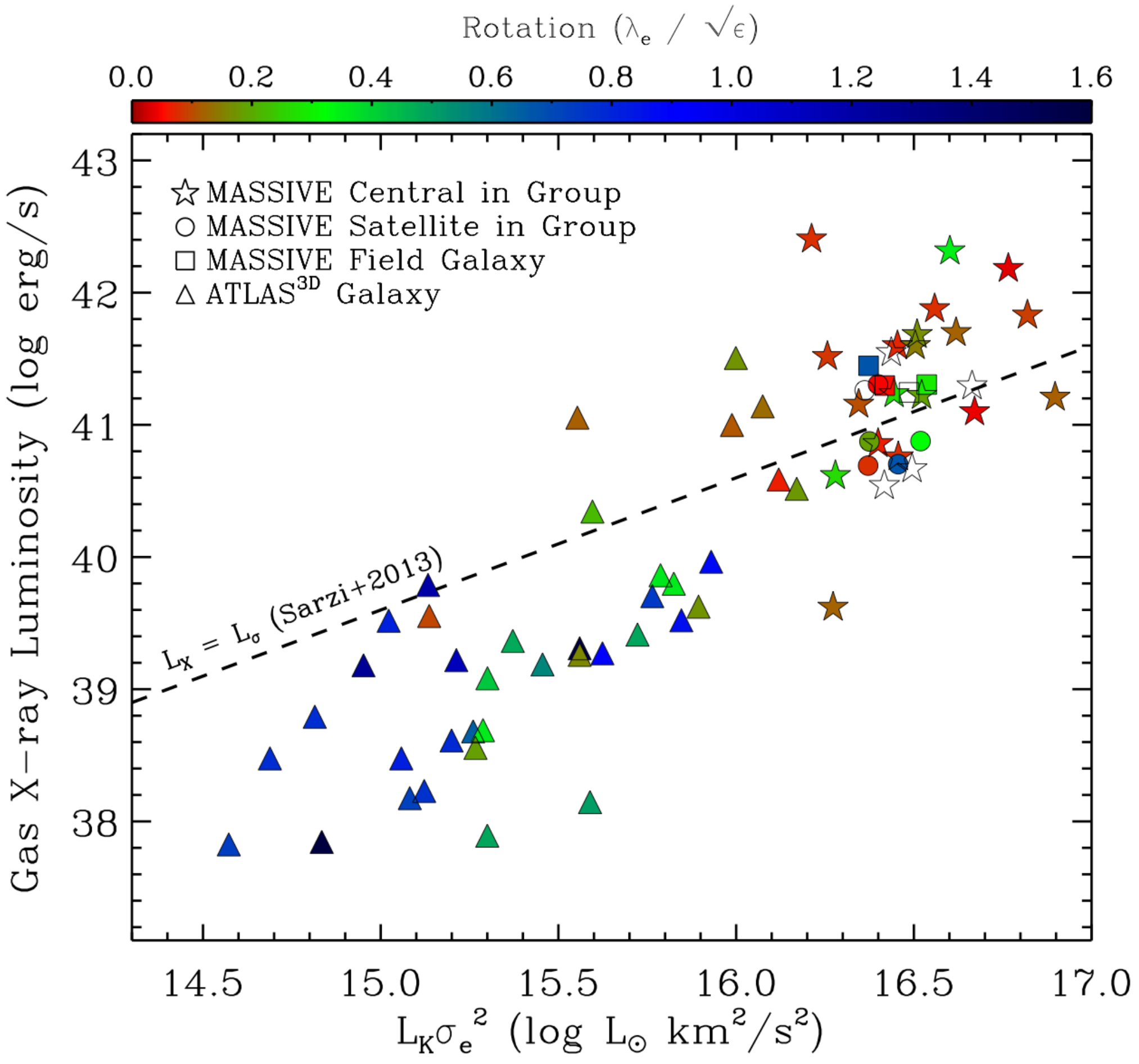}
\caption{Logarithm of 0.3--5.0~keV X-ray luminosity of the
  diffuse halo gas versus $K$-band luminosity multiplied by the square
  of the central stellar velocity dispersion (i.e., the kinetic energy
  associated with stellar motion). Symbols have the same meaning as
  previous figures. Color gradient represents the 2-D rotational
  parameter
  $\lambda_R \sim \langle R |V| \rangle / \langle R \sqrt{V^2 +
    \sigma_{*}^2}$,
  measurements are taken from Veale et al. (in prep.) and Emsellem et
  al. (2011). The nominal separation between fast and slow rotators is
  $\lambda_{R_e} = 0.31 \sqrt{\epsilon}$. Following Sarzi et
  al. (2013), we show the theoretical prediction for the X-ray gas
  being produced due to stellar mass loss material
  ($L_{\rm X,gas} = L_{\sigma} = \frac{3}{2} \dot{M} \sigma_e^2$), dashed-line.}
\label{fig:lx_lksig}
\end{figure}

\subsubsection{The role of outflows in X-ray gas heating}

A further source of heating could be due to energy imparted due to
out-flowing material, conceivably in the form of a radio jet due to
black hole accretion or a supernova driven wind. Radio inflation of
X-ray cavities shock heats and redistributes gas to large radii (see
\citealt{Finoguenov:2002aa,Forman:2005aa,mcnamara07,Kraft:2012aa,Bogdan:2014aa}). This
gas remains trapped within the wider halo, but then rains back towards
the galaxy and the incident ISM, giving rise to additional shocks as
well as being subject to gravitational heating, causing the gas to be
heated to higher temperatures (see
\citealt{Allen:2006aa,Dunn:2010aa}). Here we investigate whether the
increased gas temperatures seen in some of the sample may be due to
mechanical heating in the form of an AGN-driven radio jet.

In Fig.~\ref{fig:T_sigma}b, we show the rest-frame NRAO VLA Sky Survey
(NVSS) or Faint Images of the Radio Sky at Twenty-cm (FIRST) radio
luminosities at 1.4 GHz ($P_{\rm 1.4 GHz}$) for the ATLAS$^{\rm 3D}$
and MASSIVE galaxies measured using the integrated flux measurements
provided in the appropriate radio catalog (see Table 1). We find a
general trend of increasing radio luminosity with increasing
$\sigma_e$. Such a trend is expected given the well-known relations
between the radio-power and the mass of the stellar component and the
central black holes (e.g., \citealt{Best:2005aa}), both of which
$\sigma$ is used as a proxy for. However, we find no obvious trend
between $\beta_{\rm spec}$ and $P_{\rm 1.4 GHz}$, suggesting that
radio power alone cannot be driving the scatter in the gas
temperatures and/or that mechanical heating, specifically on the scale
of the galaxy, has a limited effect. We note that of the two sources
that exhibit temperatures in excess of that required for the gas to
escape the system (i.e., $\beta_{\rm spec} < 0.25$), one galaxy
(NGC~708) is among the most radio-luminous considered here with
$P_{\rm 1.4 GHz} \sim 4 \times 10^{22}$~W/Hz. In NGC~708 the
mechanical heating from the jet may be sufficient to expel the hot gas
from the galaxy entirely. Indeed, in-depth studies of the cluster
host, Abell~262, show a complex structure and cooling flow towards the
central region surrounding NGC~708 with X-ray cavities inflated by
radio lobes, suggesting a once powerful outflow (see
\citealt{Blanton:2004aa}). In line with these results, we find that
the X-ray emitting gas within $R_e$ is better fit by two separate
temperature components with $T_1 \sim 2.0$~keV and $T_2 \sim
1.1$~keV.
The cooler component has a substantially lower luminosity by a factor
$\sim5-7$ compared with the hot component. The more luminous component
has a similar gas temperature to that found at larger radii, which may
be the result of a recent merger between two large halos. However, we
note the presence of a secondary bright galaxy that lies at a distance
of $r< R_e$, which may also be the source of the cooler gas component.

Energy injection by supernovae (SNe) activity may induce shocks within
the incident gas, and hence, may also play a role in heating. Of the
ATLAS$^{\rm 3D}$ and MASSIVE galaxies with X-ray observations that
have detectable hot gas emission, all have star-formation rates (SFRs)
less than $\sim 0.1 M_{\odot}/{\rm yr}^{-1}$. Indeed, we show that at
low-$\sigma_e$ ($< 200 \kmps$), where SNe may play a more impactful
role in gas heating, all of the sources considered here fall
significantly below the prediction for SNe heating of
\cite{Pellegrini:2011aa} with $f =0.2$, where $f$ is the fraction of
energy imparted by Type-1a supernovae that is converted to heat.

\subsubsection{The role of galaxy rotation in X-ray gas heating}

Intrinsic flattening and/or the rotation of a galaxy is believed to
serve a role in the ability of a system to retain its hot gas (e.g.,
\citealt{Eskridge:1995aa,Ciotti:1996aa,Brighenti:1997aa}). Using the
apparent ellipticity ($\epsilon$) in conjunction with the measurement
of the system rotational support ($\lambda_e$), we parameterize the
relative rotational speed of the galaxies, and separate our sample
into slow and fast rotators (see Sections~\ref{sec:kinematics} and
~\ref{sec:atlas3d}). We follow \cite{Sarzi:2013aa} and in
Figure~\ref{fig:T_sigma}c we confirm that the fastest rotators with
$\lambda_e / \sqrt{\epsilon} \gtrsim 1.0$ are generally low $\sigma_e$
galaxies with $\sigma_e \lesssim 150$~km/s, while the slowest rotators
with $\lambda_e / \sqrt{\epsilon} \lesssim 0.2$ are the most massive
systems with $\sigma_e \gtrsim 200$~km/s (see also
\citealt{Emsellem:2011aa}). Overall, there appears to be a smooth
continuum with decreasing rotation for increasing $T_{\rm gas}$. Of
the two galaxies with $\beta_{\rm spec} < 0.25$, NGC~708 is a slow
rotator that may have its high gas temperature driven by a radio jet
(see previous sub-section), while NGC~1129 is considered a fast
rotator. Although these are only two objects, they may suggest that
the temperature of the hot gas is primarily set by the galaxy central
potential, but that a combination of other physical galaxy properties
play a secondary role (e.g., galaxy rotation; mechanical heating from
radio jets).

To probe this further, \cite{Sun:2007aa} and \cite{Sarzi:2013aa}
present the $L_{\rm X,gas}$--$L_{K}\sigma_{e}^2$ diagram, which
in-effect traces the X-ray luminosity expected due to the
thermalization of the energy from stellar mass loss material. In this
paradigm, $L_K$ provides a tracer for the mass of the stars where the
gas may originate, and $\sigma_e^2$ provides a proxy for the
gravitational heating energy. This is in analog to the
$L_B \sigma_{e}^2$ parameter introduced by
\cite{Canizares:1987aa}. Based on the ATLAS$^{\rm 3D}$ sample alone,
Sarzi et al. conclude that slow rotating early-type galaxies exhibit
X-ray luminosities\footnote{In the \cite{Sarzi:2013aa} study, the
  X-ray measurements are taken from \cite{Boroson:2011aa}.} that are
sustained by the thermalization of stellar mass loss material, such
that $L_{X} = L_{\sigma} = \frac{3}{2} \dot{M} \sigma_e^2$, while fast
rotators lie significantly below this threshold. However, and as noted
previously, the ATLAS$^{\rm 3D}$ sample contains few slow-rotators. To
predict $L_{\sigma}$, their model assumes a stellar-mass-loss rate
($\dot{M}$) set by an extrapolation of $L_K$ and the average stellar
age of the population (see \citealt{Ciotti:1991aa, Sarzi:2013aa} for
further details).

In Figure~\ref{fig:lx_lksig}, with the inclusion of a large number of
slow rotators from the MASSIVE sample, we clearly show that ({\it a})
a simplistic stellar-mass loss model $L_{\rm X,gas} = L_{\sigma}$ does
not provide a good description of the data, and ({\it b}) that
deviations from this theoretical prediction are not correlated with
the rotation properties of the galaxies. Fast and slow rotators are
clearly present both above and below the $L_{\rm X,gas}$ prediction
from stellar mass loss material. Further examining the dispersion
along the $L_{X} = L_{\sigma}$ axis, we find a strong trend with gas
temperature. The sources that lie significantly below the
$L_{X} = L_{\sigma}$ prediction comprise the coolest early-type
galaxies, while those that lie above the prediction possess the
hottest gas temperatures.\footnote{We note that if we compare the
  MASSIVE sample to the ATLAS$^{\rm 3D}$ sample using the
  \cite{Boroson:2011aa} X-ray measurements our conclusions remain
  qualitatively consistent. Our conclusions are mainly driven by the
  increased dynamic range in stellar mass afforded by our combined
  (MASSIVE$+$ATLAS$^{\rm 3D}$) galaxy sample.}

Although we find an apparent deviation from the theoretical
prediction, we cannot yet rule out that stellar mass loss dominates
the gas supply in these galaxies. Substantial uncertainty remains in
the mass loss rates. Taking empirical scalings based on individual AGB
stars yields $\sim$50\% level differences in the expected mass-loss
rates \citep{Athey:2002aa}.  However, the impact of stellar age and
metallicity on the mass-loss rates, while important, have not been
adequately calibrated. In particular, both stellar age and metallicity
provide additional parameters in this scaling. The zeroth order
relation shown here assumes a constant age, but \cite{Ciotti:1991aa}
suggest that $\dot{M}$ scales as $t_{\rm age}^{-1.3}$. In fact,
including the steadily increasing age towards more massive galaxies
would make the discrepancy between model and data worse. The other
possible second parameter is metallicity. More massive early-type
galaxies are also more metal rich (e.g.,
\citealt{Trager:2000aa,mcdermidetal2014,greeneetal2015}), which
naturally leads to more mass loss per unit mass. Thus, one possible
explanation for the mismatch between the expected luminosity due to
mass loss and that observed, simply arises due to our presumed model
of mass loss.  A third possibility is that the low-mass galaxies are
simply less effective at retaining their gas (e.g., due to more recent
star formation episode, ram-pressure stripping), while the
discrepancies at the high-mass end are caused by innacuracies in the
mass-loss models. The flattening of galaxies, particularly in slow
rotators, may also play a role here. While beyond the scope of this
work, it will be very interesting in the future to see if more
realistic models of mass loss can explain the full range of observed
X-ray luminosities.

\begin{figure}[ht]
\centering
\includegraphics[width=\linewidth]{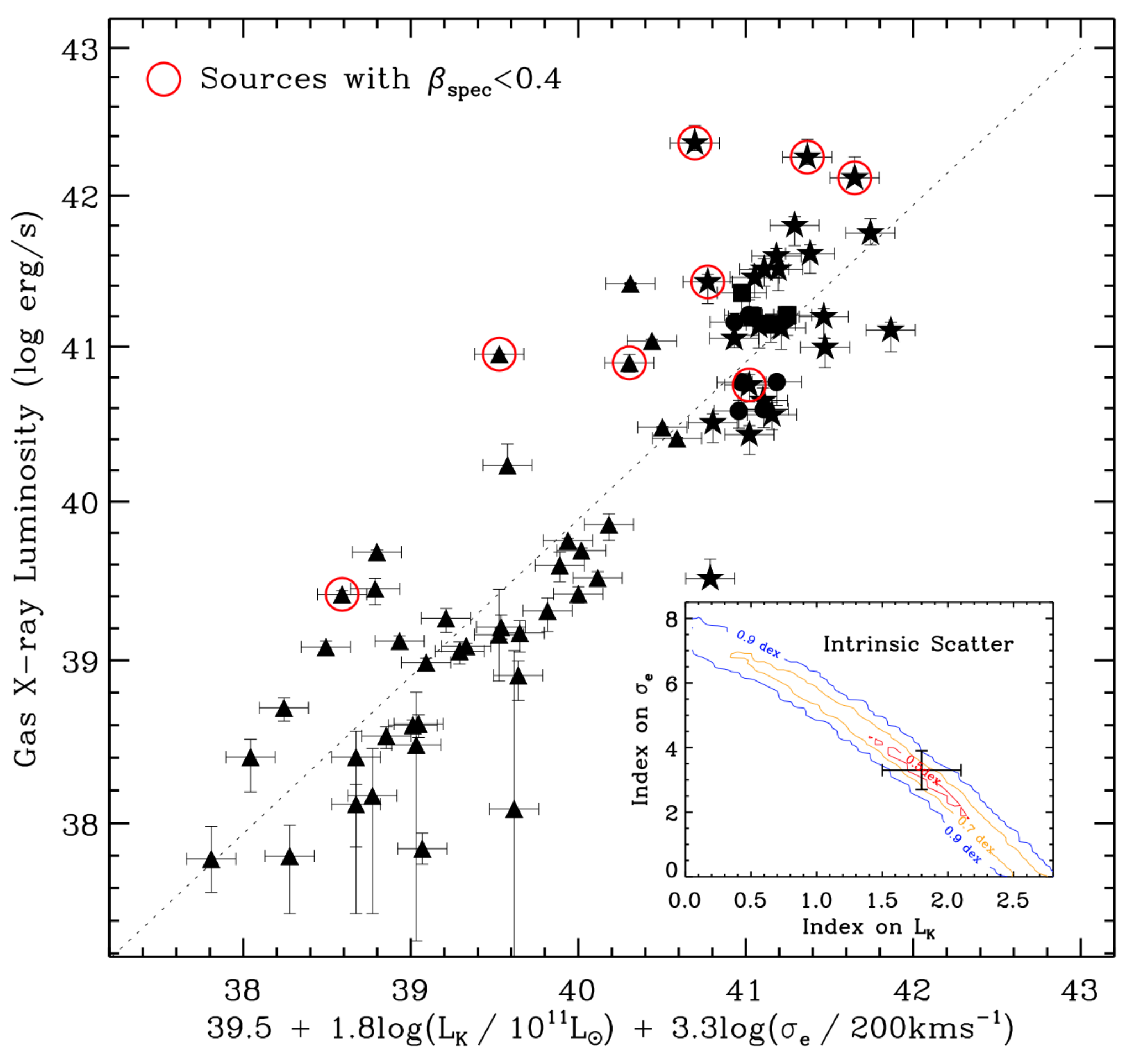}
\caption{Logarithm of 0.3--5.0~keV X-ray luminosity of the diffuse
  halo gas versus the multivariate relationship for $L_K$ and
  $\sigma_e$ described in Section~\ref{sec:finalsubsec} for all
  ellipticals in the MASSIVE and ATLAS$^{\rm 3D}$ samples. Dotted line
  represents the 1:1 relation between the empirical relation and
  $L_{\rm X,gas}$. Filled symbols have the same meaning as previous
  figures. Red open circles represent outliers in the
  $T_{\rm gas} - \sigma_e$ relation with $\beta_{\rm spec} < 0.4$.
  Inset figure shows the best fit and associated uncertainties to the
  indices in the derived $L_K^{\alpha}\sigma_e^{\gamma}$ empirical
  relation. Contours represent the median intrinsic scatter of the
  posterior draws in the derived relations as a function of the
  indices $\alpha$ and $\gamma$ (contours levels -- red:0.5~dex;
  yellow:0.7~dex; blue:0.9~dex). }
\label{fig:final_fig}
\end{figure}

\subsection{An empirical two-optical-parameter X-ray--optical
  relation}
\label{sec:finalsubsec}

$L_K \sigma_e^2$ provides a simple model for the origin of $L_{\rm X,gas}$,
harnessing $L_K$ and $\sigma_e$ as proxies for the stellar-mass
content and the gravitational heating, respectively.  However, as we
show in the previous section, this particular parameterization is
clearly a poor description of the data. Here we attempt to improve on
this scaling relation, following an empirical approach based on our
large galaxy sample. Following \cite{Sun:2007aa} and
\cite{Sarzi:2013aa}, we attempt to parameterize $L_{\rm X,gas}$ using the same
optical proxies, but now introducing free variables for the index
powers, such that $L_{\rm X,gas} \sim L_K^{\alpha}\sigma_e^{\gamma}$.

We perform a Bayesian multi-variate linear regression analysis in
logarithmic space that seeks to find the best-fit linear relation
between log($L_K \sigma_e$) and log($L_{\rm X,gas}$). This process
harnesses a modified version of the {\tt LINMIX\_ERR} algorithm (see
Section~\ref{sec:lx_opt}) to sample the multi-dimensional parameter
space while accounting for uncertainties in all three measured
parameters in order to minimize the intrinsic scatter between the
observables. The fitting process maintains flat residuals between
$L_{\rm X,gas}$ and the fit to $L_K^{\alpha}\sigma_e^{\gamma}$ (i.e.,
forcing a 1:1 relation between $L_{\rm X,gas}$ and
$L_K^{\alpha}\sigma_e^{\gamma}$). This stipulation of flat residuals
forces the most conservative estimate of the intrinsic scatter between
the variables during the fit.

We find that the best-fit empirical relation goes as
$L_{\rm X,gas} = a{\rm log}(L_K/10^{11} \Lsun) + b{\rm log}(\sigma_e /
{\rm km/s}) + c$,
where $a = 1.8 \pm 0.3$, $b = 3.3 \pm 0.6$ and $c = 39.5 \pm 0.4$ (see
Figure~\ref{fig:final_fig}) with an intrinsic scatter of
$\sim 0.41$--0.47~dex. The steep dependence on $L_K$ (an exponent
$\sim 2$) that is preferred during the fitting process likely arises
due to the wide range in $T_{\rm gas}$ at a given $L_K$ that was
observed in Figure~\ref{fig:lx_lk_sigma}a, compounded with the steep
slope of the $L_{\rm X,gas}-T_{\rm gas}$ relation found in
Section~\ref{sec:lx_t}. However, we also note that we still may not be
probing the full range in gas temperatures that may exist in the
entire galaxy population at a given $L_K$, particularly at low stellar
masses, due to selection biases related to the sources chosen for
X-ray observations.

Finally, in Figure~\ref{fig:final_fig}, we highlight the eight
sources that are $\gtrsim 2\sigma$ outliers from the best-fit
$\beta_{\rm spec} \sim 0.6$ relation (i.e., those sources with high
$T_{\rm gas}$ based on a virial origin for the gas heating). We find
that the majority are systematically shifted towards higher values of
$L_{\rm X,gas}$ based on the $L_{\rm X,gas}$ predicted from our
empirical fit to the optical properties. This may suggest an
additional form of heating or an increased gas mass that is not
directly linked to the host galaxy. Hence, the large scatter found in
the $L_{\rm X,gas}$ relations may arise due to variations in
$m_{\rm gas}$ (or $M_{\rm halo}$) at fixed stellar mass. Such systems
may exhibit significant cooling flows, particularly those that have
experienced recent mergers. Future studies aiming to examine these
changes in $m_{\rm gas}$ should endeavor to provide a complete and
unbiased X-ray view of the full galaxy population for a given optical
property.

\section{Summary} \label{sec:summary}

We have investigated the connection between the X-ray and optical
properties of massive early-type galaxies in the nearby
Universe. Using sensitive {\it Chandra} ACIS X-ray spectroscopy, we
have measured the X-ray luminosity at 0.3--5~keV and temperatures of
the hot gas entrained within the gravitational potential of the 33
sources selected from the volume-limited sample of 116 galaxies in the
MASSIVE Survey that have suitable {\it Chandra} data. These 33
galaxies are among the most massive galaxies in the Universe with
$L_K \gtrsim 3 \times 10^{11} \Lsun$ and within 108~Mpc. We combined
our MASSIVE sample with 41 lower mass elliptical galaxies that have
comparable optical integral field unit and X-ray measurements in the
ATLAS$^{\rm 3D}$ sample to investigate the X-ray--optical properties
over the largest dynamic range in $L_K$.

We summarize our conclusions as follows:

\begin{enumerate}

\item We demonstrate that all early-type galaxies follow a universal
  scaling relation going as
  $L_{\rm X,gas} \propto T_{\rm gas}^{\sim 4.5}$, when the X-ray
  properties are measured within an optical effective radius of the
  galaxy. The observed scatter of $\sim 0.5$~dex about this relation
  is consistent across the full considered parameter space. In
  addition, we further find the scatter is independent of secondary
  galaxy properties such as the galaxy rotational support or AGN
  activity ($P_{\rm 1.4 GHz}$). By careful comparison of the X-ray
  luminosities measured within optical radii and out to large radii,
  we conclude that excess X-ray luminosity observed in large aperture
  measurements of some massive ellipticals that lie within groups or
  clusters is likely the result of contamination from the large scale
  environment.

\item In accordance with prior studies, we show that the most massive
  early-type systems exhibit high X-ray luminosities that lie in the
  region of $L_{\rm X,gas} - L_K$ parameter-space roughly predicted by
  an extrapolation of the previously measured relationship, going as
  $L_{\rm X,gas} \propto L_K^{2.5-2.8}$. We demonstrate that the
  scatter about the $L_{\rm X,gas} - L_K$ relation is mostly driven by
  the X-ray temperature of the hot gas, with the hottest systems lying
  at higher $L_{\rm X,gas}$ compared to the simple
  $L_{\rm X,gas} - L_K$ prediction. We further show that similar
  qualitative inferences can be made for the galaxies when considered
  in $L_{\rm X,gas} - \sigma_e$ space. Given the significant scatter
  in both $L_{\rm X,gas} - L_K$ and $L_{\rm X,gas} - \sigma_e$,
  combined with an obvious temperature dependence, we draw no
  definitive conclusions about whether either of these relations (in
  isolation) can be considered intrinsic to early-type galaxies.

\item We find a statistically significant correlation between the
  stellar velocity dispersion and the X-ray gas temperature at the
  99.99\% level such that $T \sim \sigma_e^{1.6}$. The minimum
  observed X-ray temperature is consistent with being set directly by
  the gravitational potential of the system, with
  $\beta_{\rm spec} = \mu m_p \sigma_e^2 / kT_{\rm gas} = 1$. On
  average, we find that all galaxies experience some form of heating
  above the virial temperature, measuring a median
  $\beta_{\rm spec} \sim 0.6$.  However, for those sources with X-ray
  temperatures significantly exceeding those expected from virial
  heating alone, we find no obvious links with the large scale
  environment of the galaxy (i.e., mass of the large scale halo), the
  presence of a radio jet, or the rotation/flattening of the galaxy.

\item We probe the importance of recycled gas in explaining the X-ray
  luminosity observed in early-type galaxies. We find that the
  simplest model, such that the X-ray luminosity is produced due to
  thermalization of the energy from stellar mass loss material, fails
  to reproduce the observations when a significant number of
  slow-rotating galaxies are included. Even when including the large
  uncertainties involved with deriving the mass-loss rates, other
  physical mechanisms are still required to explain the data in hand
  (e.g., a lack of gas retention at low mass; stochastic accretion at
  high mass).

\item We investigate the existence of an empirical relationship
  between $L_{\rm X,gas}$, $L_K$ and $\sigma_e$. We calculate that the
  best fit multivariate relation is described as
  $L_{\rm X,gas} = a{\rm log}(L_K/10^{11} \Lsun) + b{\rm log}(\sigma_e
  / {\rm km/s}) + c$,
  where $a = 1.8 \pm 0.3$, $b = 3.3 \pm 0.6$ and $c = 39.5 \pm 0.4$
  with an intrinsic scatter of only $\sim 0.41$--0.47~dex. Those
  galaxies with $\beta_{\rm spec} < 0.4$ (i.e., with increased
  $T_{\rm gas}$ compared to the expected $T_{\rm virial}$) have
  systematically high-$L_{\rm X,gas}$ than that expected from the
  best-fit relation possibly due to the presence of cooling flows or
  an increased gas mass in these systems.

\end{enumerate}

\acknowledgments

We would like to thank the anonymous referee for useful and insightful
comments that allowed us to strengthen our results and conclusions.
We are thankful to Bill Forman, Christine Jones, Yuanyuan Su and Ming
Sun for helpful discussions. ADG acknowledges funding from NASA grant
AR3-14016X. Also JEG acknowledges support from the Miller Institute at
Berkeley. This survey is supported in part by NSF AST-1411945 and
AST-1411642. {\it Facilities:} \facility{Chandra (ACIS)}.

\bibliography{submitted.bbl}

\begin{thebibliography}{134}
\expandafter\ifx\csname natexlab\endcsname\relax\def\natexlab#1{#1}\fi

\bibitem[{{Allen} {et~al.}(2006){Allen}, {Dunn}, {Fabian}, {Taylor}, \&
  {Reynolds}}]{Allen:2006aa}
{Allen}, S.~W., {Dunn}, R.~J.~H., {Fabian}, A.~C., {Taylor}, G.~B., \&
  {Reynolds}, C.~S. 2006, \mnras, 372, 21

\bibitem[{{Athey} {et~al.}(2002){Athey}, {Bregman}, {Bregman}, {Temi}, \&
  {Sauvage}}]{Athey:2002aa}
{Athey}, A., {Bregman}, J., {Bregman}, J., {Temi}, P., \& {Sauvage}, M. 2002,
  \apj, 571, 272

\bibitem[{{Bell} {et~al.}(2004){Bell}, {Wolf}, {Meisenheimer}, {Rix}, {Borch},
  {Dye}, {Kleinheinrich}, {Wisotzki}, \& {McIntosh}}]{bell04}
{Bell}, E.~F., {et~al.} 2004, \apj, 608, 752

\bibitem[{{Bender} {et~al.}(1989){Bender}, {Surma}, {Doebereiner},
  {Moellenhoff}, \& {Madejsky}}]{Bender:1989aa}
{Bender}, R., {Surma}, P., {Doebereiner}, S., {Moellenhoff}, C., \& {Madejsky},
  R. 1989, \aap, 217, 35

\bibitem[{{Best} {et~al.}(2005){Best}, {Kauffmann}, {Heckman}, {Brinchmann},
  {Charlot}, {Ivezi{\'c}}, \& {White}}]{Best:2005aa}
{Best}, P.~N., {Kauffmann}, G., {Heckman}, T.~M., {Brinchmann}, J., {Charlot},
  S., {Ivezi{\'c}}, {\v Z}., \& {White}, S.~D.~M. 2005, \mnras, 362, 25

\bibitem[{{Binney}(2005)}]{binney05}
{Binney}, J. 2005, \mnras, 363, 937

\bibitem[{{Blakeslee} {et~al.}(2003){Blakeslee}, {Franx}, {Postman}, {Rosati},
  {Holden}, {Illingworth}, {Ford}, {Cross}, {Gronwall}, {Ben{\'{\i}}tez},
  {Bouwens}, {Broadhurst}, {Clampin}, {Demarco}, {Golimowski}, {Hartig},
  {Infante}, {Martel}, {Miley}, {Menanteau}, {Meurer}, {Sirianni}, \&
  {White}}]{Blakeslee:2003aa}
{Blakeslee}, J.~P., {et~al.} 2003, \apjl, 596, L143

\bibitem[{{Blanton} {et~al.}(2004){Blanton}, {Sarazin}, {McNamara}, \&
  {Clarke}}]{Blanton:2004aa}
{Blanton}, E.~L., {Sarazin}, C.~L., {McNamara}, B.~R., \& {Clarke}, T.~E. 2004,
  \apj, 612, 817

\bibitem[{{Blanton} \& {Moustakas}(2009)}]{blanton09}
{Blanton}, M.~R., \& {Moustakas}, J. 2009, \araa, 47, 159

\bibitem[{{Bogd{\'a}n} {et~al.}(2012{\natexlab{a}}){Bogd{\'a}n}, {David},
  {Jones}, {Forman}, \& {Kraft}}]{Bogdan:2012aa}
{Bogd{\'a}n}, {\'A}., {David}, L.~P., {Jones}, C., {Forman}, W.~R., \& {Kraft},
  R.~P. 2012{\natexlab{a}}, \apj, 758, 65

\bibitem[{{Bogd{\'a}n} \& {Gilfanov}(2008)}]{Bogdan:2008aa}
{Bogd{\'a}n}, {\'A}., \& {Gilfanov}, M. 2008, \mnras, 388, 56

\bibitem[{{Bogd{\'a}n} \& {Gilfanov}(2011)}]{Bogdan:2011aa}
---. 2011, \mnras, 418, 1901

\bibitem[{{Bogd{\'a}n} \& {Goulding}(2015)}]{Bogdan:2015aa}
{Bogd{\'a}n}, {\'A}., \& {Goulding}, A.~D. 2015, \apj, 800, 124

\bibitem[{{Bogd{\'a}n} {et~al.}(2012{\natexlab{b}}){Bogd{\'a}n}, {Forman},
  {Zhuravleva}, {Mihos}, {Kraft}, {Harding}, {Guo}, {Li}, {Churazov},
  {Vikhlinin}, {Nulsen}, {Schindler}, \& {Jones}}]{Bogdan:2012ab}
{Bogd{\'a}n}, {\'A}., {et~al.} 2012{\natexlab{b}}, \apj, 753, 140

\bibitem[{{Bogd{\'a}n} {et~al.}(2014){Bogd{\'a}n}, {van Weeren}, {Kraft},
  {Forman}, {Randall}, {Giacintucci}, {Churazov}, {O'Dea}, {Baum},
  {Noell-Storr}, \& {Jones}}]{Bogdan:2014aa}
---. 2014, \apjl, 782, L19

\bibitem[{{Boroson} {et~al.}(2011){Boroson}, {Kim}, \&
  {Fabbiano}}]{Boroson:2011aa}
{Boroson}, B., {Kim}, D.-W., \& {Fabbiano}, G. 2011, \apj, 729, 12

\bibitem[{{Brighenti} \& {Mathews}(1997)}]{Brighenti:1997aa}
{Brighenti}, F., \& {Mathews}, W.~G. 1997, \apjl, 486, L83

\bibitem[{{Brown} \& {Bregman}(2000)}]{Brown:2000aa}
{Brown}, B.~A., \& {Bregman}, J.~N. 2000, \apj, 539, 592

\bibitem[{{Bundy} {et~al.}(2010)}]{bundyetal2010}
{Bundy}, K., {et~al.} 2010, \apj, 719, 1969

\bibitem[{{Buote}(2000)}]{Buote:2000ab}
{Buote}, D.~A. 2000, \mnras, 311, 176

\bibitem[{{Canizares} {et~al.}(1987){Canizares}, {Fabbiano}, \&
  {Trinchieri}}]{Canizares:1987aa}
{Canizares}, C.~R., {Fabbiano}, G., \& {Trinchieri}, G. 1987, \apj, 312, 503

\bibitem[{{Cappellari} \& {Emsellem}(2004)}]{cappellariemsellem2004}
{Cappellari}, M., \& {Emsellem}, E. 2004, \pasp, 116, 138

\bibitem[{{Cappellari} {et~al.}(2011){Cappellari}, {Emsellem}, {Krajnovi{\'c}},
  {McDermid}, {Scott}, {Verdoes Kleijn}, {Young}, {Alatalo}, {Bacon}, {Blitz},
  {Bois}, {Bournaud}, {Bureau}, {Davies}, {Davis}, {de Zeeuw}, {Duc},
  {Khochfar}, {Kuntschner}, {Lablanche}, {Morganti}, {Naab}, {Oosterloo},
  {Sarzi}, {Serra}, \& {Weijmans}}]{Cappellari:2011aa}
{Cappellari}, M., {et~al.} 2011, \mnras, 413, 813

\bibitem[{{Cappellari} {et~al.}(2013){Cappellari}, {Scott}, {Alatalo}, {Blitz},
  {Bois}, {Bournaud}, {Bureau}, {Crocker}, {Davies}, {Davis}, {de Zeeuw},
  {Duc}, {Emsellem}, {Khochfar}, {Krajnovi{\'c}}, {Kuntschner}, {McDermid},
  {Morganti}, {Naab}, {Oosterloo}, {Sarzi}, {Serra}, {Weijmans}, \&
  {Young}}]{Cappellari:2013ab}
---. 2013, \mnras, 432, 1709

\bibitem[{{Cavagnolo} {et~al.}(2009){Cavagnolo}, {Donahue}, {Voit}, \&
  {Sun}}]{Cavagnolo:2009aa}
{Cavagnolo}, K.~W., {Donahue}, M., {Voit}, G.~M., \& {Sun}, M. 2009, \apjs,
  182, 12

\bibitem[{{Choi} {et~al.}(2015){Choi}, {Ostriker}, {Naab}, {Oser}, \&
  {Moster}}]{Choi:2015aa}
{Choi}, E., {Ostriker}, J.~P., {Naab}, T., {Oser}, L., \& {Moster}, B.~P. 2015,
  \mnras, 449, 4105

\bibitem[{{Ciotti} {et~al.}(1991){Ciotti}, {D'Ercole}, {Pellegrini}, \&
  {Renzini}}]{Ciotti:1991aa}
{Ciotti}, L., {D'Ercole}, A., {Pellegrini}, S., \& {Renzini}, A. 1991, \apj,
  376, 380

\bibitem[{{Ciotti} \& {Pellegrini}(1996)}]{Ciotti:1996aa}
{Ciotti}, L., \& {Pellegrini}, S. 1996, \mnras, 279, 240

\bibitem[{{Civano} {et~al.}(2014){Civano}, {Fabbiano}, {Pellegrini}, {Kim},
  {Paggi}, {Feder}, \& {Elvis}}]{Civano:2014aa}
{Civano}, F., {Fabbiano}, G., {Pellegrini}, S., {Kim}, D.-W., {Paggi}, A.,
  {Feder}, R., \& {Elvis}, M. 2014, \apj, 790, 16

\bibitem[{{Cowie} \& {Barger}(2008)}]{cowie08}
{Cowie}, L.~L., \& {Barger}, A.~J. 2008, \apj, 686, 72

\bibitem[{{Cox} {et~al.}(2006){Cox}, {Di Matteo}, {Hernquist}, {Hopkins},
  {Robertson}, \& {Springel}}]{Cox:2006ab}
{Cox}, T.~J., {Di Matteo}, T., {Hernquist}, L., {Hopkins}, P.~F., {Robertson},
  B., \& {Springel}, V. 2006, \apj, 643, 692

\bibitem[{{Crook} {et~al.}(2007){Crook}, {Huchra}, {Martimbeau}, {Masters},
  {Jarrett}, \& {Macri}}]{crooketal2007}
{Crook}, A.~C., {Huchra}, J.~P., {Martimbeau}, N., {Masters}, K.~L., {Jarrett},
  T., \& {Macri}, L.~M. 2007, \apj, 655, 790

\bibitem[{{Davis} \& {White}(1996)}]{Davis:1996aa}
{Davis}, D.~S., \& {White}, III, R.~E. 1996, \apjl, 470, L35

\bibitem[{{Davis} {et~al.}(2016){Davis}, {Greene}, {Ma}, {Pandya}, {Blakeslee},
  {McConnell}, \& {Thomas}}]{Davis:2016aa}
{Davis}, T.~A., {Greene}, J., {Ma}, C.-P., {Pandya}, V., {Blakeslee}, J.~P.,
  {McConnell}, N., \& {Thomas}, J. 2016, \mnras, 455, 214

\bibitem[{{Diehl} \& {Statler}(2007)}]{Diehl:2007aa}
{Diehl}, S., \& {Statler}, T.~S. 2007, \apj, 668, 150

\bibitem[{{Diehl} \& {Statler}(2008)}]{Diehl:2008ac}
---. 2008, \apj, 687, 986

\bibitem[{{Dunn} {et~al.}(2010){Dunn}, {Allen}, {Taylor}, {Shurkin}, {Gentile},
  {Fabian}, \& {Reynolds}}]{Dunn:2010aa}
{Dunn}, R.~J.~H., {Allen}, S.~W., {Taylor}, G.~B., {Shurkin}, K.~F., {Gentile},
  G., {Fabian}, A.~C., \& {Reynolds}, C.~S. 2010, \mnras, 404, 180

\bibitem[{{Edge} \& {Stewart}(1991)}]{Edge:1991aa}
{Edge}, A.~C., \& {Stewart}, G.~C. 1991, \mnras, 252, 428

\bibitem[{{Emsellem}(2013)}]{emsellemetal2013}
{Emsellem}, E. 2013, \mnras, 433, 1862

\bibitem[{{Emsellem} {et~al.}(2007)}]{emsellemetal2007}
{Emsellem}, E., {et~al.} 2007, \mnras, 379, 401

\bibitem[{{Emsellem} {et~al.}(2011){Emsellem}, {Cappellari}, {Krajnovi{\'c}},
  {Alatalo}, {Blitz}, {Bois}, {Bournaud}, {Bureau}, {Davies}, {Davis}, {de
  Zeeuw}, {Khochfar}, {Kuntschner}, {Lablanche}, {McDermid}, {Morganti},
  {Naab}, {Oosterloo}, {Sarzi}, {Scott}, {Serra}, {van de Ven}, {Weijmans}, \&
  {Young}}]{Emsellem:2011aa}
---. 2011, \mnras, 414, 888

\bibitem[{{Eskridge} {et~al.}(1995){Eskridge}, {Fabbiano}, \&
  {Kim}}]{Eskridge:1995aa}
{Eskridge}, P.~B., {Fabbiano}, G., \& {Kim}, D.-W. 1995, \apjs, 97, 141

\bibitem[{{Ettori} {et~al.}(2004){Ettori}, {Tozzi}, {Borgani}, \&
  {Rosati}}]{Ettori:2004aa}
{Ettori}, S., {Tozzi}, P., {Borgani}, S., \& {Rosati}, P. 2004, \aap, 417, 13

\bibitem[{{Fabbiano}(2006)}]{Fabbiano:2006ab}
{Fabbiano}, G. 2006, \araa, 44, 323

\bibitem[{{Faber} {et~al.}(1997){Faber}, {Tremaine}, {Ajhar}, {Byun},
  {Dressler}, {Gebhardt}, {Grillmair}, {Kormendy}, {Lauer}, \&
  {Richstone}}]{Faber:1997aa}
{Faber}, S.~M., {et~al.} 1997, \aj, 114, 1771

\bibitem[{{Faber} {et~al.}(2007){Faber}, {Willmer}, {Wolf}, {Koo}, {Weiner},
  {Newman}, {Im}, {Coil}, {Conroy}, {Cooper}, {Davis}, {Finkbeiner}, {Gerke},
  {Gebhardt}, {Groth}, {Guhathakurta}, {Harker}, {Kaiser}, {Kassin},
  {Kleinheinrich}, {Konidaris}, {Kron}, {Lin}, {Luppino}, {Madgwick},
  {Meisenheimer}, {Noeske}, {Phillips}, {Sarajedini}, {Schiavon}, {Simard},
  {Szalay}, {Vogt}, \& {Yan}}]{faber07}
---. 2007, \apj, 665, 265

\bibitem[{{Falc{\'o}n-Barroso} {et~al.}(2011){Falc{\'o}n-Barroso},
  {S{\'a}nchez-Bl{\'a}zquez}, {Vazdekis}, {Ricciardelli}, {Cardiel}, {Cenarro},
  {Gorgas}, \& {Peletier}}]{Falcon-Barroso:2011aa}
{Falc{\'o}n-Barroso}, J., {S{\'a}nchez-Bl{\'a}zquez}, P., {Vazdekis}, A.,
  {Ricciardelli}, E., {Cardiel}, N., {Cenarro}, A.~J., {Gorgas}, J., \&
  {Peletier}, R.~F. 2011, \aap, 532, A95

\bibitem[{{Ferrarese} {et~al.}(1994){Ferrarese}, {van den Bosch}, {Ford},
  {Jaffe}, \& {O'Connell}}]{Ferrarese:1994aa}
{Ferrarese}, L., {van den Bosch}, F.~C., {Ford}, H.~C., {Jaffe}, W., \&
  {O'Connell}, R.~W. 1994, \aj, 108, 1598

\bibitem[{{Ferrarese} {et~al.}(2006){Ferrarese}, {C{\^o}t{\'e}}, {Jord{\'a}n},
  {Peng}, {Blakeslee}, {Piatek}, {Mei}, {Merritt}, {Milosavljevi{\'c}},
  {Tonry}, \& {West}}]{Ferrarese:2006aa}
{Ferrarese}, L., {et~al.} 2006, \apjs, 164, 334

\bibitem[{{Finoguenov} \& {Jones}(2002)}]{Finoguenov:2002aa}
{Finoguenov}, A., \& {Jones}, C. 2002, \apj, 574, 754

\bibitem[{{Forman} {et~al.}(1985){Forman}, {Jones}, \&
  {Tucker}}]{Forman:1985aa}
{Forman}, W., {Jones}, C., \& {Tucker}, W. 1985, \apj, 293, 102

\bibitem[{{Forman} {et~al.}(2005){Forman}, {Nulsen}, {Heinz}, {Owen}, {Eilek},
  {Vikhlinin}, {Markevitch}, {Kraft}, {Churazov}, \& {Jones}}]{Forman:2005aa}
{Forman}, W., {et~al.} 2005, \apj, 635, 894

\bibitem[{{Furusho} {et~al.}(2003){Furusho}, {Yamasaki}, \&
  {Ohashi}}]{Furusho:2003aa}
{Furusho}, T., {Yamasaki}, N.~Y., \& {Ohashi}, T. 2003, \apj, 596, 181

\bibitem[{{Gilfanov}(2004)}]{Gilfanov:2004aa}
{Gilfanov}, M. 2004, \mnras, 349, 146

\bibitem[{{Goulding} {et~al.}(2012){Goulding}, {Forman}, {Hickox}, {Jones},
  {Kraft}, {Murray}, {Vikhlinin}, {Coil}, {Cooper}, {Davis}, \&
  {Newman}}]{goulding12b}
{Goulding}, A.~D., {et~al.} 2012, \apjs, 202, 6

\bibitem[{{Greene} {et~al.}(2015){Greene}, {Janish}, {Ma}, {McConnell},
  {Blakeslee}, {Thomas}, \& {Murphy}}]{greeneetal2015}
{Greene}, J.~E., {Janish}, R., {Ma}, C.-P., {McConnell}, N.~J., {Blakeslee},
  J.~P., {Thomas}, J., \& {Murphy}, J.~D. 2015, \apj, 807, 11

\bibitem[{{Greene} {et~al.}(2013){Greene}, {Murphy}, {Graves}, {Gunn},
  {Raskutti}, {Comerford}, \& {Gebhardt}}]{greeneetal2013}
{Greene}, J.~E., {Murphy}, J.~D., {Graves}, G.~J., {Gunn}, J.~E., {Raskutti},
  S., {Comerford}, J.~M., \& {Gebhardt}, K. 2013, \apj, 776, 64

\bibitem[{{Helsdon} \& {Ponman}(2000)}]{Helsdon:2000aa}
{Helsdon}, S.~F., \& {Ponman}, T.~J. 2000, \mnras, 315, 356

\bibitem[{{Helsdon} {et~al.}(2001){Helsdon}, {Ponman}, {O'Sullivan}, \&
  {Forbes}}]{Helsdon:2001aa}
{Helsdon}, S.~F., {Ponman}, T.~J., {O'Sullivan}, E., \& {Forbes}, D.~A. 2001,
  \mnras, 325, 693

\bibitem[{{Hill} {et~al.}(2008)}]{hilletal2008b}
{Hill}, G.~J., {et~al.} 2008, in Astronomical Society of the Pacific Conference
  Series, Vol. 399, Panoramic Views of Galaxy Formation and Evolution, ed.
  {T.~Kodama, T.~Yamada, \& K.~Aoki}, 115

\bibitem[{{Huchra} {et~al.}(2012){Huchra}, {Macri}, {Masters}, {Jarrett},
  {Berlind}, {Calkins}, {Crook}, {Cutri}, {Erdo{\v g}du}, {Falco}, {George},
  {Hutcheson}, {Lahav}, {Mader}, {Mink}, {Martimbeau}, {Schneider},
  {Skrutskie}, {Tokarz}, \& {Westover}}]{huchraetal2012}
{Huchra}, J.~P., {et~al.} 2012, \apjs, 199, 26

\bibitem[{{Humphrey} {et~al.}(2008){Humphrey}, {Buote}, {Brighenti},
  {Gebhardt}, \& {Mathews}}]{Humphrey:2008ab}
{Humphrey}, P.~J., {Buote}, D.~A., {Brighenti}, F., {Gebhardt}, K., \&
  {Mathews}, W.~G. 2008, \apj, 683, 161

\bibitem[{{Humphrey} {et~al.}(2009){Humphrey}, {Buote}, {Brighenti},
  {Gebhardt}, \& {Mathews}}]{Humphrey:2009aa}
---. 2009, \apj, 703, 1257

\bibitem[{{Humphrey} {et~al.}(2011){Humphrey}, {Buote}, {Canizares}, {Fabian},
  \& {Miller}}]{Humphrey:2011aa}
{Humphrey}, P.~J., {Buote}, D.~A., {Canizares}, C.~R., {Fabian}, A.~C., \&
  {Miller}, J.~M. 2011, \apj, 729, 53

\bibitem[{{Humphrey} {et~al.}(2006){Humphrey}, {Buote}, {Gastaldello},
  {Zappacosta}, {Bullock}, {Brighenti}, \& {Mathews}}]{Humphrey:2006aa}
{Humphrey}, P.~J., {Buote}, D.~A., {Gastaldello}, F., {Zappacosta}, L.,
  {Bullock}, J.~S., {Brighenti}, F., \& {Mathews}, W.~G. 2006, \apj, 646, 899

\bibitem[{{Humphrey} {et~al.}(2012){Humphrey}, {Buote}, {O'Sullivan}, \&
  {Ponman}}]{Humphrey:2012aa}
{Humphrey}, P.~J., {Buote}, D.~A., {O'Sullivan}, E., \& {Ponman}, T.~J. 2012,
  \apj, 755, 166

\bibitem[{{Jarrett} {et~al.}(2003){Jarrett}, {Chester}, {Cutri}, {Schneider},
  \& {Huchra}}]{jarrettetal2003}
{Jarrett}, T.~H., {Chester}, T., {Cutri}, R., {Schneider}, S.~E., \& {Huchra},
  J.~P. 2003, \aj, 125, 525

\bibitem[{{Jeltema} {et~al.}(2008){Jeltema}, {Binder}, \&
  {Mulchaey}}]{Jeltema:2008aa}
{Jeltema}, T.~E., {Binder}, B., \& {Mulchaey}, J.~S. 2008, \apj, 679, 1162

\bibitem[{{Kelly}(2007)}]{Kelly:2007aa}
{Kelly}, B.~C. 2007, \apj, 665, 1489

\bibitem[{{Khosroshahi} {et~al.}(2007){Khosroshahi}, {Ponman}, \&
  {Jones}}]{Khosroshahi:2007aa}
{Khosroshahi}, H.~G., {Ponman}, T.~J., \& {Jones}, L.~R. 2007, \mnras, 377, 595

\bibitem[{{Kim} \& {Fabbiano}(2013)}]{Kim:2013aa}
{Kim}, D.-W., \& {Fabbiano}, G. 2013, \apj, 776, 116

\bibitem[{{Kim} \& {Fabbiano}(2015)}]{Kim:2015aa}
---. 2015, \apj, 812, 127

\bibitem[{{Kormendy} \& {Bender}(1996)}]{Kormendy:1996aa}
{Kormendy}, J., \& {Bender}, R. 1996, \apjl, 464, L119

\bibitem[{{Kormendy} \& {Bender}(2012)}]{Kormendy:2012aa}
---. 2012, \apjs, 198, 2

\bibitem[{{Kormendy} {et~al.}(2009){Kormendy}, {Fisher}, {Cornell}, \&
  {Bender}}]{Kormendy:2009ab}
{Kormendy}, J., {Fisher}, D.~B., {Cornell}, M.~E., \& {Bender}, R. 2009, \apjs,
  182, 216

\bibitem[{{Kraft} {et~al.}(2012){Kraft}, {Birkinshaw}, {Nulsen}, {Worrall},
  {Croston}, {Forman}, {Hardcastle}, {Jones}, \& {Murray}}]{Kraft:2012aa}
{Kraft}, R.~P., {et~al.} 2012, \apj, 749, 19

\bibitem[{{Krajnovi{\'c}} {et~al.}(2011){Krajnovi{\'c}}, {Emsellem},
  {Cappellari}, {Alatalo}, {Blitz}, {Bois}, {Bournaud}, {Bureau}, {Davies},
  {Davis}, {de Zeeuw}, {Khochfar}, {Kuntschner}, {Lablanche}, {McDermid},
  {Morganti}, {Naab}, {Oosterloo}, {Sarzi}, {Scott}, {Serra}, {Weijmans}, \&
  {Young}}]{Krajnovic:2011aa}
{Krajnovi{\'c}}, D., {et~al.} 2011, \mnras, 414, 2923

\bibitem[{{Krajnovi{\'c}} {et~al.}(2013){Krajnovi{\'c}}, {Karick}, {Davies},
  {Naab}, {Sarzi}, {Emsellem}, {Cappellari}, {Serra}, {de Zeeuw}, {Scott},
  {McDermid}, {Weijmans}, {Davis}, {Alatalo}, {Blitz}, {Bois}, {Bureau},
  {Bournaud}, {Crocker}, {Duc}, {Khochfar}, {Kuntschner}, {Morganti},
  {Oosterloo}, \& {Young}}]{Krajnovic:2013aa}
---. 2013, \mnras, 433, 2812

\bibitem[{{Lauer} {et~al.}(1995){Lauer}, {Ajhar}, {Byun}, {Dressler}, {Faber},
  {Grillmair}, {Kormendy}, {Richstone}, \& {Tremaine}}]{Lauer:1995aa}
{Lauer}, T.~R., {et~al.} 1995, \aj, 110, 2622

\bibitem[{{Lauer} {et~al.}(2007){Lauer}, {Faber}, {Richstone}, {Gebhardt},
  {Tremaine}, {Postman}, {Dressler}, {Aller}, {Filippenko}, {Green}, {Ho},
  {Kormendy}, {Magorrian}, \& {Pinkney}}]{Lauer:2007aa}
---. 2007, \apj, 662, 808

\bibitem[{{Lehmer} {et~al.}(2007){Lehmer}, {Brandt}, {Alexander}, {Bell},
  {McIntosh}, {Bauer}, {Hasinger}, {Mainieri}, {Miyaji}, {Schneider}, \&
  {Steffen}}]{Lehmer:2007aa}
{Lehmer}, B.~D., {et~al.} 2007, \apj, 657, 681

\bibitem[{{Lehmer} {et~al.}(2014){Lehmer}, {Berkeley}, {Zezas}, {Alexander},
  {Basu-Zych}, {Bauer}, {Brandt}, {Fragos}, {Hornschemeier}, {Kalogera},
  {Ptak}, {Sivakoff}, {Tzanavaris}, \& {Yukita}}]{Lehmer:2014aa}
---. 2014, \apj, 789, 52

\bibitem[{{Loewenstein}(2000)}]{Loewenstein:2000aa}
{Loewenstein}, M. 2000, \apj, 532, 17

\bibitem[{{Loewenstein} \& {White}(1999)}]{Loewenstein:1999aa}
{Loewenstein}, M., \& {White}, III, R.~E. 1999, \apj, 518, 50

\bibitem[{{Ma} {et~al.}(2014){Ma}, {Greene}, {McConnell}, {Janish},
  {Blakeslee}, {Thomas}, \& {Murphy}}]{Ma:2014aa}
{Ma}, C.-P., {Greene}, J.~E., {McConnell}, N., {Janish}, R., {Blakeslee},
  J.~P., {Thomas}, J., \& {Murphy}, J.~D. 2014, \apj, 795, 158

\bibitem[{{Mahdavi} \& {Geller}(2001)}]{Mahdavi:2001aa}
{Mahdavi}, A., \& {Geller}, M.~J. 2001, \apjl, 554, L129

\bibitem[{{Markevitch}(1998)}]{Markevitch:1998aa}
{Markevitch}, M. 1998, \apj, 504, 27

\bibitem[{{Mathews}(1990)}]{Mathews:1990aa}
{Mathews}, W.~G. 1990, \apj, 354, 468

\bibitem[{{Mathews} \& {Brighenti}(2003)}]{Mathews:2003aa}
{Mathews}, W.~G., \& {Brighenti}, F. 2003, \araa, 41, 191

\bibitem[{{Maughan} {et~al.}(2012){Maughan}, {Giles}, {Randall}, {Jones}, \&
  {Forman}}]{Maughan:2012aa}
{Maughan}, B.~J., {Giles}, P.~A., {Randall}, S.~W., {Jones}, C., \& {Forman},
  W.~R. 2012, \mnras, 421, 1583

\bibitem[{{McDermid} {et~al.}(2014){McDermid}, {Cappellari}, {Alatalo},
  {Bayet}, {Blitz}, {Bois}, {Bournaud}, {Bureau}, {Crocker}, {Davies}, {Davis},
  {de Zeeuw}, {Duc}, {Emsellem}, {Khochfar}, {Krajnovi{\'c}}, {Kuntschner},
  {Morganti}, {Naab}, {Oosterloo}, {Sarzi}, {Scott}, {Serra}, {Weijmans}, \&
  {Young}}]{mcdermidetal2014}
{McDermid}, R.~M., {et~al.} 2014, \apjl, 792, L37

\bibitem[{{McNamara} \& {Nulsen}(2007)}]{mcnamara07}
{McNamara}, B.~R., \& {Nulsen}, P.~E.~J. 2007, \araa, 45, 117

\bibitem[{{Memola} {et~al.}(2009){Memola}, {Trinchieri}, {Wolter}, {Focardi},
  \& {Kelm}}]{Memola:2009aa}
{Memola}, E., {Trinchieri}, G., {Wolter}, A., {Focardi}, P., \& {Kelm}, B.
  2009, \aap, 497, 359

\bibitem[{{Mitchell} {et~al.}(1979){Mitchell}, {Dickens}, {Burnell}, \&
  {Culhane}}]{Mitchell:1979aa}
{Mitchell}, R.~J., {Dickens}, R.~J., {Burnell}, S.~J.~B., \& {Culhane}, J.~L.
  1979, \mnras, 189, 329

\bibitem[{{Mittal} {et~al.}(2011){Mittal}, {Hicks}, {Reiprich}, \&
  {Jaritz}}]{Mittal:2011aa}
{Mittal}, R., {Hicks}, A., {Reiprich}, T.~H., \& {Jaritz}, V. 2011, \aap, 532,
  A133

\bibitem[{{Mulchaey}(2000)}]{Mulchaey:2000ab}
{Mulchaey}, J.~S. 2000, \araa, 38, 289

\bibitem[{{Mulchaey} \& {Jeltema}(2010)}]{Mulchaey:2010aa}
{Mulchaey}, J.~S., \& {Jeltema}, T.~E. 2010, \apjl, 715, L1

\bibitem[{{Mulchaey} \& {Zabludoff}(1998)}]{Mulchaey:1998aa}
{Mulchaey}, J.~S., \& {Zabludoff}, A.~I. 1998, \apj, 496, 73

\bibitem[{{Mushotzky}(1984)}]{Mushotzky:1984aa}
{Mushotzky}, R.~F. 1984, Physica Scripta Volume T, 7, 157

\bibitem[{{Mushotzky} \& {Scharf}(1997)}]{Mushotzky:1997aa}
{Mushotzky}, R.~F., \& {Scharf}, C.~A. 1997, \apjl, 482, L13

\bibitem[{{Negri} {et~al.}(2015){Negri}, {Pellegrini}, \&
  {Ciotti}}]{Negri:2015aa}
{Negri}, A., {Pellegrini}, S., \& {Ciotti}, L. 2015, \mnras, 451, 1212

\bibitem[{{Neumann} \& {Boehringer}(1995)}]{Neumann:1995aa}
{Neumann}, D.~M., \& {Boehringer}, H. 1995, \aap, 301, 865

\bibitem[{{Ortiz-Gil} {et~al.}(2004){Ortiz-Gil}, {Guzzo}, {Schuecker},
  {B{\"o}hringer}, \& {Collins}}]{Ortiz-Gil:2004aa}
{Ortiz-Gil}, A., {Guzzo}, L., {Schuecker}, P., {B{\"o}hringer}, H., \&
  {Collins}, C.~A. 2004, \mnras, 348, 325

\bibitem[{{Osmond} \& {Ponman}(2004)}]{Osmond:2004aa}
{Osmond}, J.~P.~F., \& {Ponman}, T.~J. 2004, \mnras, 350, 1511

\bibitem[{{O'Sullivan} {et~al.}(2001){O'Sullivan}, {Forbes}, \&
  {Ponman}}]{OSullivan:2001aa}
{O'Sullivan}, E., {Forbes}, D.~A., \& {Ponman}, T.~J. 2001, \mnras, 328, 461

\bibitem[{{O'Sullivan} {et~al.}(2003){O'Sullivan}, {Ponman}, \&
  {Collins}}]{OSullivan:2003aa}
{O'Sullivan}, E., {Ponman}, T.~J., \& {Collins}, R.~S. 2003, \mnras, 340, 1375

\bibitem[{{Paggi} {et~al.}(2015){Paggi}, {Fabbiano}, {Civano}, {Pellegrini},
  {Elvis}, \& {Kim}}]{Paggi:2015aa}
{Paggi}, A., {Fabbiano}, G., {Civano}, F., {Pellegrini}, S., {Elvis}, M., \&
  {Kim}, D.-W. 2015, ArXiv e-prints

\bibitem[{{Paturel} {et~al.}(2003){Paturel}, {Petit}, {Prugniel}, {Theureau},
  {Rousseau}, {Brouty}, {Dubois}, \& {Cambr{\'e}sy}}]{patureletal2003}
{Paturel}, G., {Petit}, C., {Prugniel}, P., {Theureau}, G., {Rousseau}, J.,
  {Brouty}, M., {Dubois}, P., \& {Cambr{\'e}sy}, L. 2003, \aap, 412, 45

\bibitem[{{Peletier} {et~al.}(1990){Peletier}, {Davies}, {Illingworth},
  {Davis}, \& {Cawson}}]{Peletier:1990aa}
{Peletier}, R.~F., {Davies}, R.~L., {Illingworth}, G.~D., {Davis}, L.~E., \&
  {Cawson}, M. 1990, \aj, 100, 1091

\bibitem[{{Pellegrini}(2011)}]{Pellegrini:2011aa}
{Pellegrini}, S. 2011, \apj, 738, 57

\bibitem[{{Ponman} {et~al.}(1996){Ponman}, {Bourner}, {Ebeling}, \&
  {B{\"o}hringer}}]{Ponman:1996aa}
{Ponman}, T.~J., {Bourner}, P.~D.~J., {Ebeling}, H., \& {B{\"o}hringer}, H.
  1996, \mnras, 283, 690

\bibitem[{{Revnivtsev} {et~al.}(2007){Revnivtsev}, {Churazov}, {Sazonov},
  {Forman}, \& {Jones}}]{Revnivtsev:2007aa}
{Revnivtsev}, M., {Churazov}, E., {Sazonov}, S., {Forman}, W., \& {Jones}, C.
  2007, \aap, 473, 783

\bibitem[{{S{\'a}nchez-Bl{\'a}zquez} {et~al.}(2006){S{\'a}nchez-Bl{\'a}zquez},
  {Peletier}, {Jim{\'e}nez-Vicente}, {Cardiel}, {Cenarro},
  {Falc{\'o}n-Barroso}, {Gorgas}, {Selam}, \&
  {Vazdekis}}]{Sanchez-Blazquez:2006aa}
{S{\'a}nchez-Bl{\'a}zquez}, P., {et~al.} 2006, \mnras, 371, 703

\bibitem[{{Sanders} \& {Fabian}(2012)}]{Sanders:2012aa}
{Sanders}, J.~S., \& {Fabian}, A.~C. 2012, \mnras, 421, 726

\bibitem[{{Sarzi} {et~al.}(2013){Sarzi}, {Alatalo}, {Blitz}, {Bois},
  {Bournaud}, {Bureau}, {Cappellari}, {Crocker}, {Davies}, {Davis}, {de Zeeuw},
  {Duc}, {Emsellem}, {Khochfar}, {Krajnovi{\'c}}, {Kuntschner}, {Lablanche},
  {McDermid}, {Morganti}, {Naab}, {Oosterloo}, {Scott}, {Serra}, {Young}, \&
  {Weijmans}}]{Sarzi:2013aa}
{Sarzi}, M., {et~al.} 2013, \mnras, 432, 1845

\bibitem[{{S{\'e}rsic}(1963)}]{sersic63}
{S{\'e}rsic}, J.~L. 1963, Boletin de la Asociacion Argentina de Astronomia La
  Plata Argentina, 6, 41

\bibitem[{{Sivakoff} {et~al.}(2004){Sivakoff}, {Sarazin}, \&
  {Carlin}}]{Sivakoff:2004aa}
{Sivakoff}, G.~R., {Sarazin}, C.~L., \& {Carlin}, J.~L. 2004, \apj, 617, 262

\bibitem[{{Su} {et~al.}(2015){Su}, {Irwin}, {White}, \& {Cooper}}]{Su:2015aa}
{Su}, Y., {Irwin}, J.~A., {White}, III, R.~E., \& {Cooper}, M.~C. 2015, \apj,
  806, 156

\bibitem[{{Sun} {et~al.}(2007){Sun}, {Jones}, {Forman}, {Vikhlinin}, {Donahue},
  \& {Voit}}]{Sun:2007aa}
{Sun}, M., {Jones}, C., {Forman}, W., {Vikhlinin}, A., {Donahue}, M., \&
  {Voit}, M. 2007, \apj, 657, 197

\bibitem[{{Sun} {et~al.}(2009){Sun}, {Voit}, {Donahue}, {Jones}, {Forman}, \&
  {Vikhlinin}}]{Sun:2009aa}
{Sun}, M., {Voit}, G.~M., {Donahue}, M., {Jones}, C., {Forman}, W., \&
  {Vikhlinin}, A. 2009, \apj, 693, 1142

\bibitem[{{Thomas} {et~al.}(2005){Thomas}, {Maraston}, {Bender}, \& {Mendes de
  Oliveira}}]{Thomas:2005aa}
{Thomas}, D., {Maraston}, C., {Bender}, R., \& {Mendes de Oliveira}, C. 2005,
  \apj, 621, 673

\bibitem[{{Thomas} {et~al.}(2011)}]{thomasetal2011}
{Thomas}, J., {et~al.} 2011, \mnras, 415, 545

\bibitem[{{Trager} {et~al.}(2000){Trager}, {Faber}, {Worthey}, \&
  {Gonz{\'a}lez}}]{Trager:2000aa}
{Trager}, S.~C., {Faber}, S.~M., {Worthey}, G., \& {Gonz{\'a}lez}, J.~J. 2000,
  \aj, 119, 1645

\bibitem[{{Trinchieri} \& {Fabbiano}(1985)}]{Trinchieri:1985aa}
{Trinchieri}, G., \& {Fabbiano}, G. 1985, \apj, 296, 447

\bibitem[{{Vajgel} {et~al.}(2014){Vajgel}, {Jones}, {Lopes}, {Forman},
  {Murray}, {Goulding}, \& {Andrade-Santos}}]{Vajgel:2014aa}
{Vajgel}, B., {Jones}, C., {Lopes}, P.~A.~A., {Forman}, W.~R., {Murray}, S.~S.,
  {Goulding}, A., \& {Andrade-Santos}, F. 2014, \apj, 794, 88

\bibitem[{{van Dokkum} {et~al.}(2008)}]{vandokkumetal2008}
{van Dokkum}, P.~G., {et~al.} 2008, \apjl, 677, L5

\bibitem[{{van Dokkum} {et~al.}(2010)}]{vandokkumetal2010}
---. 2010, \apj, 709, 1018

\bibitem[{{Werner} {et~al.}(2012){Werner}, {Allen}, \&
  {Simionescu}}]{Werner:2012aa}
{Werner}, N., {Allen}, S.~W., \& {Simionescu}, A. 2012, \mnras, 425, 2731

\bibitem[{{White} {et~al.}(1997){White}, {Jones}, \& {Forman}}]{White:1997aa}
{White}, D.~A., {Jones}, C., \& {Forman}, W. 1997, \mnras, 292, 419

\bibitem[{{White} \& {Sarazin}(1991)}]{White:1991ab}
{White}, III, R.~E., \& {Sarazin}, C.~L. 1991, \apj, 367, 476

\bibitem[{{Wu} {et~al.}(1999){Wu}, {Xue}, \& {Fang}}]{Wu:1999aa}
{Wu}, X.-P., {Xue}, Y.-J., \& {Fang}, L.-Z. 1999, \apj, 524, 22

\bibitem[{{Xue} \& {Wu}(2000)}]{Xue:2000aa}
{Xue}, Y.-J., \& {Wu}, X.-P. 2000, \apj, 538, 65

\bibitem[{{York} {et~al.}(2000)}]{yorketal2000}
{York}, D.~G., {et~al.} 2000, \aj, 120, 1579

\bibitem[{{Young} {et~al.}(2011){Young}, {Bureau}, {Davis}, {Combes},
  {McDermid}, {Alatalo}, {Blitz}, {Bois}, {Bournaud}, {Cappellari}, {Davies},
  {de Zeeuw}, {Emsellem}, {Khochfar}, {Krajnovi{\'c}}, {Kuntschner},
  {Lablanche}, {Morganti}, {Naab}, {Oosterloo}, {Sarzi}, {Scott}, {Serra}, \&
  {Weijmans}}]{Young:2011aa}
{Young}, L.~M., {et~al.} 2011, \mnras, 414, 940

\end{thebibliography}


\appendix

\section{Chandra X-ray measurements of ATLAS$^{\rm 3D}$ galaxies}

To provide the most accurate and complete comparison to our X-ray
measurements of the MASSIVE galaxies, we require a sample of
lower-mass early-type galaxies that have comparable X-ray
observations. As stated in Section~\ref{sec:atlas3d} of the main
article, the ATLAS$^{\rm 3D}$ galaxy survey provides the most natural
comparison to the MASSIVE galaxies. Recently, \cite{Kim:2015aa} and
presented (constrained) X-ray gas and luminosity measurements of (49)
61 of the early-type galaxies in ATLAS$^{\rm 3D}$ that have
observations with the {\it Chandra} X-ray telescope. As stated
previously, these measurements were performed using large apertures,
which encompassed all of the detected X-ray photons that were possibly
associated with the galaxy. Given that some of these sources were in
group/cluster environments, we also require small aperture
measurements, performed on the scales of the galaxy-radius, to limit
contamination from the large-scale environment. \cite{Su:2015aa}
present X-ray measurements for a sub-sample of 37 of the 49 objects in
\cite{Kim:2015aa}. Here we present X-ray measurements performed within
$R_e$ for the remaining 12 objects, as well as a comparison between
the X-ray measurements presented here and those in \cite{Kim:2015aa}
and \cite{Su:2015aa}.

The {\it Chandra} data were retrieved, reduced, and analyzed following
precisely the same procedure outlined in
section~\ref{sec:xray_datreduc} and \ref{sec:xray_spec} of the main
article. The basic properties and the X-ray measurements performed
within $1 R_e$ apertures of the 49 galaxies from \cite{Kim:2015aa} are
presented in Table A1. For the sake of consistency, we reanalyzed each
of the 49 galaxies presented in \cite{Kim:2015aa}, and compared the
X-ray measurements within $1 R_e$ apertures to those of
\cite{Su:2015aa} and the measurements out to large radii with those of
\cite{Kim:2015aa}. Overall, our X-ray measurements were in excellent
agreement with those of \cite{Kim:2015aa}. For brevity we do not
repeat the results here, and defer the reader to
\cite{Kim:2015aa}. However, we found several inconsistencies with the
gas luminosities and temperatures between the \cite{Kim:2015aa} and
\cite{Su:2015aa} measurements. While not wholly unexpected given the
different apertures used in these analyses, at least 11 objects show
significant differences in the presented measurements, beyond typical
radial variations. Upon further analysis, we deduced that at least
eight of the galaxies presented in \cite{Su:2015aa} suffer from
significant contamination due to AGN or strong point sources (e.g.,
LMXBs) close to or at the peak X-ray emission. In many cases these
dominate the X-ray emission around 0.5--4.5~keV if the emission is not
adequately screened, resulting in erroneously high X-ray temperatures
and luminosities. We further deduced that several other low-luminosity
objects may have suffered insufficient background subtraction. Given
that we find excellent agreement for all of the \cite{Kim:2015aa}
measurements, we elect to use our own $1 R_e$ X-ray measurements for
those sources where we find significant differences between
\cite{Kim:2015aa} and \cite{Su:2015aa}.

Of the 49 galaxies in \cite{Kim:2015aa} with constrained X-ray
measurements, we remove four ATLAS$^{\rm 3D}$ objects due to overlap
with the MASSIVE sample and five objects due to unconstrained gas
measurements within $1 R_e$. Specifically, the observation of NGC~3998
suffered from significant X-ray pile-up imparted by a central
AGN. NGC~524 was performed on a sub-array that was smaller than
$1 R_e$. NGC~2778 and NGC~6278 were both undetected in diffuse
emission. NGC~1266 appears to have no extended hot gas emission. We
note that there are clearly three star-forming knots close to the
nucleus of the galaxy, and these dominate the X-ray emission, but we
detect no X-ray emission beyond these. Thus, we conservatively choose
to remove NGC~1266 from our analyses. Our final comparison
ATLAS$^{\rm 3D}$ galaxy sample contains 41 sources.

\begin{table*}
\footnotesize
\begin{center}
\setlength{\tabcolsep}{1.15mm}
\caption{The ATLAS$^{\rm 3D}$ X-ray sample\label{tbl:atlassample}}
\begin{tabular}{lccrrcccrrccc}
\tableline\tableline
\multicolumn{1}{c}{Common} &
\multicolumn{1}{c}{$\alpha_{\rm J2000}$} &
\multicolumn{1}{c}{$\delta_{\rm J2000}$} &
\multicolumn{1}{c}{$D_L$} &
\multicolumn{1}{c}{$R_e$} &
\multicolumn{1}{c}{$\sigma_e$} &
\multicolumn{1}{c}{$L_K$} &
\multicolumn{1}{c}{$M_{\rm Halo}$} & 
\multicolumn{1}{c}{$\lambda_{R_e}$} &
\multicolumn{1}{c}{$\epsilon$} &
\multicolumn{1}{c}{$L_{{\rm X,gas,}R_e}$} &
\multicolumn{1}{c}{$T_{{\rm X,gas,}R_e}$} &
\multicolumn{1}{c}{Source} \\
\multicolumn{1}{c}{Name} &
\multicolumn{2}{c}{(deg)} &
\multicolumn{1}{c}{(Mpc)} &
\multicolumn{1}{c}{($''$)} &
\multicolumn{1}{c}{($\kmps$)} &
\multicolumn{1}{c}{(log $L_{\odot}$)} &
\multicolumn{1}{c}{(log $M_{\odot}$)} &
\multicolumn{1}{c}{} &
\multicolumn{1}{c}{} &
\multicolumn{1}{c}{($10^{40} \ergps$)} &
\multicolumn{1}{c}{(keV)} &
\multicolumn{1}{c}{} \\
\multicolumn{1}{c}{(1)} &
\multicolumn{2}{c}{(2)} &
\multicolumn{1}{c}{(3)} &
\multicolumn{1}{c}{(4)} &
\multicolumn{1}{c}{(5)} &
\multicolumn{1}{c}{(6)} &
\multicolumn{1}{c}{(7)} &
\multicolumn{1}{c}{(8)} &
\multicolumn{1}{c}{(9)} &
\multicolumn{1}{c}{(10)} &
\multicolumn{1}{c}{(11)} &
\multicolumn{1}{c}{(12)}  \\

\tableline

NGC1023  &    40.1001  &      39.0633   &     11.1   &  47.9    &   166.7    &     10.8   & 13.2  &     0.391   &    0.363   &    0.048 $\pm$   0.004 &      0.36 $\pm$     0.08  & 1  \\
NGC2768  &    137.906  &      60.0372   &     21.8   &  63.1    &   198.2    &     11.2   & 12.9  &     0.253   &    0.472   &    0.73 $\pm$    0.03  &      0.32 $\pm$     0.06  & 1  \\
NGC3245  &    156.827  &      28.5074   &     20.3   &  25.1    &   209.9    &     10.8   & 12.0  &     0.626   &    0.442   &    0.14 $\pm$    0.02  &      0.31 $\pm$     0.06  & 1  \\
NGC3377  &    161.927  &      13.9856   &     10.9   &  35.5    &   128.2    &     10.4   & 13.3  &     0.522   &    0.503   &    0.007 $\pm$   0.004 &      0.26 $\pm$     0.10  & 1  \\
NGC3379  &    161.957  &      12.5816   &     10.3   &  39.8    &   185.8    &     10.8   & 13.3  &     0.157   &    0.104   &    0.008 $\pm$   0.002 &      0.29 $\pm$     0.10  & 1  \\
NGC3384  &    162.070  &      12.6293   &     11.3   &  32.4    &   130.0    &     10.6   & 13.3  &     0.407   &    0.065   &    0.007 $\pm$   0.004 &      0.25 $\pm$     0.25  & 2  \\
NGC3414  &    162.818  &      27.9750   &     24.5   &  24.0    &   231.7    &     10.9   & 12.8  &     0.073   &    0.194   &    0.10 $\pm$    0.02  &      0.69 $\pm$     0.13  & 1  \\
NGC3599  &    168.862  &      18.1104   &     19.8   &  23.4    &    63.7    &     10.2   & 13.4  &     0.282   &    0.080   &    0.019 $\pm$   0.017 &      0.10 $\pm$     0.10  & 1  \\
NGC3607  &    169.228  &      18.0518   &     22.2   &  38.9    &   206.5    &     11.1   & 13.4  &     0.209   &    0.185   &     0.26 $\pm$    0.05 &      0.41 $\pm$     0.12  & 1  \\
NGC3608  &    169.246  &      18.1485   &     22.3   &  29.5    &   169.0    &     10.7   & 13.4  &     0.043   &    0.19    &     0.36 $\pm$    0.06 &      0.39 $\pm$     0.07  & 2  \\
NGC3665  &    171.182  &      38.7628   &     33.1   &  30.9    &   216.3    &     11.3   & 12.0  &     0.410   &    0.216   &     0.92 $\pm$    0.16 &      0.32 $\pm$     0.09  & 1  \\
NGC3945  &    178.307  &      60.6756   &     23.2   &  28.2    &   182.8    &     11.0   & -     &     0.524   &    0.09    &     0.20 $\pm$    0.04 &      0.32 $\pm$     0.06  & 1  \\
NGC4026  &    179.855  &      50.9617   &     13.2   &  20.4    &   190.1    &     10.5   & 14.2  &     0.548   &    0.556   &    0.015 $\pm$   0.005 &      0.51 $\pm$     0.36  & 1  \\
NGC4036  &    180.362  &      61.8957   &     24.6   &  28.8    &   188.8    &     11.1   & -     &     0.685   &    0.555   &     0.188 $\pm$  0.037 &      0.49 $\pm$     0.24  & 1  \\
NGC4111  &    181.763  &      43.0654   &     14.6   &  12.0    &   158.9    &     10.6   & 14.2  &     0.619   &    0.582   &     0.330 $\pm$  0.019 &      0.61 $\pm$     0.06  & 1  \\
NGC4203  &    183.771  &      33.1972   &     14.7   &  29.5    &   129.1    &     10.6   & -     &     0.305   &    0.154   &    0.06 $\pm$     0.01 &      0.29 $\pm$     0.08  & 1  \\
NGC4251  &    184.535  &      28.1753   &     19.1   &  19.5    &   137.1    &     10.8   & 14.2  &     0.584   &    0.508   &    0.030 $\pm$   0.014 &      0.25 $\pm$     0.15  & 1  \\
NGC4261  &    184.847  &      5.82471   &     30.8   &  38.0    &   265.5    &     11.3   & 15.0  &     0.085   &    0.222   &      3.282 $\pm$ 0.045 &      0.79 $\pm$     0.06  & 1  \\
NGC4278  &    185.028  &      29.2806   &     15.6   &  31.6    &   212.8    &     10.8   & 14.2  &     0.178   &    0.103   &     0.154 $\pm$  0.006 &      0.32 $\pm$     0.13  & 1  \\
NGC4342  &    185.913  &      7.05394   &     23.0   &   6.6    &   242.1    &     10.4   & 15.0  &     0.528   &    0.442   &    0.041 $\pm$   0.006 &      0.57 $\pm$     0.12  & 1  \\
NGC4365  &    186.118  &      7.31752   &     23.3   &  52.5    &   221.3    &     11.2   & 15.0  &     0.088   &    0.254   &     0.42 $\pm$    0.04 &      0.49 $\pm$     0.09  & 1  \\
NGC4374  &    186.266  &      12.8870   &     18.5   &  52.5    &   258.2    &     11.3   & 15.0  &     0.024   &    0.147   &      3.866 $\pm$ 0.053 &      0.75 $\pm$     0.06  & 1  \\
NGC4382  &    186.350  &      18.1911   &     17.9   &  66.1    &   179.1    &     11.3   & 15.0  &     0.163   &    0.202   &     0.629 $\pm$  0.023 &      0.4 $\pm$      0.07  & 1  \\
NGC4406  &    186.549  &      12.9460   &     16.8   &  93.3    &   190.5    &     11.4   & 15.0  &     0.052   &    0.211   &      9.98 $\pm$   1.28 &      0.81 $\pm$     0.06  & 2  \\
NGC4459  &    187.250  &      13.9786   &     16.1   &  36.3    &   158.1    &     10.8   & 15.0  &     0.438   &    0.148   &     0.166 $\pm$  0.013 &      0.38 $\pm$     0.08  & 2  \\
NGC4473  &    187.454  &      13.4293   &     15.3   &  26.9    &   186.6    &     10.7   & 15.0  &     0.229   &    0.421   &    0.049 $\pm$   0.007 &      0.34 $\pm$      0.1  & 1  \\
NGC4477  &    187.509  &      13.6364   &     16.5   &  38.9    &   148.9    &     10.8   & 15.0  &     0.446   &    0.135   &     0.615 $\pm$  0.021 &      0.33 $\pm$     0.08  & 1  \\
NGC4494  &    187.850  &      25.7750   &     16.6   &  49.0    &   150.0    &     11.2   & 14.2  &     0.212   &    0.173   &    0.014 $\pm$    0.13 &      0.25  $\pm$    0.48  & 1  \\
NGC4526  &    188.513  &      7.69914   &     16.4   &  44.7    &   208.9    &     11.1   & 15.0  &     0.453   &    0.361   &     0.50 $\pm$    0.11 &      0.29  $\pm$    0.02  & 2  \\
NGC4552  &    188.916  &      12.5560   &     15.8   &  33.9    &   224.4    &     10.9   & 15.0  &     0.049   &    0.047   &      2.21 $\pm$   0.81 &      0.6 $\pm$      0.01  & 2  \\
NGC4596  &    189.983  &      10.1760   &     16.5   &  38.9    &   125.6    &     10.8   & -     &     0.639   &    0.254   &     0.15 $\pm$   0.014 &      0.25  $\pm$    0.12  & 1  \\
NGC4636  &    190.708  &      2.68778   &     14.3   &  89.1    &   181.6    &     11.0   & 13.7  &     0.036   &    0.094   &      11.34 $\pm$ 0.052 &      0.73  $\pm$    0.06  & 1  \\
NGC4697  &    192.150  &     -5.80085   &     11.4   &  61.7    &   169.4    &     10.9   & 13.6  &     0.322   &    0.447   &     0.232 $\pm$  0.036 &      0.32  $\pm$    0.04  & 2  \\
NGC4710  &    192.412  &      15.1655   &     16.5   &  30.2    &   104.7    &     10.6   & 13.5  &     0.652   &    0.699   &    0.030 $\pm$   0.009 &      0.14  $\pm$    0.13  & 1  \\
NGC5198  &    202.548  &      46.6708   &     39.6   &  24.0    &   201.4    &     11.0   & 10.6  &     0.061   &    0.146   &     0.182 $\pm$  0.173 &      0.45  $\pm$     0.3  & 1  \\
NGC5422  &    210.175  &      55.1645   &     30.8   &  21.4    &   157.4    &     10.7   & 13.6  &     0.600   &    0.604   &    0.017 $\pm$   0.017 &      0.15  $\pm$     0.8  & 2  \\
NGC5576  &    215.265  &      3.27105   &     24.8   &  21.9    &   155.2    &     10.9   & 12.3  &     0.102   &    0.306   &    0.036 $\pm$   0.042 &      0.49  $\pm$     0.4  & 2  \\
NGC5813  &    225.297  &      1.70197   &     31.3   &  57.5    &   210.9    &     11.4   & 13.7  &     0.071   &    0.170   &      32.07 $\pm$  0.37 &      0.71  $\pm$    0.06  & 1  \\
NGC5838  &    226.359  &      2.09936   &     21.8   &  25.1    &   276.1    &     11.0   & 13.7  &     0.521   &    0.361   &     0.332 $\pm$  0.037 &      0.4  $\pm$      0.1  & 1  \\
NGC5846  &    226.622  &      1.60564   &     24.2   &  58.9    &   223.4    &     11.4   & 13.7  &     0.032   &    0.062   &      13.79 $\pm$  0.31 &      0.72  $\pm$    0.06  & 1  \\
NGC5866  &    226.623  &      55.7633   &     14.9   &  36.3    &   157.0    &     10.9   & 12.4  &     0.319   &    0.566   &     0.1210 $\pm$ 0.008 &      0.35  $\pm$    0.09  & 1  \\

\tableline\tableline

\end{tabular}
\end{center}
{\bf Notes:-} $^{1}$Common galaxy name;
$^{2}$Galaxy co-ordinates;
$^{3}$Luminosity distance in megaparsecs;
$^{4}$Optical effective radius in arcseconds\citep{Cappellari:2013ab};
$^{5}$Velocity dispersion measured at $R_e$ in \kmps \citep{Cappellari:2011aa};
$^{6}$Logarithm of $K$-band luminosity in units of solar luminosities \citep{Cappellari:2011aa};
$^{7}$Logarithm of halo mass in units of solar masses \citep{crooketal2007};
$^{8}$2-D galaxy rotation parameter;
$^{9}$Ellipticity, such that $\epsilon = 1 - b/a$;
$^{10}$Gas X-ray luminosity measured with apertures of $R_e$ in units of $10^{40} \ergps$;
$^{11}$Gas X-ray temperature in units of keV;
$^{12}$Source for X-ray measurements: (1) this work; (2) \cite{Su:2015aa}.\\
\normalsize
\end{table*}

\end{document}